\documentclass[twocolumn]{aastex61}
\usepackage{amsmath}
\usepackage{amssymb}

\newcommand{\mpcr}{{\rm m}\mbox{{\tiny +}}{\rm cr}}
\newcommand{\icpoc}{{\rm ic}\mbox{{\tiny +}}{\rm oc}}
\received{July 28, 2016}
\revised{April 13, 2017}
\shorttitle{Mercury's Internal Structure}
\shortauthors{Margot et al.}

\begin{document}

\title{Mercury's Internal Structure}

\correspondingauthor{Jean-Luc Margot}
\email{jlm@epss.ucla.edu}

\author[0000-0001-9798-1797]{Jean-Luc Margot}
\affiliation{Department of Earth, Planetary, and Space Sciences, University of California, Los Angeles, CA 90095, USA}
\affiliation{Department of Physics and Astronomy, University of California, Los Angeles, CA 90095, USA}

\author{Steven~A.~Hauck, II}
\affiliation{Department of Earth, Environmental, and Planetary Sciences, Case Western Reserve University, Cleveland, OH 44106, USA}

\author{Erwan~Mazarico}
\affiliation{Planetary Geodynamics Laboratory, NASA Goddard Space Flight Center, Greenbelt, MD 20771, USA}

\author{Sebastiano~Padovan}
\affiliation{German Aerospace Center, Institute of Planetary Research, Berlin, 12489, Germany}

\author{Stanton~J.~Peale}
\altaffiliation{Deceased 14 May 2015}
\affiliation{Department of Physics, University of California, Santa Barbara, CA 93106, USA}

\begin{abstract}

  We describe the current state of knowledge about Mercury's interior structure.  We review the available observational constraints, including mass, size, density, gravity field, spin state, composition, and tidal response.  These data enable the construction of models that represent the distribution of mass inside Mercury.  In particular, we infer radial profiles of the pressure, density, and gravity in the core, mantle, and crust.  We also examine Mercury's rotational dynamics and the influence of an inner core on the spin state and the determination of the moment of inertia.  Finally, we discuss the wide-ranging implications of Mercury's internal structure on its thermal evolution, surface geology, capture in a unique spin-orbit resonance, and magnetic field generation.

\end{abstract}

\keywords{Mercury, interior structure, spin state, gravity field, tidal response}

\setcitestyle{notesep={; }}

\section{Introduction}
\subsection{Importance of planetary interiors}
We seek to understand the interior structures of planetary bodies
because the interiors affect planetary properties and processes in
several fundamental ways.  First, a knowledge of the interior informs
us about a planet's makeup and enables us to test hypotheses related
to planet formation.  Second, interior properties dictate the thermal
evolution of planetary bodies and, consequently, the history of
volcanism and tectonics on these bodies.  Many geological features are
the surface expression of processes that take place below the surface.
Third, the structure of the interior and the nature of the
interactions among inner core, outer core, and mantle have a
profound influence on the evolution of the spin state and the response
of the planet to external forces and torques.  These processes dictate
the planet's tectonic and insolation regimes and also affect its
overall shape.  Finally, interior properties control the generation of
planetary magnetic fields, and, therefore, the development of
magnetospheres.

Four of the six primary science objectives of the {MESSENGER}
mission~\citep{solo01} rely on an understanding of the planet's
interior structure.  These four mission objectives pertain to the high
density of Mercury, its geologic history, the nature of its magnetic
field, and the structure of its core.

\subsection{Objectives} 
An ideal representation of a planetary interior would include the
description of physical and chemical quantities at every location
within the volume of the planetary body at every point in time.  Here,
we focus on a description of Mercury's interior at the current epoch.
For a description of the evolution of the state of the planet over
geologic time, see Chapter 19.  Because our ability to
specify properties throughout the planetary volume is limited, we
simplify the problem by assuming axial or spherical symmetry.
Specifically, we seek self-consistent depth profiles of density,
pressure, and temperature, informed by observational constraints
(radius, mass, moment of inertia, composition).  The solution requires
the use of equations of state and assumptions about material properties,
both
guided by laboratory data.  We compute the bulk modulus and
thermal expansion coefficient as part of the estimation process, and
we use the profiles to compute other rheological properties, such
as viscosity and additional elastic moduli.
Finally, we use our models to numerically evaluate the planet's tidal
response and compare it with observational data.
Our models of the interior structure are relevant to a wide range of
problems, but Mercury's unusual insolation and thermal patterns
violate our symmetry assumptions.  These assumptions must be lifted
for certain applications that require precise temperature distributions.

Our primary objective is to provide a family of simplified models of
Mercury's interior that satisfy the currently available observational
constraints.  A secondary objective is to select, among these models,
a recommended model that matches all available constraints.  This
model may be considered a Preliminary Reference Mercury Model (PRMM),
evoking a distant connection with its venerable Earth
analog~\citep{dzie81}.

\subsection{Available observational constraints}
All of our knowledge about Mercury comes from Earth-based
observations, three Mariner 10 flybys, three MESSENGER flybys, and the
four-year orbital phase of the MESSENGER mission.  In the absence of
seismological data, our information about the interior comes primarily
from geodesy, the study of the gravity field, shape, and spin state of
the planet, including solid-body tides.  We will also draw on
constraints derived from the surface expression of global contraction
and observations of surface composition, with the caveat that the
composition at depth may be substantially different from that inferred
for surface material.  The structure of the magnetic field and its
dynamo origin can also be used to inform interior models.

\subsection{Outline}
The primary observational constraints
(Sections~\ref{sec-rot}--\ref{sec-spin}) are used to develop two- and
three-layer structural models (Sections~\ref{sec-2}).  We then add
compositional constraints (Section~\ref{sec-comp}) and develop
multi-layer models (Section~\ref{sec-N}).  We examine the tidal
response of the planet (Section~\ref{sec-tides}) and the influence of
an inner core (Section~\ref{sec-inner}).  We conclude with a
discussion of a representative interior model (Section~\ref{sec-prmm})
and implications (Section~\ref{sec-implications}).

\section{Rotational dynamics}
\label{sec-rot}
In his classic 1976 paper, Stanton J. Peale described the effects of a molten core
on the dynamics of Mercury's rotation and proposed an ingenious method
for measuring the size and state of the core~\citep{peal76}.  Most of
our knowledge about Mercury's interior structure can be traced to
Peale's ideas and to the powerful connection between dynamics and
geophysics.  We review aspects of Mercury's rotational dynamics that
are relevant to determining its interior structure.  \citet{peal88}
provided a more extensive review.

\subsection{Spin-orbit resonance}
Radar observations by \citet{pett65} revealed that the spin period of
Mercury differs from its orbital period.  To explain the radar
results, \citet{colo65} correctly hypothesized that Mercury rotates on
its spin axis three times for every two revolutions around the Sun.
Mercury is the only known planetary body to exhibit a 3:2 spin-orbit
resonance~\citep{colo66,gold66aj}.

\subsection{Physical librations}
Peale's observational procedure allows the detection of a molten core
by measuring deviations from the mean resonant spin rate of the
planet.  As Mercury follows its eccentric orbit, it experiences
periodically reversing torques due to the gravitational influence of
the Sun on the asymmetric shape of the planet.  The torques affect
the rotational angular momentum and cause small deviations of the spin
frequency from its resonant value of 3/2 times the mean orbital
frequency.  The resulting oscillations in longitude are called
physical librations, not to be confused with optical librations, which
are the torque-free oscillations of the long axis of a uniformly
spinning body about the line connecting it to a central body.  Because
the forcing and rotational response occur with a period $P \sim$88
days dictated by Mercury's orbital motion, these librations have been
referred to as forced librations.  This terminology is not universally
accepted \citep[e.g.,][]{bois95} and
loses meaning 
when the amount of angular momentum exchanged between spin and orbit
is not negligible \citep[e.g.,][]{naid15}.  We will instead refer to
these librations as 88-day librations, in part to distinguish them
from librations with longer periods.

The amplitude $\phi_0$ of the 88-day librations for a solid Mercury
can be written as~\citep{peal72,peal88}
\begin{equation}
    \phi_0 = \frac{3}{2}\frac{(B-A)}{C} \left(1-11e^2+\frac{959}{48}e^4+... \right),
\label{eq-phi}
\end{equation}
where $A < B < C$ are principal moments of inertia and $e$ is the
orbital eccentricity, currently $\sim$0.2056~\citep[e.g.,][]{star15cmda}.
This equation encapsulates the fact that the gravitational torques are
proportional to the difference in equatorial moments of inertia
$(B-A)$.  The polar moment of inertia $C$ appears in the denominator
as it represents a measure of the resistance to changes in rotational
motion.
If the mantle is decoupled from a molten core that does not
participate in the 88-day librations, then the moment of inertia in
the denominator must be replaced by $C_{\mpcr}$, the value appropriate
for the mantle and crust.  \citet{peal76} noted that $C_{\mpcr}/C \simeq 0.5$,
suggesting that a measurement of the amplitude of the 88-day
librations can be used to determine the state of the core if $(B-A)$
is known.  This result holds over a wide range of core-mantle coupling
behaviors~\citep{peal02, ramb07}.

\subsection{Cassini state}

\citet{peal69,peal88} formulated general equations for the motion of
the rotational axis of a triaxial body under the influence of
gravitational torques.  He wrote these equations in the context
of an orbit that precesses at a fixed rate around a reference plane
called the {\em Laplace plane}, extending and refining earlier work by
\citet{colo66}.
These equations generalize Cassini's laws and describe the dynamics of
the Moon, Mercury, Galilean satellites, and other bodies.  In the case
of Mercury, the gravitational torques are due to the Sun, and the
$\sim$300\,000-year precession of the orbit is due to the effect of
external perturbers, primarily Jupiter, Venus, Saturn, and Earth.

\begin{figure}[phtb]
  \centering
          \includegraphics[width=\columnwidth]{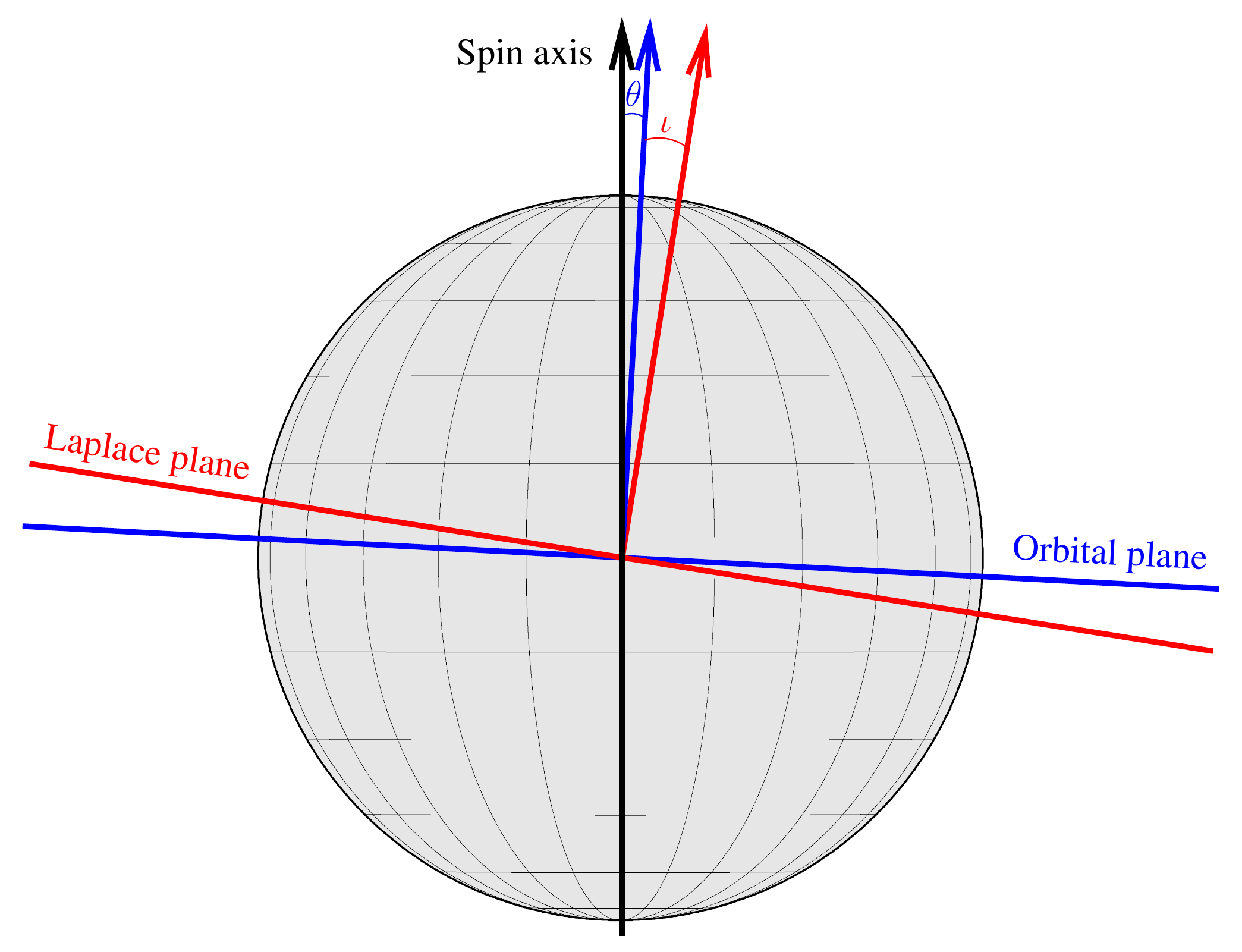}
	\caption{Geometry of Cassini state 1: the three vectors
          representing spin axis orientation (black), normal to the
          orbital plane (blue), and normal to the Laplace plane (red)
          remain coplanar as the orbit precesses around the Laplace
          plane with a $\sim$300\,000-year period.  The inclination of
          Mercury's orbit with respect to the Laplace plane is
          represented by the angle $\iota$, which is shown to scale.
          The tilt of Mercury's spin axis with respect to the orbit
          normal is the obliquity $\theta$, which is shown with an
          exaggeration factor of 100 for clarity.}
\label{fig-axes} 
\end{figure}

On the basis of these theoretical calculations, \citet{peal69,peal88}
predicted that tidal evolution would carry Mercury to a Cassini state,
in which the spin axis orientation, orbit normal, and normal to the
Laplace plane remain coplanar (Figure~\ref{fig-axes}).
Specifically, he predicted that Mercury would reach Cassini state 1,
with an obliquity near zero degrees.
Numerical simulations~\citep{bill05,yseb06,peal06,bois07} and analytical
calculations~\citep{dhoe08} support these predictions.

In a Cassini state, the obliquity has evolved to a value where the
spin precession period matches the orbit precession
period~\citep{glad96}.  Because the spin precession period and the
gravitational torques depend on moment of inertia differences, there
is a powerful relationship between the obliquity of a body in a
Cassini state and its moments of inertia.  \citet{peal76,peal88} wrote
\begin{equation}
K_1(\theta)\left( \frac{C-A}{C}\right) + K_2(\theta)\left(\frac{B-A}{C} \right) = K_3(\theta),
\label{eq-ki}
\end{equation}
where $K_1,K_2,K_3$ are functions of the obliquity $\theta$ that
involve the orbital eccentricity, inclination with respect to the
Laplace plane, mean motion, spin rate, and precession rate.  In
this equation, the appropriate moment of inertia in the denominator is
that of the entire planet, even if the core is molten, because it is
hypothesized that the core follows the mantle on the $\sim$300\,000-year
time scale of the orbital precession.

If we can confirm that Mercury is in a Cassini state, a measurement of
the obliquity becomes extremely valuable: it provides a direct
constraint on moment of inertia differences and, in combination with
degree-2 gravity information, on the polar moment of inertia.  A free
precession of the spin axis about the Cassini state could, in
principle, compromise the determination of the obliquity.  However,
such free precession would require a recent excitation because the
corresponding damping timescale is $\sim$10$^5$ y~\citep{peal05}.

\subsection{Polar moment of inertia}
\label{sec-moi}

Absent seismological data, the polar moment of inertia is arguably the
most important quantity needed to quantify the interior structure of
a planetary body.  \citet{peal76,peal88} showed that it is possible to
measure the polar moment of inertia $C$ by combining the obliquity
with two quantities related to the gravity field.  The gravity field
of a body of mass $M$ and radius $R$ can be described with spherical
harmonics~\citep[e.g.,][]{kaul00}.  The second-degree coefficients
$C_{20}$ and $C_{22}$ in
the spherical harmonic expansion are related to the moments of
inertia, as follows:
\begin{equation}
    C_{20}=-\frac{(C-(A+B)/2)}{MR^2},
\label{eq-c20}
\end{equation}
\begin{equation}
    C_{22}=\frac{(B-A)}{4MR^2}.
\label{eq-c22}
\end{equation}
Combining equations (\ref{eq-ki}), (\ref{eq-c20}), and (\ref{eq-c22}), we find
\begin{equation}
  \frac{C}{MR^2} = (-C_{20} + 2 C_{22}) \frac{K_1(\theta)}{K_3(\theta)}  +
4C_{22} \frac{K_2(\theta)}{K_3(\theta)},
\label{eq-moi}
\end{equation}
which provides a direct relationship between the obliquity, gravity
harmonics, and polar moment of inertia for bodies in Cassini state
1.

To complete Peale's argument, we determine the polar moment of inertia
of the core, which can be done if the core is molten and does not
participate in the 88-day librations.
To do so, we write the identity
\begin{equation}
\frac{C_{\mpcr}}{C} = \left(\frac{C_{\mpcr}}{B-A}\right) \left( \frac{B-A}{MR^2} \right) \left( \frac{MR^2}{C} \right),
\label{eq-cmc}
\end{equation}
which yields the moment of inertia of the mantle and crust
$C_{\mpcr}$ and, therefore, the moment of inertia of the core $C_{\rm c} = C-C_{\mpcr}$.
Two spin state quantities and two gravity quantities 
provide all the information necessary to determine these values.  A
measurement of the libration amplitude $\phi_0$ provides a direct
estimate of the first factor on the right-hand side of equation
(\ref{eq-cmc}) via equation (\ref{eq-phi}).  A measurement of the gravitational harmonic $C_{22}$
provides a direct estimate of the second factor.  Measurements of the
obliquity, $C_{20}$, and $C_{22}$ yield an estimate of the third
factor via equation (\ref{eq-moi}).

The four quantities $\phi_0, \theta, C_{20}, {\rm and}\ C_{22}$ identified by
\citet{peal76,peal88} thus provide a powerful probe of the interior
structure of the planet.

\subsection{Orbital precession}
Implementing Peale's procedure requires precise knowledge of Mercury's
orbital configuration.  Whereas the mean motion and orbital
eccentricity have been determined from centuries of observations,
relatively little attention had been paid to the orientation of the
Laplace plane and the orbital precession rate.  \citet{yseb06} used a
Hamiltonian approach and numerical fits to ephemeris data to determine
these ancillary quantities.  They showed that the Laplace plane
orientation varies due to planetary perturbations on $\sim$10 ky
timescales, and they defined an {\em instantaneous Laplace plane}
valid at the current epoch for the purpose of identifying the position
of the Cassini state and interpreting spin-gravity data.

\citet{yseb06} gave the coordinates of the normal to the instantaneous
Laplace plane in ecliptic and equatorial coordinates at epoch J2000.0
as
\begin{equation}
\lambda_{\rm inst} =  66.6^{\circ}, %
\beta_{\rm inst}    = 86.725^{\circ},\\
\end{equation}
\begin{equation}
{\rm RA}_{\rm inst} = 273.72^{\circ},   
{\rm DEC}_{\rm inst} = 69.53^{\circ},       
  \label{eq-laplace}
\end{equation}
where $\lambda$ is ecliptic longitude, $\beta$ is ecliptic latitude, RA
is right ascension, and DEC is declination.  The uncertainty in the
determination is of order $1^{\circ}$, but the orientation of the
narrow error ellipse is such that it can affect the
interpretation of the spin state data only at a level that is well below
that due to measurement uncertainties.

The inclination of Mercury's orbit with respect to the instantaneous
Laplace plane and the orbit precession rate about that plane at the
current epoch are $\iota=8.6^\circ$ and
$\dot\Omega=-0.110^\circ/{\rm century}$,
respectively~\citep{yseb06}.  We will use both of these quantities to
estimate Mercury's interior structure in Sections~\ref{sec-2} and
\ref{sec-N}.  \citet{star15cmda} performed an independent analysis and
confirmed the values of \citet{yseb06}, including the orientation of
the instantaneous Laplace plane, the inclination $\iota$, and the
precession rate
$\dot\Omega$.
\citet{dhoe09} used a Hamiltonian approach
and found an instantaneous Laplace plane orientation that differs from
our preferred value by 1.4$^\circ$.

\section{Gravity constraints}
\label{sec-grav}
\subsection{Methods}

We are interested in measuring the masses and sizes of planetary
bodies because bulk density is a fundamental indicator of composition.
In multi-planet systems, masses can be estimated by observing the
effects of mutual orbital perturbations, manifested as variations in
orbital elements or variations in transit times.  Another common
mass measurement technique is to determine the orbit of natural satellites.

The most precise mass estimates are obtained by radiometric tracking
of a spacecraft while it is in close proximity to the body of
interest, typically by using the onboard telecommunications system and
a network of ground-based radio telescopes.  The geodetic observations
are then used to obtain a spherical harmonic expansion of the gravity
field and to reconstruct the spacecraft trajectory with high fidelity.
In addition to providing high-precision mass estimates, this technique
enables the measurement of the spherical harmonic coefficients
$C_{20}$ and $C_{22}$, which provide important constraints on interior
structure (Section~\ref{sec-moi}).

In the following sections, we describe gravity results obtained from tracking
the Mariner 10 spacecraft at a frequency of 2.3 GHz (S-band) during
three flybys in 1974--1975 and the MESSENGER spacecraft at frequencies
of 7.2 GHz uplink and 8.4 GHz downlink (X-band) during the flybys and
orbital phase of the mission.

\subsection{Mass and density results}
\label{sec-mass}

The mass, size, and density of Mercury were known with remarkable
precision prior to the exploration of the planet by spacecraft.  After
adding radar measurements to two centuries of optical observations,
\citet{ash71} fit planetary ephemerides and determined Mercury's mass
to 0.25\% fractional uncertainty.  They found a value of $6025000 \pm
15000$ in inverse solar masses, i.e., $M = (3.300 \pm 0.008) \times
10^{23}\ {\rm kg}$, which is almost identical to the modern estimate.
Using this measurement and the radar estimate of the average
equatorial radius that was available at the time, $R = (2\,439 \pm 1)$
km, it was apparent that Mercury's bulk density was anomalously high,
with $\rho= (5\,430 \pm 15)\ {\rm kg\, m}^{-3}$. On the basis of their
density calculation, \citet{ash71} concluded that Mercury must be
substantially richer in heavy elements than Earth.  The
pre-Mariner 10 estimates of mass, size, and density remain in excellent
agreement with the MESSENGER results, but spacecraft data have enabled
a reduction in uncertainties by a factor of $\sim$50.

\citet{howa74} analyzed the tracking data from the first flyby of
Mercury by Mariner 10 and obtained a gravitational parameter $GM =
(2.2032 \pm 0.0002) \times 10^{13} {\rm m}^{3} {\rm s}^{-2}$, where
$G$ is the gravitational constant.  Analysis of data from all three
Mariner 10 flybys yielded $GM = (2.203209 \pm 0.000091) \times 10^{13}
{\rm m}^{3} {\rm s}^{-2}$\citep{ande87}.  From more than three
years of orbital tracking data of MESSENGER, \citet{maza14} obtained
$GM = (2.203187080 \pm 0.000000086) \times 10^{13} {\rm m}^{3} {\rm
  s}^{-2}$, estimated from a gravity field solution to degree and
order 50.
An independent analysis to degree and order 40 by \citet{verm16}
yielded $GM = (2.203187404 \pm 0.000000090) \times 10^{13} {\rm m}^{3}
{\rm s}^{-2}$.  When translating the MESSENGER values to a mass
estimate, the majority of the uncertainty comes from the
$5\times10^{-5}$ uncertainty in the gravitational constant.  With $G =
(6.67408 \pm 0.00031) \times 10^{-11} {\rm m}^3 {\rm kg}^{-1} {\rm
  s}^{-2}$ \citep{mohr16},
the current best estimate of the mass of Mercury is
\begin{equation}
M=(3.301110 \pm 0.00015) \times 10^{23}\ {\rm kg}.
\end{equation}

From a combination of laser altimetry \citep{zube12} and radio
occultation data, \citet{perr15} determined Mercury's average radius
to be 
\begin{equation}
R = (2\,439.36 \pm 0.02)\ {\rm km},
\end{equation}
although the stated radius uncertainty may be optimistic given the
sparse sampling of the southern hemisphere.  
The corresponding bulk density is
\begin{equation}
  \rho = (5\,429.30 \pm 0.28)\ {\rm kg\, m}^{-3}.
\label{eq-5429}
\end{equation}

Mercury's bulk density is similar to that of Earth, $\rho_\oplus = 5514\ {\rm
  kg\, m}^{-3}$, despite the different sizes of the two bodies.  The
pressure $P$ at the center of a homogeneous sphere scales as
$P \propto \rho^2 R^2$, so materials in Earth's interior are
more compressed (i.e., denser) than those in Mercury's interior.  If
we assume that both planets are made of a combination of a light
component (i.e., silicates) and a heavy component (i.e., metals), we
can infer from their similar densities and differing sizes that
Mercury has a larger metallic component, as recognized by
\citet{ash71}.

\subsection{$C_{20}$ and $C_{22}$ results}

The first measurements of the $C_{20}$ and $C_{22}$ gravity
coefficients were obtained from Mariner 10 data recorded during one
equatorial flyby with $\sim$700 km minimum altitude and one polar
flyby with $\sim$300 km minimum altitude.  \citet{ande87} determined
$C_{20} = (-6.0 \pm 2.0) \times 10^{-5}$ and $C_{22} = (1.0 \pm 0.5)
\times 10^{-5}$.
These values have large fractional uncertainties because there were
only two favorable flybys, but the values are consistent with the most
recent MESSENGER results~\citep{maza14,verm16}.
With the normalization that is commonly used in geodetic
studies~\citep[][p.7]{kaul00}, the Mariner 10 values can also be expressed as
$\bar{C}_{20} = C_{20} / \sqrt{5} = (-2.68 \pm 0.9) \times 10^{-5}$
and $\bar{C}_{22} = C_{22} / \sqrt{5/12} = (1.55 \pm 0.8) \times
10^{-5}$, where the overbar indicates normalized coefficients.

The next opportunity for measurements arose from the three MESSENGER
flybys of Mercury in 2008--2009.  However, the equatorial geometry of
these flybys did not provide adequate leverage to measure $C_{20}$
accurately.  Because the Mariner 10 tracking data have been lost, it
was not possible to perform a joint solution including both equatorial
and polar flybys.  For these reasons, \citet{smit10} cautioned that
their recovery of $\bar{C}_{20} = (-0.86 \pm 0.30) \times 10^{-5}$
might not be reliable.  However, the equatorial geometry was suitable
for an accurate estimate of $\bar{C}_{22} = (1.26 \pm 0.12) \times
10^{-5}$.

Data acquired during the orbital phase of the MESSENGER mission
provided significantly better sensitivity and lower uncertainties.
\citet{smit12} analyzed the first six months of data ($>$300 orbits)
and found $\bar{C}_{20} = (-2.25 \pm 0.01) \times 10^{-5}$ and
$\bar{C}_{22} = (1.25 \pm 0.01) \times 10^{-5}$, where the error bars
represent a calibrated uncertainty that is about 10 times the formal
uncertainty of the fit.  An independent analysis of the same data by
\citet{geno13} confirmed these results.  More recently, \citet{maza14}
analyzed three years of data (2275 orbits) and estimated a gravity
field solution to degree and order 50.  This solution yielded an
order-of-magnitude improvement in the calibrated uncertainties in
$C_{20}$ and $C_{22}$: $\bar{C}_{20} = (-2.2505 \pm 0.001) \times
10^{-5}$ and $\bar{C}_{22} = (1.2454 \pm 0.001) \times 10^{-5}$.  An
independent analysis by \citet{verm16} confirmed these values to
better than 0.4\%.

The unnormalized quantities that we use in equations
(\ref{eq-c20}--\ref{eq-cmc}) are based on the \citet{maza14} values:
$C_{20} = (-5.0323 \pm 0.0022) \times 10^{-5}$ and $C_{22} = (0.8039
\pm 0.0006) \times 10^{-5}$.  The $J_2/C_{22}=-C_{20}/C_{22}$ value of
6.26 is distinct from the equilibrium value of 7.86 for a body in a
3:2 spin-orbit resonance with the current value of the orbital
eccentricity \citep{mats09}, indicating that Mercury is not in
hydrostatic equilibrium.

\subsection{$k_2$ results}
\label{sec-k2obs}

In addition to the static gravity field, \citet{maza14} also solved
for the time-variable degree-2 potential which captures the tidal
forcing due to the Sun.  The tidal forcing is parameterized by the Love
number $k_2$ (Section~\ref{Sec__TidesGovEq}).  \citet{maza14} obtained an estimate of $k_2 = 0.451 \pm
0.014$. However, because of potential mismodeling and systematic
effects in the analysis, they could not rule out a wider range of
values ($0.43-0.50$).  The preferred value of \citet{verm16} is $k_2 =
0.464 \pm 0.023$.  They, too, encountered a wider range of best-fit
values ($0.420-0.465$) in various trials.  The weighted mean of these
two estimates is $k_2 = 0.455 \pm 0.012$.  These estimates are within
the expected range from theoretical studies
\citep{vanh03, vanh07, rivo09} and from predictions of interior models
informed by MESSENGER data and Earth-based radar data~\citep{pado14}.

\section{Spin-state constraints}
\label{sec-spin}
Most of the quantities necessary to implement Peale's method of
probing Mercury's interior were known when he wrote his paper in 1976.
The mass, size, and density had been determined to $<1\%$ precision
prior to the arrival of Mariner 10, the data from which confirmed and
improved the ground-based estimates (Section~\ref{sec-grav}).  Values
of the second-degree gravity coefficients $C_{20}$ and $C_{22}$ had
also been determined, albeit with substantial uncertainties.  In
contrast, there were no satisfactory measurements of the spin state.
Librations had not been detected, and the best spacecraft
determination of the orientation of the rotation axis had a 50\% error
ellipse of $\pm2.6^{\circ}$ by $\pm6.5^{\circ}$~\citep{klaa76}, about
three orders of magnitude short of the required precision.
\citet{peal76} speculated that measurement of the obliquity and
libration angles ($\theta$ and $\phi_0$) would ``almost certainly require rather sophisticated
instrumentation on the surface of the planet.''  Fortunately, the
measurements were obtained with Earth-based instruments as well as
instruments aboard the MESSENGER orbiter.

\subsection{Methods}

Three observational methods have been used to measure Mercury's spin
state: Earth-based radar observations, joint analysis of MESSENGER
laser altimetry tracks and stereo-derived digital terrain models, and
MESSENGER radio tracking observations.  All three yielded estimates of
Mercury's obliquity, but
only the first two have yielded libration measurements so far.
Another important distinction between these methods is that the first
two measure the spin state of the rigid outer part of the planet,
i.e., the lithosphere, whereas the gravity-based analyses are
sensitive to the rotation of the entire planet.

\begin{figure*}[phtb]
  \centering
    \includegraphics[width=6.5in]{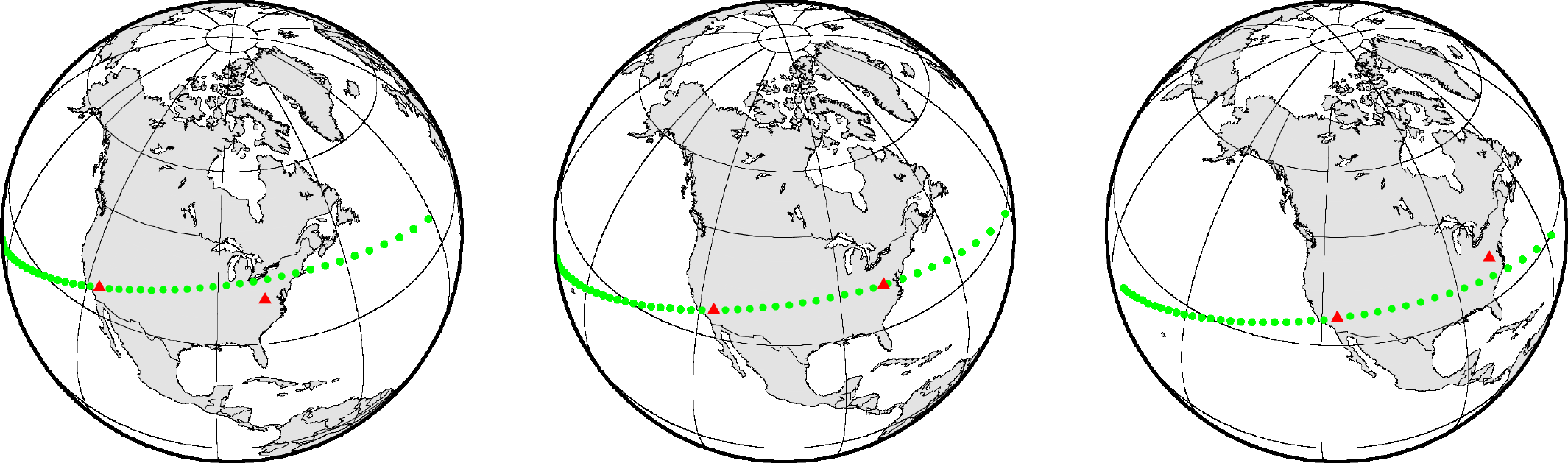}
	\caption{Radar echoes from Mercury sweep over the surface of
          the Earth.  Diagrams show the trajectory of the speckles one
          hour before (left), during (center), and one hour after
          (right) the epoch of maximum correlation.  Echoes from two
          receive stations (red triangles) exhibit a strong
          correlation when the antennas are suitably aligned with the
          trajectory of the speckles (green dots shown with a
          1-s time interval).  From \citet{marg12jgr}.}
\label{fig-path} 
\end{figure*}

The spin state of Mercury can be characterized to high precision with
an Earth-based radar technique that relies on the theoretical ideas of
\citet{holi88,holi92}.
He showed that radar echoes from solid planets can display a high
degree of correlation when observed by two receiving stations with
appropriate positions in four-dimensional space-time.  Normally each
station observes a specific time history of fluctuations in the echo
power (also known as {\em speckles}), and the signals recorded at
separate antennas do not correlate.  But during certain times on
certain days of the year, the antennas become suitably aligned with
the speckle trajectory, which is tied to the rotation of the observed
planet (Figure~\ref{fig-path}).
During these brief ($\sim$10--20 s) time intervals a cross-correlation
of the two echo time series yields a high score at a certain value of
the time lag ($\sim$5--10 s).  The {\em epoch} at which the high
correlation occurs provides a strong constraint on the orientation of
the spin axis.  The {\em time lag} at which the high correlation
occurs provides a direct measurement of the spin rate.
\citet{marg07,marg12jgr} illuminated Mercury with monochromatic
radiation (8560 MHz, 450 kW) from the Deep Space Network (DSN)
70-m
antenna in Goldstone, California (DSS-14), and recorded the speckle
patterns as they swept over two receiving stations (DSS-14 and the
100-m antenna in Green Bank, West Virginia).  They obtained
measurements of the instantaneous spin state of Mercury at 35 epochs
between 2002 and 2012, from which they inferred both obliquity and
libration angles.

\citet{star15grl} combined imaging~\citep{hawk07} and laser
altimetry~\citep{cava07} data obtained by MESSENGER during orbital
operations to independently measure the spin state of Mercury.  The
basic idea is to produce digital terrain models (DTMs) from stereo
analysis of the imaging data and to coregister the laser altimetry
profiles to the DTMs~\citep{star15pss}.  During the coregistration
step, a rotational model is adjusted in a way that minimizes the
radial height differences between the two data sets.  This adjustment
enables the recovery of the spin axis orientation, which yields the
value of the obliquity.  It also enables the recovery of the amplitude
of the physical librations because the laser profiles sample the
topography of the surface at different phases of the libration cycle.
In practice, \citet{star15grl} produced 165 individual gridded DTMs
from thousands of images of the surface.  Their DTMs cover
$\sim$50\% of the northern hemisphere of Mercury with a grid spacing
of 222 m/pixel, an effective horizontal resolution of 3.8 km, and an
average height error of 60 m.
For the coregistration step, they used 2325 laser profiles from three
years of Mercury Laser Altimeter (MLA) observations.  The laser
altimetry data have a spacing between footprints that varied between
170 m and 440 m and a nominal ranging accuracy of 1 m.

The third method for estimating the spin state of Mercury is to adjust
a rotational model of the planet during analysis of the radio tracking
data (Section~\ref{sec-grav}).  \citet{maza14} and \citet{verm16}
analyzed three years of radio science data and produced estimates of
the spin axis orientation.
The detection of the physical librations with this technique is
possible, but measuring the libration amplitude accurately remains
challenging.

\subsection{Obliquity results}
\label{sec-obliquity}
Analysis of the Earth-based radar data yielded an estimate of the
obliquity $\theta = (2.042 \pm 0.08)\ {\rm arcminutes}$, where the
adopted one-standard-deviation uncertainty corresponds to 5
arcseconds~\citep{marg12jgr}.  Remarkably, the analysis of the
spacecraft imaging and laser altimetry data, a completely independent
data set, yielded an almost identical (0.6\%) estimate of $(2.029 \pm
0.085)\ {\rm arcminutes}$, with similar
uncertainties~\citep{star15grl}.  The weighted mean of these two
estimates is $\theta = (2.036 \pm 0.058)$ arcminutes.

The best-fit spin axis orientation at epoch J2000.0 from analysis of
the radar data is at equatorial coordinates (281.0103$^\circ$,
61.4155$^\circ$) and ecliptic coordinates (318.2352$^\circ$,
82.9631$^\circ$) in the corresponding J2000 frames~\citep{marg12jgr}.
The MESSENGER DTM and laser altimetry results are within 0.8 arcseconds, at
equatorial coordinates (281.0098$^\circ$, 61.4156$^\circ$) and
ecliptic coordinates (318.2343$^\circ$,
82.9633$^\circ$)~\citep{star15grl}.

\begin{figure}[phtb]
	\centering
                \includegraphics[width=\columnwidth]{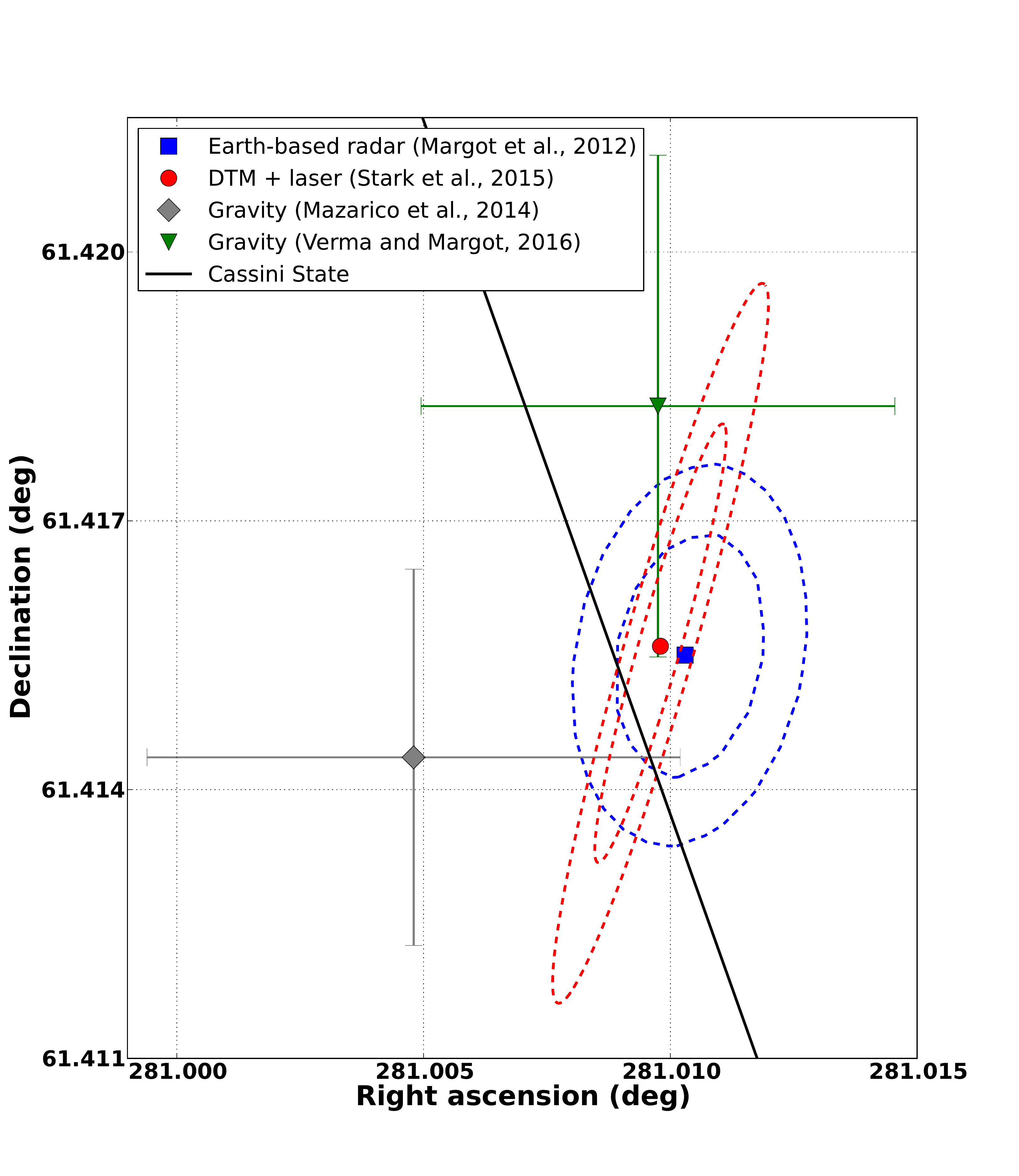}
	  \caption{Orientation of the spin axis of Mercury obtained by
          three different techniques.  The Earth-based radar results
          and the MESSENGER DTM and laser altimetry results are shown
          with contours representing the 1- and 2-standard deviation
          uncertainty regions.  The gravity results are shown with
          error bars representing the formal uncertainties of the fit
          multiplied by 10.  The oblique line shows the predicted
          location of Cassini state 1 at epoch J2000.0 from the
          analysis of \citet{yseb06}. Points to the left and right of
          the line lead and lag the Cassini state,
          respectively.}
        \label{fig-cassini} 
\end{figure}

Radio science tracking data can be used to estimate the orientation of
the axis about which Mercury's gravity field rotates, which is not
necessarily aligned with the axis about which the lithosphere rotates.
\citet{maza14} and \citet{verm16} used this technique and reported
obliquities of $(2.06 \pm 0.16)\ {\rm and\ } (1.88 \pm 0.16)\ {\rm
  arcminutes}$, respectively.  These results are consistent with those
obtained by \citet{marg12jgr} and \citet{star15grl}, 
albeit with
uncertainties that are twice as large (Figure~\ref{fig-cassini}).

\citet{marg07} provided observational evidence that Mercury is in or
very near Cassini state 1, an important condition for the success of
Peale's procedure.  The current best-fit values place the radar-based
and MESSENGER-based poles within 2.7 and 1.7 arcseconds of the Cassini
state, respectively (Figure~\ref{fig-cassini}), confirming that
Mercury closely follows the Cassini state.  There are several possible
interpretations for the imperfect agreement: (1) given the 5--6
arcsecond uncertainty in spin axis orientation, Mercury may in fact be
in the exact Cassini state, (2) Mercury may also be in the exact
Cassini state if our knowledge of the location of that state is
incorrect, which is possible because it is difficult to determine the
exact Laplace pole orientation, (3) Mercury may lag the exact Cassini
state by a few arcseconds, (4) Mercury may lead the exact Cassini
state, although this seems less likely on the basis of the evidence at
hand.  Measurements of the offset between the spin axis orientation
and the Cassini state location have been used to place bounds on
energy dissipation due to solid-body tides and core-mantle
interactions in the Moon~\citep{yode81,will01moon}.  However, the
interpretation of an offset from the Cassini state at Mercury is
complicated by the influence of various core-mantle coupling
mechanisms~\citep{peal14} and the presence of an inner
core~\citep{peal16}.

\subsection{Libration results}
\label{sec-librations}

Analysis of Earth-based radar observations obtained at 18 epochs
between 2002 and 2006 yielded measurements of Mercury's instantaneous
spin rate that revealed an obvious libration signature with a period
of 88 days~\citep{marg07}.  From these data and the Mariner 10
estimate of $C_{22}$ in equation (\ref{eq-cmc}), it was possible to
show with 95\% confidence that $C_{\mpcr}/C$ is smaller than
unity.  These results provided direct observational evidence that
Mercury has a molten outer core~\citep{marg07}.  Measurements of Mercury's
magnetic field prior to the radar observations had provided
inconclusive suggestions about the nature of Mercury's core.  A dynamo
mechanism involving motion in an electrically conducting molten outer core
was the preferred explanation~\citep{ness75,stev83rpp}, but alternative
theories that did not require a liquid core, such as remanent
magnetism in the crust, could not be ruled out~\citep{step76,ahar04}.

\begin{figure}[h!]
  \begin{center}
            \includegraphics[width=\columnwidth]{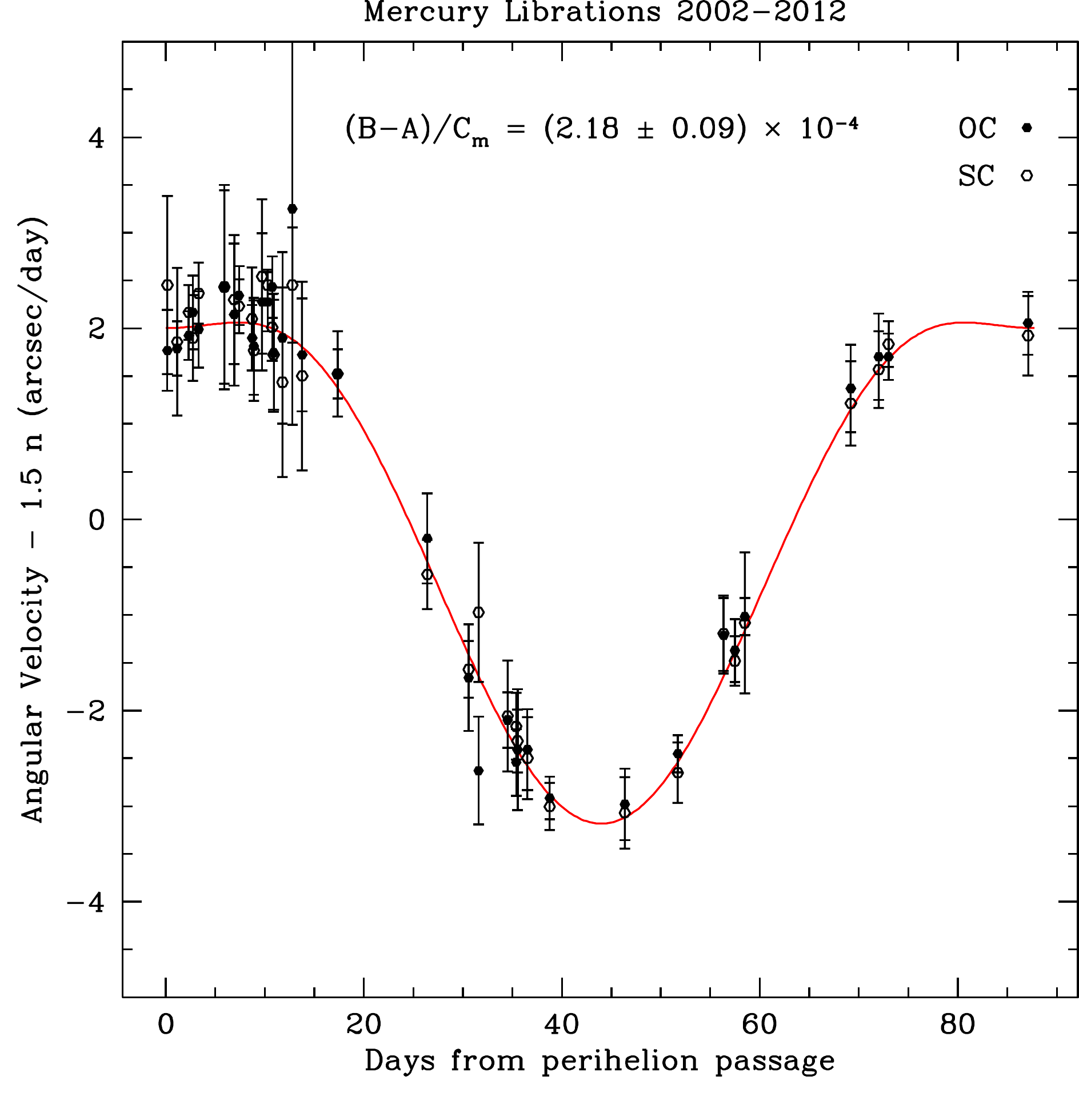}
	  \caption{Mercury 88-day librations revealed by 35
          instantaneous spin rate measurements obtained with
          Earth-based radar between 2002 and 2012.  The vertical axis
          represents deviations of the angular velocity from the exact
          resonant rate of 3/2 times the mean orbital motion $n$.  The
          measurements with their one-standard-deviation errors are
          shown in black.  OC and SC represent measurements in two
          orthogonal polarizations (opposite-sense circular and
          same-sense circular, respectively).  A numerical integration
          of the torque equation is shown in red.  The flat top on the
          angular velocity curve near pericenter is due to the
          momentary retrograde motion of the Sun in the body-fixed
          frame and corresponding changes in the torque.  The
          amplitude of the libration curve is determined by a
          one-parameter least-squares fit to the observations, which
          yields a value of $(B-A)/C_{m+cr} = (2.18 \pm 0.09) \times
          10^{-4}$. From \citet{marg12jgr}.}
        \label{fig-88d} 
  \end{center}
\end{figure}

Earth-based radar observations continued during the flyby and orbital
phases of MESSENGER.  By 2012, measurements at 35 epochs had been
obtained (Figure~\ref{fig-88d}).
One can fit a libration model~\citep{marg09cmda} to these data and
derive the value of $(B-A)/C_{\mpcr}$.
\citet{marg12jgr} found a value of $(B-A)/C_{\mpcr} = (2.18 \pm 0.09)
\times 10^{-4}$, which corresponds to a libration amplitude $\phi_0$ of $(38.5
\pm 1.6)$ arcseconds, or a longitudinal displacement at the equator of
450 m.

\citet{star15grl} analyzed three years of MESSENGER DTM and laser
altimetry data and found a libration amplitude of $(38.9 \pm 1.3)$
arcseconds, which corresponds to $(B-A)/C_{\mpcr} = (2.206 \pm 0.074)
\times 10^{-4}$.  This estimate is in excellent agreement (1\%) with
the Earth-based radar value, giving confidence in the robustness of
the results obtained by two independent techniques.  The weighted
means of these estimates are $(B-A)/C_{\mpcr} = 2.196 \pm 0.057$ and
$\phi_0 = (38.7 \pm 1.0)$ arcseconds.

\subsection{Average spin rate}

\begin{figure}[phtb]
  \centering
        \includegraphics[width=\columnwidth]{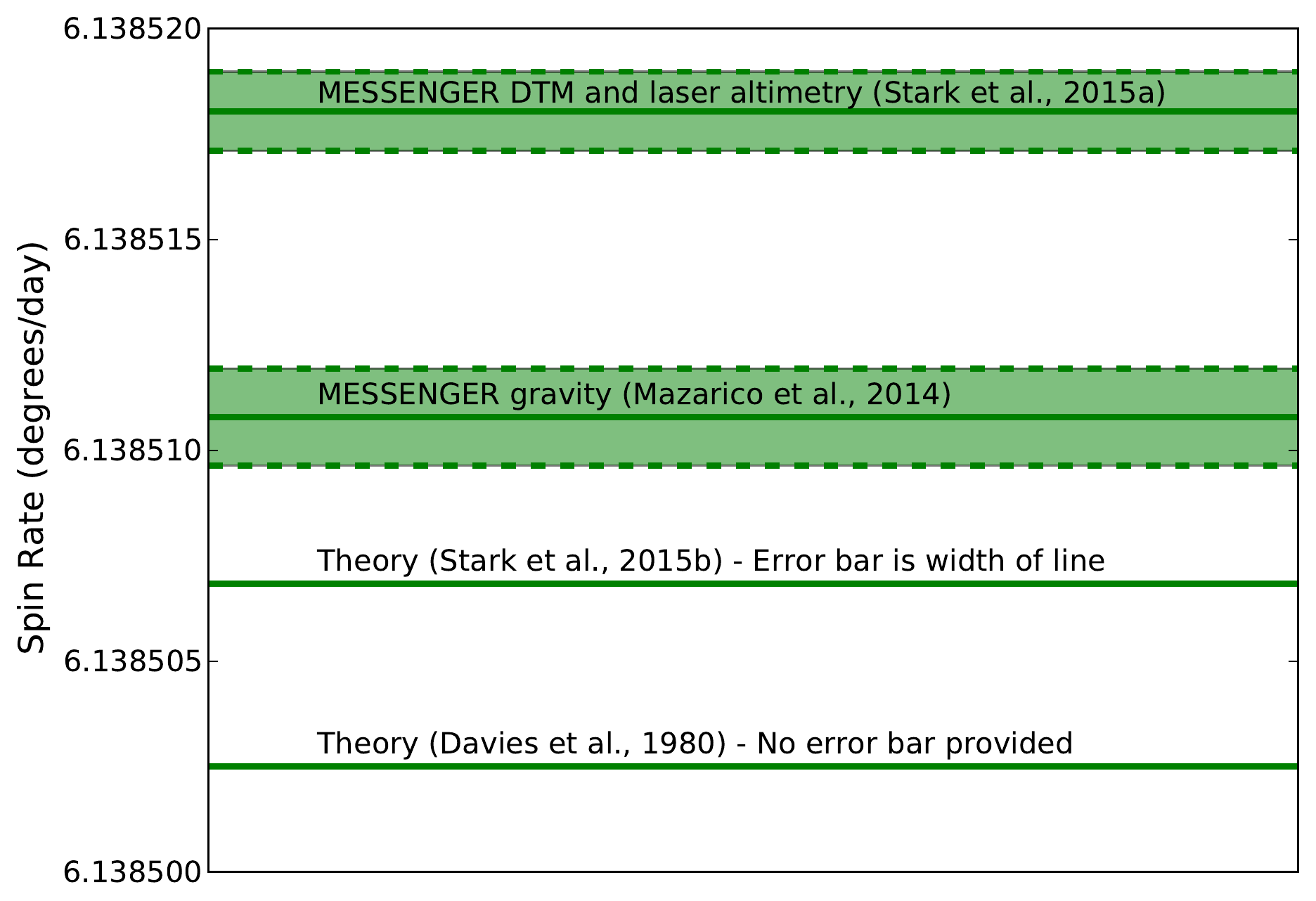}
          \caption{Theoretical and observational estimates of Mercury's mean
          resonant spin rate.  The \citet{davi80} value was
          adopted in the latest report of the International Astronomical
          Union Working Group on Cartographic Coordinates and Rotational
          Elements~\citep{arch11}.}
        \label{fig-spins}
\end{figure}

Questions remain about the precise spin behavior of Mercury, both in
terms of its average spin rate and the presence of additional
libration signatures.  There are reasons to believe that longitudinal
librations with periods of 2--20 y exist, either because of planetary
perturbations~\citep{peal07,dufe08,peal09,yseb10} or because of internal
couplings and
forcings~\citep{veas11,dumb11,vanh12,yseb13,koni13,dumb13}.  However,
the addition of long-term libration components to the rotational model
was not found to improve fits to the 2002--2012 radar
data~\citep{marg12jgr,yseb13}.  The duration of the MESSENGER data
sets is not sufficiently long to detect a long-term libration
signature, for which the primary period is expected to be $\sim$12 y.
Therefore, \citet{maza14} and \citet{star15grl} did not attempt to fit
for long-term librations.  Instead, they obtained estimates of
Mercury's average spin rate over the time span of the MESSENGER
mission.  Their estimates differ substantially from one another and
from the expected mean resonant spin rate (Fig.~\ref{fig-spins}).  One
possible explanation for the discrepancy between theoretical and
observational estimates is that the MESSENGER estimates are based on a
3- or 4-year period that represents only a small fraction of the
long-term libration cycle.

\section{Two- and three-layer structural models}
\label{sec-2}

\subsection{Governing equations}
The bulk density $\rho = M/V$ of a planetary body of mass $M$ and
volume $V$ is an important indicator of composition, but it contains
no information about the radial distribution of the material in the
interior.  Because we seek to calculate the radial density profile
$\rho(r)$, we write expressions for the mass and bulk density of a
spherically symmetric body of radius $R$ that highlight the mass contributions
from concentric spherical shells of width $dr$:
\begin{equation}
M=4\pi\int_0^R{\rho(r)r^2dr},
\label{eq-mass}
\end{equation}
\begin{equation}
\rho=\frac{3}{R^3}\int_0^R{\rho(r)r^2dr}.
\label{eq-rho}
\end{equation}
We write similar expressions for the polar moment of
inertia $C$ and its normalized value $\widetilde C$:
\begin{equation}
C=\frac{8\pi}{3}\int_0^R{\rho(r)r^4dr},
\label{eq-c}
\end{equation}
\begin{equation}
\widetilde C=\frac{C}{MR^2}=\frac{2}{\rho R^5}\int_0^R{\rho(r)r^4dr}.
\label{eq-rhomoi}
\end{equation}

We first consider a two-layer model where a mantle with constant
density $\rho_{\rm m}$ overlays a core with constant density
$\rho_{\rm c}$ and radius $R_{\rm c}$.  In a gravitationally stable
configuration, $\rho_{\rm c}>\rho_{\rm m}$.  We use equations
(\ref{eq-rho}) and (\ref{eq-rhomoi}) to derive the analytical
expressions for bulk density and normalized moment of inertia for this
two-layer model:
\begin{eqnarray}
\rho&=&\rho_{\rm c}\alpha^3 + \rho_{\rm m}\left(1-\alpha^3\right),\label{eq-rho2L}\\
\widetilde C&=&\frac{2}{5}\left[\frac{\rho_{\rm c}}{\rho}\alpha^5 + \frac{\rho_{\rm m}}{\rho}\left(1-\alpha^5 \right)\right],
\label{eq-moi2L}
\end{eqnarray}
where we have have used $\alpha = R_{\rm c}/R$ for ease of notation.
This system is underdetermined, because there are three unknowns
($\rho_{\rm c}$, $\rho_{\rm m},$ and $R_{\rm c}$) and only two
observables ($\rho$ and $\widetilde C$).  Even in the case of an
oversimplified two-layer model, it is not possible to find a solution
without making an additional assumption or securing an additional
observable.  For example, one could proceed by making an educated
guess about the density of the mantle from measurements of the
composition of the surface.  A more rigorous approach is to obtain an
additional observable that depends directly on the density of the
mantle.  We rely on Peale's procedure and the fact that Mercury is in
a Cassini state (Section~\ref{sec-obliquity}) to provide such an
observable, the polar moment of inertia of the mantle plus crust as given by
equation (\ref{eq-cmc}).  For the two-layer model, this expression
reduces to
\begin{eqnarray}
\frac{C_{\mpcr}}{C}&=&\frac{\rho_{\rm m}\left(1-\alpha^5 \right)}{\rho_{\rm c}\alpha^5 + \rho_{\rm m}\left(1-\alpha^5 \right)}.
\label{eq-cmc2L}
\end{eqnarray}

\subsection{Moment of inertia results}
\label{sec-moires}
Peale's formalism (Section~\ref{sec-moi}) enabled a determination of
Mercury's polar moment of inertia.  \citet{marg12jgr} combined
measurements of the obliquity and librations with gravity data and
found $\widetilde C=0.346 \pm 0.014$.  \citet{star15grl} also measured
$\theta$ and $\phi_0$, and found $\widetilde C=0.346 \pm 0.011$.
A uniform density sphere has $\widetilde C=0.4$, and a body with a
density profile that increases with depth has $\widetilde C<0.4$.  The
Moon, with $\widetilde C \simeq 0.393$ \citep{will96}, is nearly
homogeneous, whereas the Earth, with $\widetilde
C=0.3307$~\citep{will94}, has a substantial concentration of dense
material near the center.  Likewise, Mercury's $\widetilde C$ value
suggests the presence of a dense metallic core.

The moment of inertia of Mercury's mantle and crust is also available
from spin and gravity data (Equation~\ref{eq-cmc}).  \citet{marg12jgr}
found $C_{\mpcr}/C=0.431 \pm 0.025$ and \citet{star15grl} found
$C_{\mpcr}/C=0.421 \pm 0.021$.

Weighted means of the \citet{marg12jgr} and \citet{star15grl} results
provide the most reliable estimates to date of the moments of inertia.
We find
\begin{equation}
\widetilde C = \frac{C}{MR^2} = 0.346 \pm 0.009,
\label{eq-0346}
\end{equation}
\begin{equation}
\frac{C_{\mpcr}}{C} = 0.425 \pm 0.016.
\label{eq-0425}
\end{equation}
An error budget similar to that computed
by \citet{peal81,peal88} demonstrates that the dominant sources of
uncertainties in the moment of inertia values can be attributed to
spin quantities.  Uncertainties arising from gravitational harmonics,
tides, and orbital elements are at least an order of magnitude smaller
\citep{noye13,bala17}.  Further improvements to our knowledge of
Mercury's moments of inertia therefore require better estimates of
obliquity and libration amplitude.  Such improved estimates may also
enable a determination of the tidal quality factor $Q$ \citep{bala17}.

\subsection{Two-layer model results}
\begin{figure}[h]
   \begin{center}
     \includegraphics[width=\columnwidth]{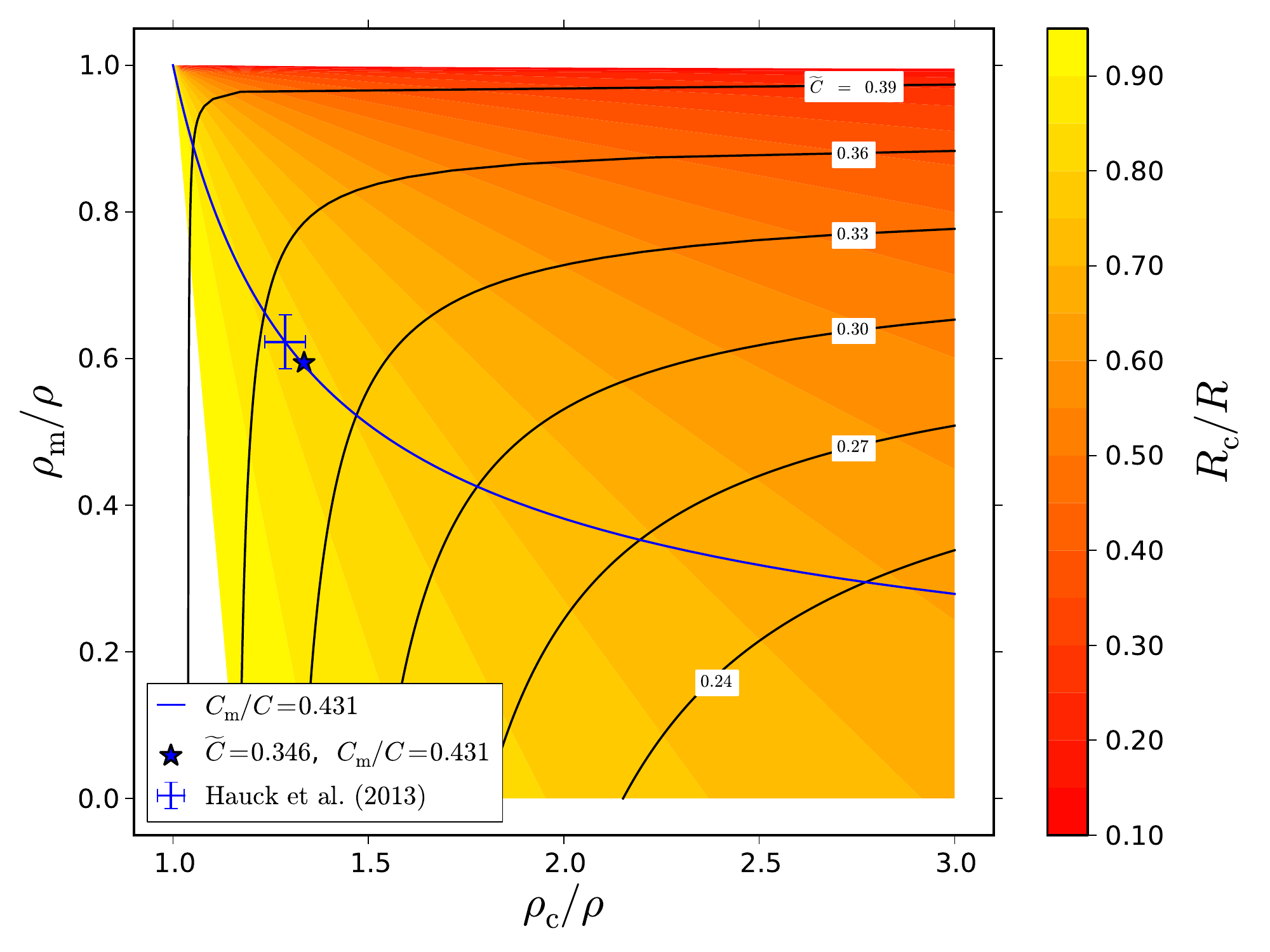}
\caption{Mantle density versus core density showing the consistency of
  the two-layer model results (star) with those of more elaborate,
  multi-layer models (error bars).  The position of the star
  corresponds to values of $\widetilde C=0.346$ and
  $C_{\mpcr}/C=0.431$~\citep{marg12jgr}.  Error bars correspond to the
  one-standard-deviation intervals for $\rho_{\rm c}/\rho$ and
  $\rho_{\rm m}/\rho$ obtained by \citet{hauc13}.  The background
  color map indicates the value $R_{\rm c}/R$ in the two-layer model.
  Black curves illustrate models with various values of the normalized
  moment of inertia $\widetilde C$.  The blue curve traces the locus
  of two-layer models with $C_{\mpcr}/C=0.431$.}
      \label{Fig__2LMod}
   \end{center}
\end{figure}

Using equations (\ref{eq-rho2L}--\ref{eq-cmc2L}) and estimates of bulk
density (\ref{eq-5429}), $\widetilde C$ (\ref{eq-0346}), and $C_{\rm
  m+cr}/C$ (\ref{eq-0425}),
we infer
\begin{align}
  R_{\rm c}/R      &=&0.8209,\ i.e., \quad R_{\rm c}   &=&2\,002\ {\rm km},\  \\ 
  \rho_{\rm c}/\rho&=&1.3344,\ i.e., \quad \rho_{\rm c}&=&7\,245\ {\rm kg\, m}^{-3},\  \\
  \rho_{\rm m}/\rho&=&0.5861,\ i.e., \quad \rho_{\rm m}&=&3\,182\ {\rm kg\, m}^{-3}.
\label{Eq__2LMerc}
\end{align}
The results obtained with the two-layer model
are within one standard deviation of the results of more elaborate,
multi-layer models that take into account mineralogical, geochemical,
and rheological constraints on the composition and physical properties
of the interior \citep[Section~\ref{sec-N}]{hauc13,rivo13}.  Figure
\ref{Fig__2LMod} illustrates the consistency of the two-layer solution
(star) and of the multi-layer models of \citet{hauc13} (error bars).
The two-layer model results are also consistent with results from
multi-layer models that consider the total contraction of the planet
\citep{knib15}.

All points shown on Figure \ref{Fig__2LMod} are consistent with
Mercury's bulk density $\rho$.  Knowledge of the normalized moment of
inertia $\widetilde C$ restricts acceptable models to a black,
constant-$\widetilde C$ curve.  The resulting degeneracy corresponds to
the underdetermined system of equations (\ref{eq-rho}) and
(\ref{eq-rhomoi}).  Knowledge of the moment of inertia of the mantle
further restricts acceptable models to the blue curve.  The intersection of the
$\widetilde C=0.346$ black curve (not shown) and of the $C_{\rm
  m+cr}/C=0.431$ blue curve yields the two-layer model solution.

Although three observables ($\rho$, $\widetilde C$, and $C_{\mpcr}/C$)
can be used to reliably estimate the parameters of a two-layer model
(core size, core density, and mantle density), they provide no
information about additional phenomena related to the origin,
evolution, and present physical state of the planet (e.g.,
mineralogical composition of the mantle, composition of the core,
presence of a solid inner core).  Additional insight can be obtained
with more elaborate three-layer and multi-layer models.

\subsection{Three-layer models}

We now consider a three-layer model with core, mantle, and crust of
density $\rho_{\rm cr}$.  We express the core and mantle radii as
fractions of the planetary radius, $\alpha = R_{\rm c}/R$ and $\beta =
R_{\rm m}/R$.  With this notation, we can write the bulk density,
moment of inertia, and the moment of inertia of the outer solid shell
as follows:
\begin{eqnarray}
\rho&=&\rho_{\rm c}\alpha^3 + \rho_{\rm m}\left(\beta^3-\alpha^3\right) + \rho_{\rm cr}\left(1-\beta^3\right),\label{eq-rho3L}\\
\widetilde C&=&\frac{2}{5}\left[\frac{\rho_{\rm c}}{\rho}\alpha^5 + \frac{\rho_{\rm m}}{\rho}\left(\beta^5-\alpha^5 \right) + \frac{\rho_{\rm cr}}{\rho}\left(1-\beta^5 \right)\right],\label{eq-rhomoi3L}\\
\frac{C_{\mpcr}}{C}&=&\frac{\rho_{\rm m}\left(\beta^5-\alpha^5 \right)+\rho_{\rm c}\left(1 -\beta^5 \right) }{\rho_{\rm c}\alpha^5 + \rho_{\rm m}\left(\beta^5-\alpha^5\right)+\rho_{\rm c}\left(1- \beta^5\right)}.\label{Eq__CmC3L}
\end{eqnarray}
This system of equations has 5 unknowns and 3 observables.  If we
assume a crustal thickness value $h_{\rm cr}$ (i.e., $\beta$) and a
crustal density value $\rho_{\rm cr}$, the system of equations
(\ref{eq-rho3L})-(\ref{Eq__CmC3L}) can be solved.  The thickness of
the crust of Mercury has been estimated from the combined analysis of
gravity and topography data \citep{maza14,pado15,jame15}.  The
density of the crust $\rho_{\rm cr}$ can be estimated
from the measured composition of the surface of Mercury \citep[e.g.,][]{pado15}.

We use the results of \citet{pado15} and consider two end-member
cases: a crust that is low-density and thin ($\rho_{\rm cr}=2\,700$
${\rm kg\, m}^{-3}$, $h_{\rm cr} = 17$ ${\rm km}$) and a crust that is
high-density and thick ($\rho_{\rm cr}=3\,100$ ${\rm kg\, m}^{-3}$,
$h_{\rm cr} = 53$ ${\rm km}$).  Compared with the two-layer model, the
inferred radius of the core is almost unaffected by the inclusion of
the crust, and the densities of the mantle and core change by less
than 1\%.  This result can be explained by the small volume of the
crust and the fact that its density is lower than that of the
underlying layers.  Consequently, the presence of the crust does not
change the values of $\rho,$ $\widetilde C$, and $C_{\rm m+cr}/C$
appreciably.

Another possible three-layer model includes a solid inner core, a
liquid outer core and a mantle.  However, the composition of the core
is not well constrained, and the system of equations
(\ref{eq-rho3L})--(\ref{Eq__CmC3L}) cannot be solved.  To make further
progress, we build multi-layer models (Section~\ref{sec-N}) that
include additional, indirect constraints from the observed composition
of the surface (Section~\ref{sec-comp}) and from assumptions about
interior properties guided by laboratory experiments.  We then
incorporate constraints that arise from the measurement of planetary
tides (Section~\ref{sec-tides}).

\section{Compositional constraints}
\label{sec-comp}

Measurements of the surface chemistry of Mercury by the MESSENGER
spacecraft have provided important information on the composition of
the
interior (e.g., Chapter 2).  Observations by the X-Ray Spectrometer
(XRS) and Gamma-Ray and Neutron Spectrometer (GRNS) instruments have
demonstrated that Mercury's surface has a
low ($<$2.5 wt~\%) abundance
of iron \citep[][Chapter 2]{nitt11, evan12, weid14}.  This surface abundance,
if also reflective of the mantle concentration of Fe~\citep{robi01},
implies that the bulk density of the mantle is only modestly higher
than those of the
magnesium end-members of the
likely major minerals, e.g., orthopyroxene enstatite
with a density of 3\,200 ${\rm kg\,
  m}^{-3}$~\citep{smyt95}.  From the application of a normative
mineralogy to the measured surface elemental
abundances~\citep{weid15}, \citet{pado15} inferred grain densities for
the crust of Mercury between 3\,000 and 3\,100 ${\rm kg\, m}^{-3}$,
a result driven primarily by the low Fe abundance.
In addition to the low surface Fe
abundance,
Mercury has relatively large concentrations of sulfur in surface
materials~\citep[Chapter 2]{nitt11}.  When taken with the Fe
observations, the measured S abundance of $\sim$1.5--2.3 wt~\% in the
crust
implies strongly chemically reducing conditions (i.e., oxygen
fugacities 2.6 to 7.3 log$_{10}$ units below the iron-w\"ustite buffer) in
Mercury's interior during the
partial melting that yielded these
materials~\citep{nitt11,mccu12,zolo13}.  This inference is consistent
with some pre-MESSENGER
expectations~\citep[e.g.,][]{wass88,burb02,mala10}.  Two consequences
of such reducing conditions are that, during global differentiation, S
is more soluble in silicate melts that later crystallize as sulfides
within the dominantly silicate material, and Si is more soluble in
metallic Fe that segregates to the core.  As a result, a wide range of
core compositions has been considered when investigating Mercury's
internal structure.  The pressure, temperature, and compositional
conditions relevant to Mercury's core have been tabulated by
\citet[][]{rivo09} and \citet{hauc13}.

As Mercury's large bulk density has long implied, the planet has a
large metallic core dominated by Fe that is likely alloyed with one
or more lighter elements.  Previous investigations focused on S as the
major alloying element
for Mercury's core~\citep[e.g.,][]{stev83icar,schu88,hard01,vanh03,hauc07,rine08,rivo09,dumb15}
because of its cosmochemical abundance and the greater availability of
thermodynamic data.
Sulfur has a strong effect on the density of Fe alloys, much greater
than silicon or carbon for a given abundance.
Additionally, S can lower the melting point of Fe alloys by hundreds
of K, which is important for maintaining a liquid outer core, and it
is relatively insoluble in solid Fe, the crystallizing phase in
Fe-rich Fe--S systems.  The latter property is important because it
leads to a nearly pure Fe inner core and an outer core that is
progressively enriched in S as a function of inner core growth.

For the most chemically reduced end-members of Mercury's inferred
interior compositions, it is likely that Si is the primary, or sole,
light alloying element in the metallic core.  Alloys of Fe and Si have
a markedly different behavior from Fe--S alloys in that they
display a solid solution
with a narrow phase loop, i.e., a
narrow region between solidus and liquidus curves at high
pressure~\citep{kuwa04}.
As a consequence, compositional differences between the potential
solids and liquids in the core are much more limited, and thus density
contrasts across the inner core boundary are smaller than for
Fe--S core compositions.  Silicon also has a smaller effect on the
density and compressibility of Fe--Si alloys than does S, with the
consequence that more Si than S is required to achieve the same
density reduction relative to pure Fe.  Data on the equation of state
of solid Fe--Si alloys are more plentiful than for liquid Fe--Si
alloys, particularly at higher pressures, though the data are
sufficient to construct models of Mercury's internal
structure~\citep{hauc13}. Due to the narrow phase loop and more
limited melting point depression induced by Si in Fe alloys
\citep[e.g.,][]{kuwa04}, inner core growth could be more extensive in
Fe--Si systems than
in S-bearing core alloys.

Over the range of inferred oxygen fugacities of 2.6 to 7.3 log$_{10}$
units below the iron-w\"ustite buffer for Mercury's interior,
an alloy of Fe with both S and Si is
likely in the core \citep{mala10,smit12,hauc13,namu16core}. Indeed, metal-silicate
partitioning experiments motivated by the surface compositions
measured by MESSENGER indicate that S and Si are likely both present
in materials that make up Mercury's core~\citep{chab14,namu16core}.
Unfortunately, data for the thermodynamic and thermoelastic properties
of ternary alloys at high pressure are more limited than for their
binary end-members.  Experiments on the behavior of super-liquidus
Fe--S--Si alloys have demonstrated large fields of two-liquid
immiscibility~\citep[e.g.,][]{sanl04,mora10} with separate S-rich and
Si-rich liquids at pressures relevant to Mercury's outermost core.
Such immiscibility, if present in Mercury's core, would lead to a
separation of phases with more S-rich liquids at the top of the core
and Si-rich liquids deeper.  In this situation, it is possible to
assume end-member behavior in two separate compositional layers within
the core and calculate properties separately for each
layer~\citep[e.g.,][]{smit12,hauc13}.  However, liquid immiscibility
in this system at higher pressures requires rather substantial amounts
of both Si and S, which may or may not be
appropriate.
Experiments by \citet{chab14} indicate a trade-off between Si and S in
Mercury's metallic core that only minimally overlaps with current
understanding of the Fe--S--Si liquid-liquid immiscibility phase
field.  Those results suggest that a mixture of Fe, S, and Si may be
more likely.  More recent work by \citet{namu16core}, however,
suggests that Mercury's core conditions may belong to the immiscible
liquid field.  In this case, Mercury's core may contain enough S for
an FeS layer that is anywhere from negligibly thin to 90 km thick,
depending on bulk S content of the planet.  Regardless, the range of
likely compositions for Mercury's core lies somewhere between an Fe-Si
end-member and a (possibly segregated) mix of Fe, Si, and S.

\section{Multi-layer structural models}
\label{sec-N}

We now wish to construct internal structure models with many layers in
order to better match the gravity, spin state, and compositional
constraints.
We extend the approach of the two- and three-layer models
(Section~\ref{sec-2}) to N-layer models with the goal of reproducing
both discontinuous and continuous variations in density with depth.
Such variations are expected on the basis of pressure-induced changes
in the density of materials.  For each material, an equation of state
(EOS) describes the density as a function of pressure, temperature,
and composition.  Pressure variations inside Mercury's core require an
EOS, but the range of pressures expected across Mercury's thin
silicate shell is relatively small.  As a result, some models do
not include an EOS for the silicate layer
\citep{hauc07,hauc13,smit12,dumb15}, although some models do
\citep{hard01,rine08,rivo09,rivo13,knib15}.
Multi-layer models provide an opportunity to reduce some of the
non-uniqueness of simpler models through application of
knowledge of the interior (e.g., potential core compositions)
\citep{hauc13,rivo13}.
They also enable investigations related to the structure of the core
\citep{hauc13,dumb15,knib15}
and the implications for the planet's thermal evolution and magnetic
field generation.

\subsection{Elements of the model}
Like two- and three-layer models, N-layer models consist of a series
of layers defined by their composition and physical state.  In contrast
to simpler models, most of the geophysically defined layers in N-layer
models are further subdivided into hundreds or thousands of sublayers.
The sublayers provide for a smoother variation of density within the
geophysically defined layers.  Sublayer properties are functionally
defined by the relevant EOS \citep{hauc13,rivo13}.

\begin{figure}[htb]
  \centering
        \includegraphics[width=\columnwidth]{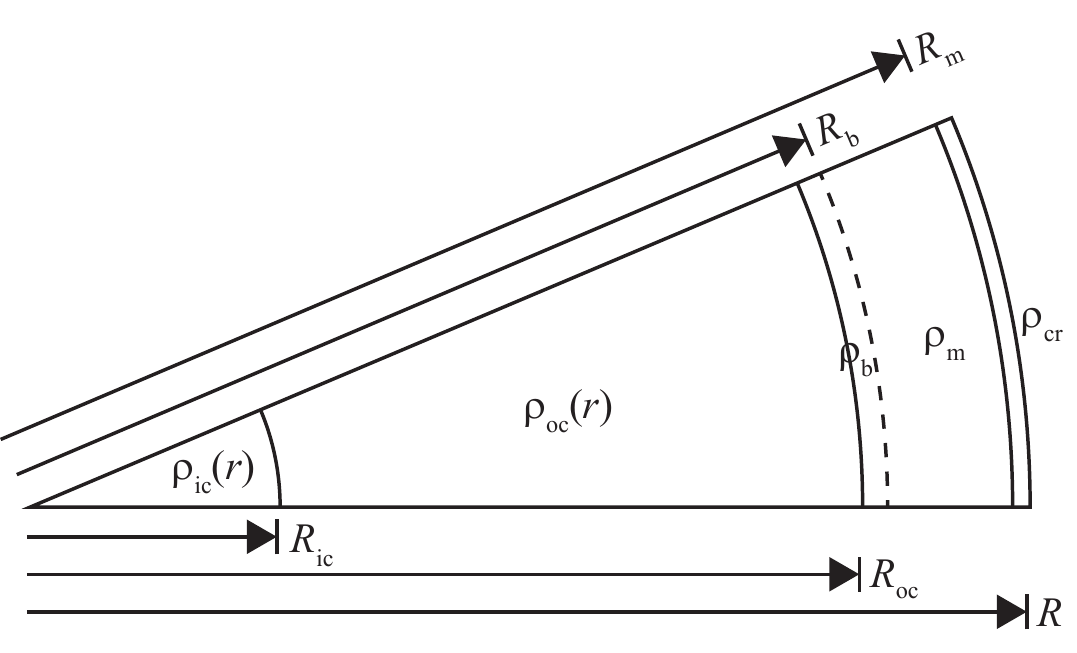}  
        \caption{Schematic representation of the internal layers of
          Mercury in models with detailed sub-layering aimed at
          capturing density variations due to changes in pressure,
          temperature, and composition with depth.  Specific radii
          mark the transitions between layers, as follows: $R_{\rm
            ic}$ between solid inner core and liquid outer core,
          $R_{\rm oc}$ between liquid outer core and the solid outer
          shell of the planet, $R_{\rm b}$ between a
          compositionally distinct
          layer at the base of the mantle and the overlying mantle,
          and $R_{\rm m}$ between mantle and crust.
          The radius of the planet is $R$.
          The radially varying densities of the inner core and outer
          core are $\rho_{\rm ic}(r)$ and $\rho_{\rm oc}(r)$,
          respectively.  The constant densities of any basal layer,
          mantle, and crust are $\rho_{\rm b}$, $\rho_{\rm m}$,
          and $\rho_{\rm cr}$, respectively.
        }
\label{fig-layers} 
\end{figure}

The basic internal organization of N-layer models is illustrated in
Figure~\ref{fig-layers}.  The metallic core is divided into a solid
inner core and a liquid outer core.  Core densities vary according to
the EOS.  The solid outer portion of the planet is divided into one or
more solid outer layers, most commonly with densities that are
constant throughout their depth extent.  Several models employ a traditional
division of the solid outer shell into a crust and a mantle
\citep{hauc13,rivo13,dumb15,knib15}.
Here, as did \citet{hauc13}, we define up to three layers within the
solid outermost portion of the planet: a basal layer at the bottom the
mantle, a mantle, and a silicate crust.
The presence of a basal layer was suggested
as a way to reconcile the low amounts of Fe observed at the planet's
surface with the high bulk density of Mercury's outer solid shell
inferred from spin and gravity data \citep{smit12,hauc13}.
Evidence for deep compensation of domical swells on Mercury
\citep{jame15}
also suggests that compositional variations deep within the solid
outer shell are present, at least regionally.

\subsection{Governing equations}
Any internal structure model for Mercury must be
consistent with three quantities: the bulk density of the planet, the
normalized moment of inertia
$\widetilde C$, and the fraction of the moment of inertia
attributed to the librating, solid outer shell of the planet
$C_{\mpcr}/C$.
This fraction is defined by 
\begin{equation}
  \frac{C_{\mpcr}}{C} + \frac{C_{\rm c}}{C} = 1,
  \label{eq-cmcrc}
\end{equation}
where $C_{\rm c}/C$ is the fraction of the moment of inertia
attributed to the core.
The moment of inertia of the core $C_{\rm c}$ is calculated from
Equation (\ref{eq-c}) integrated from the center of the planet to the
core-mantle boundary ($r=R_{\rm oc}$ in Figure~\ref{fig-layers}).
The moment of inertia of the mantle plus crust $C_{\mpcr}$ can be
determined from
integration of
Equation (\ref{eq-c}) from $r=R_{\rm oc}$ to $r=R$.

The EOSs that describe density variations with depth depend on the
pressure and temperature of the materials.  The pressure is a function
of the overburden:
\begin{equation}
      P(r) = \int_{r}^R \rho(x) g(x) dx,
  \label{eq-pressure}
\end{equation}
and depends on the local gravity
inside a sphere of radius $r$:
\begin{equation}
  g(r) = \frac{G}{r^2} M(r) = \frac{G}{r^2} 4 \pi \int_{0}^r \rho(x) x^2 dx.
  \label{eq-g}
\end{equation}

Equations (\ref{eq-pressure}) and (\ref{eq-g}) must be solved along
with Equations (\ref{eq-mass}) and (\ref{eq-c}) for the mass and polar
moment of inertia of Mercury.
Closing
the set of four equations (\ref{eq-mass}, \ref{eq-c},
\ref{eq-pressure}, \ref{eq-g}), optionally augmented by Equation
(\ref{eq-cmcrc}), requires determination of the density as of a
function of radius in the planet.  Most models of Mercury's interior
are based on a third-order Birch-Murnaghan EOS \citep{poir00}:
\begin{eqnarray}
  P(r) & = & \frac{3K_0}{2}\left[\left(\frac{\rho(r)}{\rho_0}\right)^{\frac{7}{3}}-\left(\frac{\rho(r)}{\rho_0}\right)^{\frac{5}{3}} \right] \nonumber \\ 
  & \times &   \left[ 1+\frac{3}{4} \left(K'_0-4\right)\left\{\left(\frac{\rho(r)}{\rho_0}\right)^{\frac{2}{3}}-1\right\} \right] \nonumber \\
  & + & \alpha_0 K_0 (T(r)-T_0 ),
  \label{eq-eos}
\end{eqnarray}
where $T(r), T_0, \rho_0, K_0, K'_0$, and $\alpha_0$ are the local and
reference temperatures, the reference density, the isothermal bulk
modulus and its pressure derivative, and the reference volumetric
coefficient of thermal expansion, respectively.  The density, bulk
moduli, and thermal expansivity
are parameters for which ranges are determined from laboratory experiments
and first-principles calculations.   Values 
were given by, e.g., \citet{hauc13}.
The last term on the right relates to the increase in volume with
increasing temperature.

The temperature as a function of radius can be determined for a
conductive or convective mode of heat transfer.  Most models
for Mercury's core are based on the latter assumption.
In the case of a thoroughly convective layer, the material is assumed
to follow an adiabatic temperature gradient,
\begin{equation}
	\frac{\partial T}{\partial P} = \frac{\alpha(T,P)T}{\rho(T,P) C_P},	
  \label{eq-dtdp}
\end{equation}
where $\alpha$ is the volume thermal expansion coefficient and $C_P$
is the specific heat at constant pressure. 

\subsection{Methods}

Investigations of Mercury's interior with N-layer models take the form
of a basic parameter space study.  The most fundamental parameter
decision is the choice of core alloying elements because of their
considerable influence on melting behavior (Section~\ref{sec-comp})
and because the core occupies such a large fraction of the planet.
The relative amounts of Fe and light elements
are not known,
such that broad ranges of possible core compositions tend to be
considered.  Indeed, \citet{hard01}
considered all S contents from 0 wt \% S (pure Fe) to 36.5 wt \% S
(pure FeS troilite).  Most investigations in the post-MESSENGER era
have used more limited compositional ranges.
Other parameters considered include the thickness of the crust and the
densities or density profiles of the crust and mantle.
The treatment of any crystallized solid layers within the metallic
core represents another important modeling decision. Several models
compare thermal gradients with an assumed, generally simplified, melting
curve gradient for the core alloy \citep[e.g.,][]{rivo13,dumb15}.
The intent is to develop a self-consistent prescription for the
density structure of the core that includes the appropriate EOS for
the regions of the core that are solid, liquid, or in the process of
crystallizing from the top down \citep[e.g.,][]{dumb15}.
This approach is most straightforward for Fe--S alloys because of their
well-studied
thermodynamic properties.  However, these simplified phase diagrams
tend to be based solely on eutectic compositions and do not account
for mixing behavior that may be non-ideal \citep{chen08}.
In addition, the melting relationships for Fe--Si and Fe--S--Si
compositions are not well known.  For these reasons, other studies
consider the full range of possible solid inner core sizes (from zero to the
entire core), irrespective of specific melting curves
\citep{smit12,hauc13}.

With the constraints on Mercury's interior limited to the planetary
radius, mass, and the moment of inertia parameters $C/MR^2$ and
$C_{\mpcr}/C$, knowledge of the planet's interior is necessarily non-unique.
However, through a judicious set of assumptions regarding the
composition of the interior and
an exploration of parameter space, 
it is possible to place important constraints on Mercury's internal
structure.
\citet{hauc13} and \citet{rivo13} employed Monte Carlo and Bayesian
inversion approaches, respectively, in order to estimate the structure
of Mercury's interior and to quantify the robustness of the most
probable solution.  One apparent difference in their approaches is
that
\citet{hauc13} included estimated uncertainties in the material
parameters in the EOS of core material in addition to uncertainties in
bulk density and moments of inertia, whereas \citet{rivo13} included
only the latter but considered depth-dependent density profiles for
the mantle.
Regardless of the details of the modeling and numerical approaches,
several studies have converged on a common set of fundamental outcomes
describing the internal structure of Mercury.

In assessing the agreement between interior models and observational
constraints, we use a metric based on the fractional root mean square difference, defined as
\begin{equation}
  {\rm RMS} = \left[\frac{1}{2}\sum_{i=1}^2 \left(\frac{O_i-C_i}{O_i}\right)^2\right]^{1/2},
\label{eq-fracrms}
\end{equation}
where $O$ and $C$ are observed and computed values, respectively, and
the index $i$ represents the two observables $C/MR^2$ and $C_{\rm
  m+cr}/C$.

\subsection{Results}
\label{sec-Nres}
Knowledge of the moment of inertia of a planet provides an integral
measure of the distribution of density with radius.  For Mercury,
knowledge of the fraction of the polar moment of inertia due to the
solid outer portion of the planet places further constraints on that
density distribution.  Still, taken together, the bulk density of the
planet, $C/MR^2$, and $C_{\mpcr}/C$ represent a modest set of
constraints on a body within which properties vary considerably with
depth.  As a result, N-layer models, which describe the internal
density variation more precisely than the two- and three-layer models,
are generally limited to describing a rather modest set of layers
well.
The most robust determinations include the bulk density of the solid,
outermost planetary shell that overlies the liquid portion of the
core, the bulk density of everything beneath that solid layer, and the
location of the boundary between these two layers
\citep{hauc07,hauc13,smit12,rivo13,dumb15}.
Although models based on the
moments of inertia generally do
not resolve the thickness of the crust or the density difference
between the crust and mantle, studies of gravity and topography at
higher-order harmonics do provide estimates of the crustal thickness
and its regional variations
\citep[][Chapter 3]{smit12,jame15,pado15}.

\begin{figure*}[h!]
	\centering
                \includegraphics{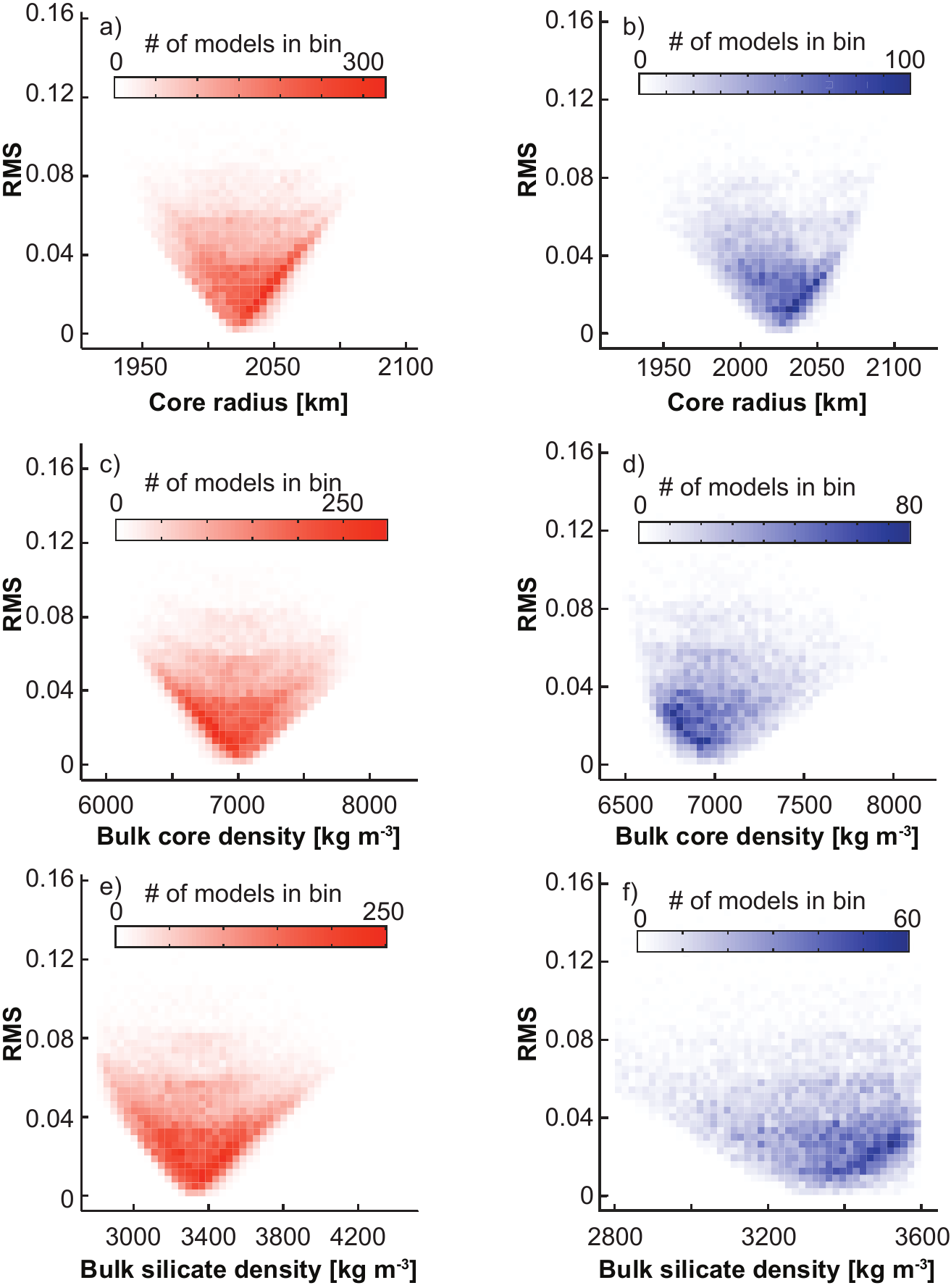}
        \caption{Two-dimensional histograms summarizing N-layer
          internal structure models of Mercury with Fe--Si core
          compositions based on the approach of \citet{hauc13} and
          current best estimates of $C/MR^2$ and $C_{\mpcr}/C$
          (Section~\ref{sec-moires}).  The left column (a, c, and e)
          represents models that include a three-layer silicate shell
          with a crust, mantle, and denser solid layer at the base
          of the mantle.  The right column (b, d, and f) represents
          models that include a two-layer silicate shell with a crust
          and mantle.  Shown are the recoveries of the radius of the
          top of the liquid outer core (a and b), bulk density of the
          metallic core (c and d), and bulk density of the silicate,
          solid, outermost shell of the planet (e and f).  The
          vertical axes show the goodness of fit expressed as a
          fractional root mean square difference (Equation
          \ref{eq-fracrms}).  Si contents of the metallic core in
          these models vary from 0 to 17 wt \% \citep{hauc13}.}
\label{fig-YY} 
\vspace{6cm}
\end{figure*}

The parameter of perhaps greatest interest regarding Mercury's
interior is the location of the boundary between the liquid outer core
and the solid outer shell.  A similar answer is obtained with a wide
variety of possible compositional models for Mercury's core: models
with both more and less S than the Fe--S eutectic composition
\citep{hauc13,rivo13,knib15},
models that include Fe--Si alloys \citep{hauc13},
and models that include combinations of S, Si, and Fe
\citep{hauc13}.
Across all these models, the top of Mercury's liquid core has
generally been estimated to be between 400 and 440 km beneath the surface
with an estimated one-standard-deviation uncertainty of less than 10\%
of that value.  Figure~\ref{fig-YY} illustrates a selection of results
for the internal structure of Mercury with the Fe--Si core composition
model results of \citet{hauc13}.
Interestingly, recent measurements of magnetic induction within
Mercury are consistent with the top of the core being 400--440 km
beneath the surface
(Chapter 5).

The bulk densities of the material above and below the transition
between the liquid core and outermost shell are also well established
across a broad range of assumed core compositions and modeling
approaches \citep[e.g.,][]{hauc13,rivo13}.
The bulk density of the core material has been found to be distributed
in the range 6\,750--7\,540 ${\rm kg\, m}^{-3}$, with central values
falling in the interval 6\,900--7\,300 ${\rm kg\, m}^{-3}$ and one-standard-deviation uncertainties of less than 5\% of the central value
(Hauck et al., 2013; Rivoldini and Van Hoolst, 2013). The bulk density
of the solid outermost shell of Mercury is distributed in the range
3\,020--3\,580 ${\rm kg\, m}^{-3}$, with central values falling in the
interval 3\,200--3\,400 ${\rm kg\, m}^{-3}$ and one-standard-deviation
uncertainties of approximately 6\% of the central value.

One of the more intriguing proposals for the structure of Mercury's
interior is the idea that a solid FeS layer could stably form at the
core-mantle boundary.  From a chemical standpoint, this layer
originates in the core and resides at the top of the core.  From a
mechanical standpoint, however, a solid layer resides at the bottom of
the mantle (Figure~\ref{fig-layers}).  The solid FeS layer hypothesis
resulted from two observations.  First, the chemically reducing
conditions observed at the surface, if pertinent to the bulk of the
planet, imply that Si will increasingly partition into the core with
decreasing oxygen fugacity.  At the pressures of Mercury's core-mantle
boundary, Fe--S--Si liquids separate into two liquid phases over a
broad range of compositions \citep{mora10}.
\citet{hauc13}
estimated from the FeS (IV) EOS that the solid phase was less dense
than the residual liquid and could float rather than sink.  Second,
the best-fitting models (e.g., those with the lowest RMS values in
Figure~\ref{fig-YY}, but not necessarily with the highest histogram values)
tend to have bulk densities for the solid outermost shell of Mercury
that are larger than $\sim$3\,200 ${\rm kg\, m}^{-3}$, the approximate
density expected for Fe-poor to Fe-absent mantle minerals such as
forsterite
and enstatite.
For these reasons,
\citet{hauc13} investigated both the situation with and without an FeS
layer.  However, the one-standard-deviation uncertainty in the outer
shell bulk density is $\sim$200 ${\rm kg\, m}^{-3}$ and permits a wide
array of possible density configurations, with and without a solid FeS
layer at the top of the core.  Furthermore, recent calculations by
\citet{knib15} question whether solid FeS is capable of floating at
the top of the core, thus potentially preventing a substantial FeS layer
from forming at the core-mantle boundary.
Additional work on the EOS of solid FeS IV at the appropriate
conditions is warranted.

\begin{figure}[htb]
\centering
\includegraphics[width=\columnwidth]{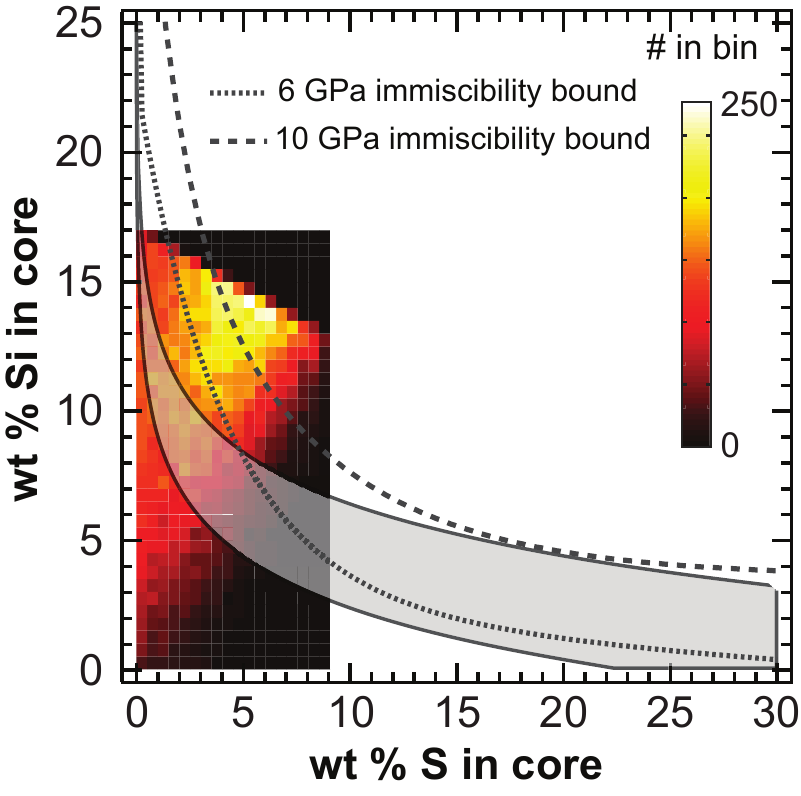}
\caption{Representation of bulk core S and Si contents in a subset of
  the internal structure models of \citet{hauc13}.  All models shown
  have an Fe--S--Si core composition and a solid FeS layer at the base
  of the mantle, and only models that match the $C/MR^2$ and $C_{\rm
    m+cr}/C$ constraints (Section~\ref{sec-moires}) are shown.  The
  two-dimensional histogram indicates the relative number of
  successful models at each bulk core composition.  Immiscibility
  limits in the Fe--S--Si system at two different pressures are shown by the dotted and dashed
  lines.  Compositions to the right of these lines result in
  immiscible Fe--S-rich and Fe--Si-rich liquids at the indicated
  pressure.  The gray region illustrates predicted bounds on core
  composition from metal-silicate partitioning experiments under the
  assumption that the S content at the surface of Mercury yields
  reliable constraints on core composition \citep{chab14}.  The lower
  and upper boundaries of the gray region represent the expected core
  compositions for Mercury-like compositions for representative
  surface S contents of 1 wt \% and 4 wt \%, respectively.}
\label{fig-ZZ} 
\end{figure}

Recently, experiments investigating the partitioning of S and Si
between silicate and metallic melts for Mercury-like compositions
\citep{chab14} have provided an opportunity to examine more closely
the nature of the core-mantle boundary region.  Figure~\ref{fig-ZZ}
illustrates a comparison of the bulk core compositions of the internal
structure models of \citet{hauc13} containing a possible solid FeS
layer at the top of the core with the predicted ranges of core
compositions compatible with MESSENGER geochemical observations of the
surface \citep{chab14}.  Also shown are the limits on compositions in
the Fe--S--Si system that display liquid-liquid immiscibility at the
relevant pressures of 6 and 10 GPa.  Compositions to the right of the
immiscibility limit curves display immiscibility and are prone to
phase separation at the given pressure.  While the majority of core
compositions in the Fe--S--Si models of \citet{hauc13} are consistent
with the segregation of
Fe--S-rich liquids at the top of Mercury's core, the general lack of
overlap of recent geochemical predictions of possible core
compositions with the immiscibility limits \citep{chab14} suggests
that liquid-liquid phase separation may not be preferred.  The further
consequence, of course, is that the conditions for crystallization of
an FeS phase at the top of the core appear less likely than the
immiscibility limits alone previously suggested.  However, as is
apparent from Figure~\ref{fig-ZZ}, the preferred core compositions of
\citet{chab14} and the most probable models that match the density and
moment of inertia parameters do not generally overlap.
There are several possible explanations for the discrepancy.  First,
it may be that the surface abundance of S cannot yield reliable
insights about core composition, either because the surface abundance
is not representative of the planet's bulk silicate composition, or
because chemical equilibrium was not satisfied during core formation.
Second, it is possible that a modeling approach not investigated so
far is required, e.g., a single, miscible Fe--S--Si liquid phase,
rather than two fully separated Fe--S and Fe--Si phases.  Third, it is
possible that the partitioning behavior observed at atmospheric
pressure by \citet{chab14} is not representative of core conditions.
Indeed, a recent geochemical experimental study with differing
silicate compositions and at slightly higher pressures \citep{namu16core}
suggests that the mantle may contain more S than the surface rocks.
In that case, the bulk core S content may be larger 
and the core conditions may belong to the immiscibility field.
However, that conclusion and the thickness of any possible FeS layer
depend strongly on Mercury's bulk S content.

Understanding the existence and size of an inner core on Mercury is a
critical goal because an inner core influences several aspects of the
planet's evolution, including magnetic field generation (Chapters 5
and 19), global contraction (Chapters 10 and 19), and rotational state
(Section~\ref{sec-inner}).  However, the size of the inner core is
difficult to quantify, for two reasons.  First, the density contrast
across the inner-outer core boundary is modest
\citep[e.g.,][]{hauc13,rivo13}.  Second, the inner core comprises only
a small fraction of the mass and density distribution of the planet.
Indeed, models with assumptions about the melting relationships of the
core can typically place only upper limits on the size of the inner
core, and these upper limits are large.  In models with core
concentrations of S exceeding a few wt \%, upper limits are $\sim$1\,450
km, i.e., $R_{\rm ic}/R \lesssim 0.6$ \citep{rivo13,dumb15,knib15}.
Upper limits as high as 1\,700--1\,800 km can be reached in models with
low core concentrations of S~\citep{rivo13,dumb15,knib15}.
Growth of an inner core to that size over the past $\sim$4 billion
years is likely incompatible with the inferred amount of global
contraction of the planet from measurements of tectonic structures on
the surface (Section~\ref{sec-inner}, Chapter 19).  Models without an
assumed core melting relationship constraint do not place strong
limits on the size of the solid inner core, although there is a slight
preference for models with an inner core radius less than $\sim$60\%
of the core radius or $\sim$50\% of the planetary
radius~\citep{hauc13}.  Additional constraints on the size of the
inner core are discussed in Section~\ref{sec-inner}.

\section{Tidal response}
\label{sec-tides}

Additional insights about Mercury's interior structure can be gained
by measuring the deformation that the planet experiences as a result
of periodic tidal forces.  These measurements are informative because
the response of a planet to tides is a function of the density,
rigidity (i.e., shear modulus), and viscosity of the subsurface
materials.  Tidal measurements have been used to support the
hypothesis of a liquid core inside Venus \citep{Konopliv1996} and Mars
\citep{Yoder2003}, and that of a global liquid ocean inside Titan
\citep{Iess2012}.
In principle, high-precision measurements of the tidal response can be
used to rule out models that are otherwise compatible with the density
and moment of inertia constraints (Section \ref{sec-N}).  When a
global liquid layer is present, the tidal response is largely
controlled by the strength and thickness of the outer solid shell
\citep[e.g.,][]{Moore2000}.  Because Mercury has a molten outer core
(Section~\ref{sec-spin}) and because the thickness of the outer solid
shell is known (Sections \ref{sec-2} and \ref{sec-N}), tidal
measurements enable investigations of the strength of the outer solid
shell.  This strength depends primarily on the mineralogy and thermal
structure of the shell.

\subsection{Tidal potential Love number $k_2$}\label{Sec__TidesGovEq} %

The tidal perturbation generated by the Sun on Mercury simultaneously
modifies the shape of the planet and the distribution of matter inside
the planet.  As a result of the redistribution of mass, solar tides
also modify Mercury's gravitational field.  From the standard
expansion of the gravitational field in spherical harmonics
\citep[e.g.,][]{kaul00}, the largest component of the tidal potential
is a degree-2 component $\Phi_2$ proportional to the mass of the Sun
and with a long axis that is aligned with the Sun-Mercury line.  The
additional potential $\phi_{2t}$ resulting from the deformation of the
planet in response to the tidal potential is parameterized by the tidal
potential Love number $k_2$:
\begin{equation}
\phi_{2t}=k_2\:\Phi_2.\label{Eq__k2}
\end{equation}
The tidal component with the largest amplitude has a period
$P_{\rm m}=87.9693$ days \citep{vanh03}, corresponding to Mercury's orbital
period.  The Love number $k_2$ is a function of
$\omega, \rho(r), \mu(r),$ and $\eta(r)$, where $\omega=2\pi / P_{\rm m}$ is the
known forcing frequency and $\rho(r)$, $\mu(r)$, and $\eta(r)$ are the
density, rigidity, and viscosity profiles.  With the appropriate
profiles, the Love number $k_2$ at the frequency $\omega$ can be
calculated by solving the equations of motion inside the planet.
These equations consist of three second-order equations that can be
transformed into a system of six first-order linear differential
equations in radius through a spherical harmonic decomposition in
latitude and longitude \citep{Alterman1959}.  We solve these equations
with a slightly modified version of the propagator matrix method
\citep[e.g.,][]{Sabadini2004}, as described by \citet{Wolf1994} and by
\citet{Moore2000,Moore2003}.

\subsection{Rheological models}\label{Sec__Rheology}

The rheological response of solid materials is elastic, viscoelastic,
or viscous, depending primarily on pressure, temperature, grain size,
and timescale of the process under consideration.  Other dependencies
include melt fraction and water content.  Earth's mantle has a
quasi-elastic response on the short timescales associated with seismic
waves and a fluid-like response on the long timescales of mantle
convection.

The Maxwell rheological model is the simplest model that captures
behavior on both short and long timescales.  It is completely defined
by two parameters, the unrelaxed (i.e., corresponding to an impulsive
or infinite-frequency perturbation) rigidity $\mu_{\rm U}$ and the
dynamic viscosity $\eta.$ The Maxwell time, defined as
\begin{equation}
\tau_{\rm M}=\frac{\eta}{\mu_{\rm U}},
\end{equation}
is a timescale that separates the elastic regime (forcing period
$\ll\tau_{\rm M}$) from the fluid regime (forcing period $\gg\tau_{\rm
  M}$).
This simple rheological model is sufficiently accurate to describe the
crust of Mercury.  The crust is cold and responds elastically
($\tau_{\rm M,crust}=10^5$ y).  We treat the liquid outer core as an
inviscid fluid.  We also use the Maxwell model to represent the
rheology of the inner core, which, if present, has a negligible effect
on the tidal response \citep{pado14}.  However, the Maxwell model
fails to capture the response of the mantle at tidal frequencies
\citep[e.g., ][]{Efroimsky2007,Nimmo2012}, because it does not provide
a good fit to laboratory and field data in the low-frequency
seismological range.

We adopt the Andrade pseudo-period rheological model to estimate the
response of Mercury's mantle to tidal forcing
\citep{Jackson2010,pado14}.  In this model, the ratio of strain to
stress, or inverse rigidity, is represented by a complex compliance.
The expressions for the real (R) and imaginary (I) parts of the
dynamic compliance in the Andrade model are \citep{Jackson2010}:
\begin{eqnarray}
J_{\rm R}&=&J_{\rm U}\left\{1+\beta^*\Gamma\left(1+n\right)\omega_{\rm a}^{-n}\cos\left(\frac{n\pi}{2}\right)\right\},\label{Eq__JAndradeR}\\
J_{\rm I}&=&J_{\rm U}\left\{\beta^*\Gamma\left(1+n\right)\omega_{\rm a}^{-n}\sin\left(\frac{n\pi}{2}\right)+\frac{1}{\omega_{\rm v}\tau_{\rm M}}\right\}. \label{Eq__JAndradeI}
\end{eqnarray}
$J_{\rm U}$ is the unrelaxed
compliance, $\Gamma$ is the gamma function, and $n$,
$\beta^*=\beta/J_{\rm U}$, and $\tau_{\rm M}=\eta J_{\rm U}$ are
related to parameters appearing in the Andrade creep function
$J(t)=J_{\rm U} + \beta t^n + t/\eta.$ The pressure ($P$), temperature
($T$), and grain size ($d$) dependencies are introduced through the
pseudo-period master variable $X_{\rm B}=2\pi/\omega_{\rm a,v}$:
\begin{eqnarray}
  X_{\rm B} &=& T_0 \left(\frac{d}{d_{\rm Ref}}\right)^{-m_{\rm a,v}} \nonumber \\
           & & \times \exp\left[\left(\frac{-E}{R}\right)\left(\frac{1}{T}-\frac{1}{T_{\rm Ref}}\right)\right] \nonumber \\
           & & \times \exp\left[\left(\frac{-V}{R}\right)\left(\frac{P}{T}-\frac{P_{\rm Ref}}{T_{\rm Ref}}\right)\right],\label{Eq__MasterVar}
\end{eqnarray}
where $T_0$ is the period of the applied forcing (in this case the
period of the primary tidal component), $m_{\rm a}$ ($m_{\rm v}$) is
the grain size exponent for anelastic (viscous) processes, and $R$ is
the gas constant.  $P_{\rm Ref}$, $T_{\rm Ref}$, and $d_{\rm Ref}$
indicate reference values (Table~\ref{tab-RheolMod}).  The unrelaxed
shear modulus $\mu_{\rm U}=1/J_{\rm U}$ is itself dependent on
pressure and temperature, which we characterize by a simple Taylor
expansion truncated at linear terms: $\mu_{\rm U}\left(P,T\right)=\mu_{\rm U}^{\rm Ref} +
(\partial \mu/\partial P)(P-P_{\rm Ref})+ (\partial \mu/\partial T)(T
- T_{\rm Ref})$.  The frequency-dependent shear modulus $\mu(\omega)$,
quality factor $Q(\omega)$, and viscosity $\eta(\omega)$ are all
obtained from the dynamic compliance, as follows
\citep{Jackson2010,pado14}:
\begin{eqnarray}
\mu\left(\omega\right)&=&\left[J^2_{\rm R}\left(\omega\right)+J^2_{\rm I}\left(\omega\right)\right]^{-1/2},\\
Q\left(\omega\right)&=&\frac{J_{\rm R}\left(\omega\right)}{J_{\rm I}\left(\omega\right)},\\
\eta\left(\omega\right)&=&\frac{1}{\omega_0 J_{\rm I}\left(\omega\right)},\label{Eq__Etaomega}
\end{eqnarray}
where $\omega_0=2\pi/T_0.$ Our choice of model parameters is described in
Table \ref{tab-RheolMod} and Section \ref{Sec__Methodology}.

\begin{table*}
\begin{center}  
\caption{Rheological models for the interior of Mercury.}
\begin{tabular}{rlclr}
\hline\hline
Layer & Model & Parameter & Definition & Value \\  \hline
Crust & Maxwell & & & \\
      &  &  $\mu_{\rm U}$ & Unrelaxed rigidity & 55 GPa           \\
      &  &  $\eta$  & Dynamic viscosity & $10^{23}$ Pa s \\ \hline
 Mantle & Andrade$^{\rm a}$ & & &  \\
 	&  & $\mu_{\rm U}^{\rm Ref}$ & Unrelaxed rigidity$^{\rm b}$ & $59-71$ GPa \\
 	&  & $T_{\rm b}$     & Mantle basal temperature$^{\rm c}$ &$1\,600-1\,850$ K  \\
	&  & $n$          & Andrade creep coefficient & 0.3                 \\
	&  & $\beta^*$ & Andrade creep parameter & 0.02            \\
	&  & $P_{\rm Ref}$     & Reference pressure  & 0.2 GPa      \\
	&  & $T_{\rm Ref}$      & Reference temperature & 1\,173 K      \\
	&  & $d_{\rm Ref}$      & Reference grain-size       &  $3.1$ $\mu{\rm m}$  \\
	&  & $d$                   & Grain size		    &    $1\,{\rm mm}-1\,{\rm cm}$     \\
	&  & $m_{\rm a},$ $m_{\rm v}$                 & Grain size exponents        &  $1.31,$ $3$ \\
	&  & $V$          & Activation volume & $10^{-5}$ m$^3$mol$^{-1}$ \\
	&  & $E_{\rm B}$      & Activation energy & $303\times10^{3}$ kJ mol$^{-1}$ \\ \hline
 FeS  & Andrade$^{\rm d}$ & & & \\ \hline
Outer core & Inviscid fluid & & &  \\
        &  & $\mu_{\rm U}$ & Unrelaxed rigidity & 0 Gpa   \\ 
        &  & $\eta$   & Dynamic viscosity & 0 Pa s \\ 
\hline
Inner core & Maxwell & & & \\
	&  & $\mu_{\rm U}$ & Unrelaxed rigidity & $100$ GPa  \\
&  & $\eta$  & Dynamic viscosity & $10^{20}$ Pa s  \\
\hline
\end{tabular}
\end{center}
\tablecomments{}
\tablenotetext{a}{The fixed parameters of the Andrade model are based on the results of \citet{Jackson2010}.}
\tablenotetext{b}{The nominal value depends on the adopted mineralogy (Table~\ref{Tab__MantleMods}).}
\tablenotetext{c}{We report $T_{\rm b}$ because the relevant temperature in equation (\ref{Eq__MasterVar}) is controlled by $T_{\rm b}$.}
\tablenotetext{d}{The FeS layer is assumed to have the same rheology as that of the base of the mantle.}
\label{tab-RheolMod}
\end{table*}

Our choices of Andrade model parameter values
(Table~\ref{tab-RheolMod}) are based on data obtained at periods
smaller than $10^{3}$ s \citep{Jackson2010}, whereas the main tide of
Mercury has a period $>10^6$ s.  The extrapolation to long time scales
can be validated to some extent by two considerations.  First, we
verified that equation (\ref{Eq__Etaomega}) yields viscosity values at
long timescales ($>10$ My) that fall within the interval for
convective viscosities commonly assumed in terrestrial mantle convection
simulations ($10^{20}-10^{23}$ Pa~s).  Second, we verified that, at
timescales appropriate for glacial rebound on Earth ($\sim$10$^{4}$ y), the
predicted viscosity values ($10^{20}-10^{21}$ Pa~s) compare favorably
with those inferred from geodynamical data
\citep[e.g.,][]{Kaufmann2000}.

The choice of Andrade model parameter values
(Table~\ref{tab-RheolMod}) is also based on laboratory data for
olivine \citep{Jackson2010}, whereas we apply the model to a variety
of mineralogies (Table \ref{Tab__MantleMods}).  This
extrapolation to other mineralogies is not strictly correct,
especially for mantle models in which olivine is not the dominant
phase.  However, the Andrade model has been successfully applied to
the description of dissipation in rocks, ices, and metals
\citep[e.g.,][and references therein]{Efroimsky2012}.  The broad
applicability of the model over a wide range of physical and chemical
properties suggests that the model can provide an adequate description
of the rheology of silicate minerals.

Recent results of laboratory experiments and thermodynamic simulations
based on Mercury surface compositions \citep{VanderKaaden2016,namu16}
suggest an 
olivine-rich source for both the northern smooth plains and the
high-Mg region of the intercrater plains and heavily cratered terrain.
These results are in accord with an olivine-dominated mineralogy for
the mantle of Mercury and further support our model parameter choices.

\subsection{Methods}\label{Sec__Methodology}

We restrict our analysis to interior models that are compatible with
the observed bulk density $\rho$, moment of inertia $C,$ and moment of
inertia of the solid outer shell $C_{\mpcr}$.  By design, the
subset of interior models has distributions of $\rho$, $C,$ and
$C_{\mpcr}$ that are approximately Gaussian with means and standard
deviations that match the nominal values of the observables and their
one-standard-deviation errors.
The mean density $\rho$ has a Gaussian distribution with mean and
standard deviation equal to 5430 ${\rm kg\, m}^{-3}$ and 10 ${\rm kg\, m}^{-3}$, 
respectively.  For $C$ and $C_{\mpcr}$, we choose Gaussian
distributions with means and standard deviations defined by the
observed values and errors (Section~\ref{sec-moires}).

We treat the interior of Mercury as a series of spherically symmetric,
incompressible layers characterized by density, thickness, rigidity,
and viscosity.  We start with the density profiles calculated by
\citet{hauc13}, but we replace the $\sim$1000 layers that characterize
the core in these models with two homogeneous layers representing the
solid inner core and the liquid outer core.  This simplification is
warranted because the tidal response is dominated by the presence of a
liquid outer core and is largely independent of the detailed density
structure of the core.  It reduces the computational cost by about three 
orders of magnitude and introduces only a small ($<$2\%) error in the
estimated value of $k_2$.  This error is smaller than the variations
induced by the unknown mineralogy and thermal state of the mantle
\citep{pado14}.

For the core of Mercury, we focus on the Si-bearing subset of models
analyzed by \citet{hauc13}, because this subset is most consistent
with the chemically reducing conditions inferred from surface materials
(Section~\ref{sec-comp}).  We also consider the subset of models with
a solid FeS~V layer included at the base of the mantle (\citet{hauc13}
and Section~\ref{sec-N}).  For the silicate mantle of Mercury, we
consider six mineralogical models based on the works of \citet{rivo09}
and \citet{mala10} (Table~\ref{Tab__MantleMods}).

\begin{table*}
\begin{center}
    \caption{Mineralogical models for the mantle of Mercury.}
\begin{tabular}{l|rrrrrrrr|c}  \hline\hline
Model & Grt & Opx & Cpx & Qtz & Spl & Pl  & Mw & Ol  & $\mu_{\rm U}^{\rm Ref}\:(\rm{GPa})$ \\  \hline
CB    & --    &   66 &   4   &  22  &  4  &  4 &  --  &    --   &  59\\
EH    &   -- &  78  &  2    &   8   &  --   &12 &  -- &   --     &  65\\ 
MA    &   23&  32  &   15 &  --    &  --  &  -- & --   &  30    &  69\\ 
TS    & 25&   -- &   -- &   -- &  8 &   -- &    2 & 65    &71\\
MC    &  15& 50  &  9  &   -- &   -- &   -- &    -- &26 & 68\\
EC    &  1 &  75 &  7  &  17 & -- &   --  &   -- &   --  & 60 \\\hline
\end{tabular}
\end{center}
\tablecomments{The adopted model mineralogies resemble those of enstatite
  chondrites (EC and EH), Bencubbin-like chondrites (CB), metal-rich
  chondrites (MC), a refractory-volatile model (TS), and a model based
  on fractionation processes in the solar nebula (MA).  For details,
  see \citet[][CB and EH]{mala10}, \citet[][MA]{Morgan1980},
  \citet[][TS and MC]{Taylor2003}, and \citet[][EC]{wass88}.  The
  central part of the table gives the mineralogical content in terms
  of the vol.\% of its components, from \citet[][CB, EH]{mala10} and
  \citet[][MA, TS, MC, EC]{rivo09}.  Mineral abbreviations follow
  \citet{Siivola2007}: Garnet (Grt), Orthopyroxene (Opx),
  Clinopyroxene (Cpx), Quartz (Qtz), Spinel (Spl), Plagioclase (Pl),
  Merwinite (Mw), Olivine (Ol).
The composite rigidity $\mu_{\rm U}^{\rm Ref}$ is obtained with Hill's
expression \citep{watt76} at $T=T_{\rm Ref}$ and $P=P_{\rm Ref}$.}
\label{Tab__MantleMods}
\end{table*}

Our use of the Andrade model (Section \ref{Sec__Rheology}) for the
rheological properties of the mantle requires knowledge of the radial
profiles of unrelaxed rigidity $\mu_{\rm U}$, temperature $T,$ and
pressure $P$ in the outer solid shell.  For each of the six
mineralogical models, we compute a composite rigidity $\mu_{\rm
  U}^{\rm Ref}$ (Table~\ref{Tab__MantleMods}) with Hill's expression
\citep{watt76} at $T=T_{\rm Ref}$ and $P=P_{\rm Ref}$.  The pressure
profile in the outer solid shell is obtained by evaluating the
overburden pressure as a function of depth.  The temperature in the
mantle is computed by solving the static heat conduction equation with
heat sources in spherical coordinates \citep[e.g.,][]{Turcotte2002} in
the mantle and crust.  For the crust, we adopted the surface value of
the heat production rate $H_0=2.2\times10^{-11}\:\rm{W\;kg^{-1}}$
\citep{Peplowski2012}.  For the mantle, we used a value of $H_0/2.5,$
which is compatible with the enrichment factor derived by
\citet{Tosi2013}.  Temperature profiles are fairly insensitive to the
value of the thermal conductivity: we used a value $k_m = 3.3\, {\rm
  Wm}^{-1} {\rm }K^{-1}$ but confirmed that a value of $k_m = 5\, {\rm
  Wm}^{-1} {\rm }K^{-1}$ yields essentially the same results.  We establish
two boundary conditions: the temperature at the surface of Mercury
$T_{\rm S}$ and the temperature at the base of the mantle $T_{\rm b}$.
The latter provides the primary control on the temperature
profile. $T_{\rm S}$ is set to $440$ K, a value obtained with an
equilibrium temperature calculation.  Both \citet{rivo13} and
\citet{Tosi2013} indicate $T_{\rm b}$ values in the range 1\,600--1\,900~K.
We define two end-member profiles: a cold-mantle profile with $T_{\rm
  b}=1\,600$ K and a hot-mantle profile with $T_{\rm b}=\,1\,850$ K.  A
larger value of $T_{\rm b}$ (e.g., 1\,900 K) would result in partial
melting at the base of the mantle according to the peridotite solidus
computed by \citet{Hirschmann2000}.  We did not consider the presence
of partial melting.

There is a scarcity of laboratory data for FeS V, which is the phase
relevant at the pressure and temperature conditions at the bottom of
the mantle of Mercury \citep{Fei1995}.  We consider the effects of the
FeS layer only in the cold-mantle case ($T_{\rm b}=1\,600$ K), because
at higher temperatures the FeS would be liquid \citep[see the phase
  diagram given by ][]{Fei1995}.  We model the rheological response of this
layer by assuming that it has the same rheological properties as those
at the base of the mantle.  This assumption
results in a lower bound on the $k_2$ estimates because we expect the
viscosity of this layer to be lower than that of the silicate layer.
The viscosity scales as the exponential of the inverse of the
homologous temperature (i.e., the ratio of the temperature of the
material to the solidus temperature) \citep[e.g.,][]{Borch1987}.  At
$T=1\,600$ K, the homologous temperature of the FeS~V is larger than
that of the silicates.  In addition, the unrelaxed rigidity of FeS~V
is likely to be smaller than that for mantle material
because the rigidity of troilite (or FeS I, the phase at
standard pressure and temperature) is 31.5 GPa \citep{Hofmeister2003}.

We apply our calculations to five different models (nominal, cold and
stiff, hot and weak, FeS-layer, and 1-mm grain size).  Given the
1\,600--1\,850~K range for the basal mantle temperature and 59--71~GPa
range for the unrelaxed rigidity of the mantle, we define a nominal
model with $T_{\rm b}=1\,725$~K and $\mu_{\rm U}=65$~GPa. Changes in
basal mantle temperature and unrelaxed rigidity have similar but
opposite effects on the tidal response.  Accordingly, we define two
end-member models: a cold and stiff mantle model with $T_{\rm
  b}=1\,600$~K and $\mu_{\rm U}=71$~GPa and a hot and weak mantle model
with $T_{\rm b}=1\,850$~K and $\mu_{\rm U}=59$~GPa.  Our fourth model is
a cold mantle model ($T_{\rm b}=1\,600$~K) with nominal rigidity
($\mu_{\rm U}=65$~GPa) and an FeS layer at the bottom of the mantle.
In all of these four models, we use a nominal grain size $d=1$~cm,
a value compatible with the estimated grain size in the mantles of
the Moon and Mars \citep{Nimmo2012,Nimmo2013}.  Our fifth and last
model is a variation of the nominal model in which we consider a grain
size of $d=1$ mm.  Model parameters are summarized in Table
\ref{Tab__k2Mod}.

\begin{table}
\begin{center}  
  \caption{Characteristics of five mantle models for the estimation of Mercury's tidal response.}
\begin{tabular}{l r r r l}\hline\hline
  Model & $\mu_{\rm U}$, GPa & $T_{\rm b}$, K & $d$, mm & ${\rm FeS}?$ \\ \hline    
   Nominal             &  65 & 1\,725 & 10  &  no \\
   Cold and stiff      &  71 & 1\,600 & 10  &  no \\
   Hot and weak        &  59 & 1\,850 & 10  &  no \\
   FeS layer           &  65 & 1\,600 & 10  &  yes \\
   1-mm grain size     &  65 & 1\,725 &  1  &  no \\ \hline
\end{tabular}
\end{center}
\tablecomments{Model names correspond to those in Figure \ref{Fig__k2Mod}.} 
\label{Tab__k2Mod}
\end{table}

Our procedure for evaluating the Love number $k_2$ and corresponding
uncertainties is as follows.  For each of the five cases described
in Table \ref{Tab__k2Mod}, we use approximately $6\times10^{4}$
density profiles from the previously identified subsets of models from
\citet{hauc13}.  For each profile, we construct an interior model and
calculate the value of $k_2$.  We then fit a Gaussian distribution to
the $\sim$6$\times10^{4}$ calculated $k_2$ values, as was done by \citet{pado14}.
We report the Love number and associated error as the mean and
standard deviation of the Gaussian fit.  Our values differ somewhat
from those of \citet{pado14} because we incorporated the most recent
estimates of the moments of inertia in this work (Equations \ref{eq-0346} and
\ref{eq-0425}).

\subsection{Results}

Our Love number calculations for models with a molten outer core
yield values $k_2 \simeq 0.5$.  However, for models with a
completely solid core, we found $k_2$ values that are approximately an
order of magnitude smaller.  Measurements of Mercury's tidal response
(Section~\ref{sec-k2obs}) therefore confirm the presence of a molten
outer core.

\begin{figure}[h!]
   \begin{center}
               \includegraphics[width=\columnwidth]{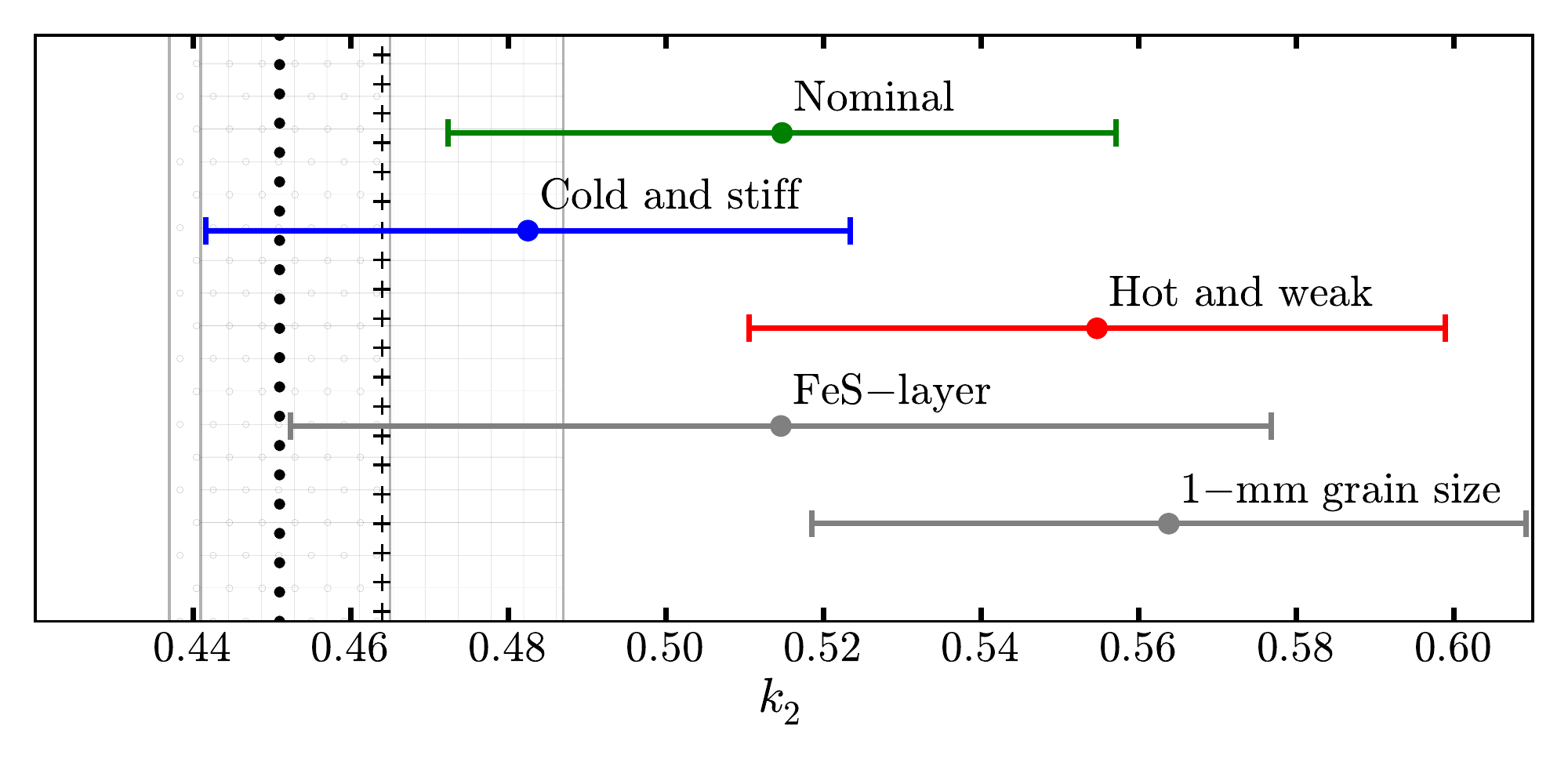}     
           \caption{Calculated values of the tidal Love number $k_2$
             for five models constructed under different assumptions about the
             rheological properties and physical structure of the
             outer solid shell of Mercury (Table \ref{Tab__k2Mod}).
             The vertical lines and hatch patterns represent two
             independent determinations of $k_2$ and associated
             one-standard-deviation uncertainties measured by
             radio tracking of the MESSENGER spacecraft.  Dot symbols
             correspond to the results of \citet[][]{maza14}, and plus symbols
               correspond to the results \citet[][]{verm16}.}
      \label{Fig__k2Mod}
   \end{center}
\end{figure}

Our results also show that the tidal response is enhanced by higher
mantle basal temperatures and by lower mantle rigidities (Figure
\ref{Fig__k2Mod}).

The comparison of our calculated values with the $k_2$ value measured
by \citet{maza14} indicates that the observed tidal signal is more
compatible with cold, rigid mantle models (Figure \ref{Fig__k2Mod}).
The $k_2$ value measured by \citet{verm16} admits a wider range of
models but still favors models with a cold and stiff mantle or a
subset of the FeS-layer models.  It is likely that models with an FeS
layer at the bottom of the mantle and high mantle rigidity ($\mu_{\rm U} =
71$ GPa) would also be compatible with $k_2$ measurements, but there
are questions about the plausibility of such a
layer~\citep[][Section~\ref{sec-Nres}]{knib15}.

The conclusion drawn from the modeling of the tidal Love number seems
robust with respect to details of the thermal model.  For instance,
consideration of a surficial regolith layer with low thermal
conductivity increases the temperature in the interior, which results
in larger $k_2$ model values and further favors a cold and stiff
mantle.  Consideration of a higher solidus temperature and $T_{\rm
  b}>1\,850$~K would also strengthen the conclusion that Mercury's
mantle is likely cold and stiff.  Unfortunately, the robustness of the
conclusion is undermined because of the large standard deviations
associated with the modeled $k_2$ values and because the actual $k_2$
value may extend beyond the range given by the one-standard-deviation
uncertainties.  The overlap in simulated $k_2$ values for the five
mantle models implies that even a more precise $k_2$ measurement would
not be sufficient to identify a unique model at this time.  However, a
reduction in uncertainties of both the measured Love number and
moments of inertia will narrow the range of mantle models that are
compatible with observations.

\section{Influence of solid inner core}
\label{sec-inner}

Torques between layers in Mercury's interior can influence the spin
state.  \citet{peal14} derived the behavior of Mercury's spin axis
orientation under the influence of a variety of core-mantle torques,
including gravitational, tidal, magnetic, topographic, viscous, and
pressure torques.  They showed that tidal torques are small in
comparison to magnetic and topographic torques, which are themselves
small compared with viscous torques.  These dissipative torques would
drive the mantle spin away from the Cassini state if it were not for
the action of a pressure torque between the outer core and the mantle.
The pressure torque is due to fluid pressure at the core-mantle
boundary, which is not spherically symmetric because of its
hydrostatic, approximately ellipsoidal shape.  In the absence of an inner
core, the pressure torque dominates the spin axis evolution and drives
the mantle spin close to the Cassini state position.

\citet{peal16} considered the additional torques due to an inner core.
Their theoretical formalism is general and applicable to other
planets, including Earth.  The shape of the inner core is distorted by
the non-radial gravitational field, and a gravitational torque between
inner core and mantle develops.  The relationship between the observed
obliquity and the moment of inertia (Equation \ref{eq-moi}), which is
based on solar torques, must be modified to account for this
additional torque.  If the inner core is small ($R_{\rm ic}/R <
0.35$), the mantle spin follows the Cassini state orientation
sufficiently closely that the moment of inertia determination is not
compromised.  However, if the inner core size exceeds 35\% of the
planetary radius, the additional torque would drag the mantle spin
away from the Cassini state by an amount that exceeds the current
observational uncertainty of 5 arcseconds, and the polar moment of
inertia would have to be re-evaluated.  In the presence of an inner
core, the obliquity of the mantle spin axis corresponds to a smaller
polar moment of inertia than that inferred from the situation with no
inner core.  This change in the value of the moment of inertia can be
evaluated for a variety of interior models by tracking the evolution
of the spin under the action of all relevant torques and enforcing the
requirement that the mantle spin axis orientation remains within the
uncertainty region of the radar observations.  \citet{peal16}
performed this calculation for a variety of inner core sizes and inner
core densities.  They found that the required adjustment to the value
of the moment of inertia increases with both inner core density and
inner core size.  For an inner core density of $9\,300 {\ \rm kg\,
  m}^{-3}$, they found corrected values of $C/MR^2 = 0.346, 0.343,
0.330, 0.327, {\rm and } 0.323$ for inner core sizes of $R_{\rm ic}/R = 0.0, 0.3,
0.4, 0.5, {\rm and } 0.6$, respectively~\citep{peal16}.  Because 85\% of the
best-fit interior models (Section \ref{sec-prmm}) have inner core
densities below $9\,300 {\ \rm kg\, m}^{-3}$,
the corrections identified by \citet{peal16} likely represent upper
bounds on any necessary adjustment to the moment of inertia due to the
presence of an inner core.

Because of the possible impact of an inner core on the determination
of Mercury's moment of inertia~\citep{dumb13,peal16}, it is important
to place bounds on the size of the inner core.  We review six lines of
evidence.  (1) \citet{peal16} found that in models with inner cores larger
than $R_{\rm ic}/R = 0.3$, the inferred mantle densities were only
$\sim$3\,000${\ \rm kg\, m}^{-3}$.  Such low mantle densities are
difficult to explain because they are lower than those of materials
that likely dominate Mercury's Fe-poor interior, such as Mg-rich
olivine and Mg-rich orthopyroxene, which have densities of
$3\,200-3\,300 {\ \rm kg\, m}^{-3}$.  If the density information is a
reliable indicator, the calculations of \citet{peal16} suggest an
inner core size $R_{\rm ic}/R \leq 0.3$.  (2) A similar conclusion can be
reached by examining the distribution of internal structure models.
We find that 65\% of models that provide the best fit to existing
observations (Section \ref{sec-prmm}) have a small inner core ($R_{\rm
  ic}/R < 0.35$).
(3) Independent constraints on inner core size arise from the fact
that planetary contraction is due in part to inner core
solidification.  The observed planetary contraction of 7~km
\citep[Chapter 10]{byrn14} suggests that the inner core size does not
exceed 800--1\,000 km, i.e., $R_{\rm ic}/R \lesssim 0.4$,
~\citep{grot11,dumb15}.  \citet{knib15} found upper bounds as large as
$R_{\rm ic}/R \lesssim 0.7$ for certain values of model parameters,
but they did not consider the effects of mantle contraction, which may
amount for about half of the planetary contraction~\citep{Tosi2013}.  (4)
Simulations of Mercury's magnetic field provide another indicator
about the size of the inner core.  Dynamo models that can reproduce
the observed features of Mercury's magnetic field \citep{cao14} favor
small inner cores ($R_{\rm ic}/R_{\rm oc} < 0.5$, i.e., $R_{\rm ic}/R
< 0.4$).  (5) \citet{dumb15} further argued that, in some situations,
the dynamics of snow formation in the fluid core would place an upper
limit on the inner core radius of 650~km ($R_{\rm ic}/R < 0.27$).
(6) Finally, several authors have noted that a large inner core $(R_{\rm
  ic}/R > 0.4)$ would produce detectable signatures in the librations of
the planet \citep{veas11,dumb11,vanh12}, but such signatures have not
been detected to date.  There is considerable interest in improving
measurements of the longitudinal librations in an attempt to place
bounds on the size of Mercury's inner
core~\citep{veas11,dumb11,vanh12}, although it is not clear that the
precision of the current measurement techniques would enable a
detection of the inner core signature.

To summarize, there is some circumstantial evidence that Mercury's
inner core is small ($R_{\rm ic}/R \lesssim 0.35$) and that the
existing estimate of $C/MR^2 =0.346 \pm 0.009$ remains valid.
However, no direct measurements of the inner core size exist, which
reduces our confidence in the knowledge of Mercury's moment of
inertia.  Improved measurements of the librations or direct
measurements of the inner core size will be required to eliminate the
uncertainty.  One approach would be to deploy seismometers on the
surface and measure seismic signals triggered by tides, internal
activity, explosive charges, or impacts.

\section{Representative model}
\label{sec-prmm}

The observational evidence from spin, tidal, and compositional
observations, summarized in Table~\ref{tab-summary}, favors a Mercury
interior model with a core composition dominated by Fe-Si and with a
small or no solid FeS layer.
Therefore, models in which the core is treated as an Fe-Si end-member
are likely representative of Mercury's interior.

\begin{table*}
\begin{center}  
\caption{Summary of observational constraints used for the calculation of internal structure models.}
\label{tab-summary}
\begin{tabular}{llrrc} \hline \hline
Parameter  & Symbol &  Value & Uncertainty & Unit\\
\hline
Mass & $M$             & 3.301110    & 0.00015 & $10^{23}$ kg\\
Radius & $R$             & 2\,439.36   & 0.02 & km\\
Density & $\rho$          & 5\,429.30   & 0.28 & ${\rm kg\, m}^{-3}$\\
Gravity spherical harmonic & $C_{20}$         & -5.0323 & 0.0022 & $10^{-5}$\\
Gravity spherical harmonic & $C_{22}$         &  0.8039 & 0.0006 & $10^{-5}$\\
Tidal Love number & $k_2$           & 0.455 & 0.012  & \\
Obliquity & $\theta$          & 2.036 & 0.058  & arcminutes\\
Amplitude of longitude librations & $\phi_0$           & 38.7 & 1.0  & arcseconds\\
\hline
Moment of inertia factor & $C/MR^2$          & 0.346 & 0.009 & \\
Moment of inertia of mantle and crust & $C_{\mpcr}/{C}$    & 0.425 & 0.016 & \\
Crustal thickness & $h_{\rm cr}$     &  35--53 & & km\\
Crustal density & $\rho_{\rm cr}$ & 2\,700--3\,100 & & ${\rm kg\, m}^{-3}$\\
\hline
  \end{tabular}
\end{center}
\tablecomments{The first eight values are direct measurements.  The
  remaining four values are derived quantities that rely on a variety
  of assumptions.  These assumptions, described below, are justified
  considering the data obtained to date and our knowledge of
  terrestrial planets.  However, additional data are required to fully
  verify the validity of some of these assumptions.  Moment of inertia
  assumptions: (1) Mercury is in Cassini state 1, (2) core does not
  follow mantle on the 88-day timescale of longitude librations, (3)
  core does follow mantle on the 300\,000-year timescale of orbital
  precession, (4) $R_{\rm ic}/R < 0.35$.  Crustal thickness
  assumptions: (1) filtering of gravity and topography data is
  effective in isolating the crustal signal, (2) compensation of
  topography is well approximated by Airy isostasy.  Crustal density
  assumptions: (1) elemental abundances derived from X-ray
  fluorescence measurements sampling the uppermost 100 $\mu$m of the
  surface are applicable to the entire crust, (2) normative mineralogy
  derived from elemental abundances correctly captures crustal
  minerals, (3) porosity of the crust does not exceed 12\%.}
\end{table*}

Our preferred models include bounds on crustal thickness and density.
Analyses of gravity-to-topography ratios suggest an average crustal
thickness of $35 \pm 18$ km \citep{pado15} and $>38$ km
\citep{jame15}.  We combine these bounds into a preferred crustal
thickness in the range 35--53 km (Table~\ref{tab-summary}).  The grain
density of crustal material can be determined from a normative
mineralogy, which itself is guided by observations of elemental
abundances at the surface of Mercury~\citep{weid14}.  With this
approach, \citet{pado15} obtained grain densities of 3\,014 ${\rm kg\,
  m}^{-3}$ and 3\,082 ${\rm kg\, m}^{-3}$ for the northern smooth
plains and for heavily cratered terrain and intercrater plains,
respectively.  If we take into account porosity values of up to 12\%
as observed on the Moon \citep{wiec13}, our preferred crustal
densities are in the range 2\,700--3\,100 ${\rm kg\, m}^{-3}$
(Table~\ref{tab-summary}).

We updated the analysis of \citet{hauc13} to conform to the radius and
density values listed in Table~\ref{tab-summary}.  In addition, we
specified an initial crustal thickness in the range 0--70 km, a
crustal density in the range 2\,700--3\,100 ${\rm kg\, m}^{-3}$, and a
core Si content in the range 0--17 wt \%.  This analysis yielded
1\,016\,236 Fe-Si interior models with considerable scatter in
structural properties.  From these models, one can extract a random
sample of models for which the distributions of ${C}/{MR^2}$ and
${C_{\mpcr}}/{C}$ values match the observed values and corresponding
one-standard-deviation uncertainties (Table~\ref{tab-summary}).  We
further restricted the set of preferred models to those that provide
the closest agreement to the observed values of ${C}/{MR^2}$ and
${C_{\mpcr}}/{C}$.  All 1\,479 models in this subset have RMS $<$
0.005, where the RMS metric is described by equation
(\ref{eq-fracrms}). These 1\,479 best-fit models constitute a family
of representative models that can be used to illustrate the remaining
scatter in the values of Mercury's internal structure parameters
(Table~\ref{tab-stats}).  Among the subset of models that provide the
closest match to observational data, the radius of Mercury's core,
$R_{\rm oc}=2024 \pm 9$ km, is determined with $<$0.5\% precision and
represents 83\% of the radius of the planet.

\begin{table*}
  \caption{Statistical properties of interior structure model parameters and corresponding PRMM values.} %
  \label{tab-stats}
  \begin{center}
  \begin{tabular}{|l|rrrrr|rr||r|}
    \hline
 Parameter             & minimum      &       1$^{\rm st}$ quartile      &      median  &      3$^{\rm rd}$ quartile       &      maximum      &      mean     &    std. dev. & PRMM\\
\hline %
${C}/{MR^2}$           &     0.34430 &      0.34523 &      0.34596 &      0.34670 &      0.34771 &      0.34597 &   0.00089         &  0.34573\\
${C_{\mpcr}}/{C}$       &     0.42294 &      0.42418 &      0.42496 &      0.42578 &      0.42712 &      0.42497 &   0.00102         &  0.42482\\
\hline 
$R_{\rm ic}$            &     0.01877 &       310.780 &       623.280 &       1003.60 &       1790.82 &       666.577 &       420    &  369.433\\
$R_{\rm oc}$            &     2009.31 &       2016.69 &       2021.30 &       2029.62 &       2062.56 &       2023.66 &       9.09   &  2015.48\\
$R_{\rm m}$             &     2369.37 &       2385.60 &       2401.37 &       2419.32 &       2439.35 &       2402.61 &       19.9   &  2401.20\\
\hline
$\rho_{\rm ic}$         &     7368.25 &       8295.31 &       8659.58 &       8991.33 &      10214.90 &       8652.52 &       488    &  8215.62\\
$\rho_{\rm oc}$         &     5937.29 &       6775.76 &       7010.49 &       7087.14 &       7187.97 &       6909.98 &       237    &  7109.73\\
$\rho_{\rm m}$          &     3206.19 &       3288.90 &       3333.75 &       3388.10 &       3593.18 &       3343.35 &       71.8   &  3278.98\\
$\rho_{\rm cr}$         &     2700.28 &       2807.00 &       2898.57 &       3006.28 &       3099.78 &       2903.03 &       116    &  2979.19\\
$\rho_{\icpoc}$         &     6671.42 &       6976.74 &       7053.32 &       7102.67 &       7190.40 &       7034.32 &       88.3   &  7116.54\\
$\rho_{\mpcr}$          &     3198.01 &       3255.43 &       3286.49 &       3327.32 &       3531.21 &       3295.84 &       53.0   &  3247.21\\                        
$\rho$                 &     5428.34 &       5429.11 &       5429.30 &       5429.52 &       5430.53 &       5429.32 &      0.31    &  5429.66\\ 
\hline
$M_{\rm ic}$            &    2.588$\times 10^{08}$ &     1.101$\times 10^{21}$ &     8.962$\times 10^{21}$ &     3.582$\times 10^{22}$ &     1.773$\times 10^{23}$ &   2.288$\times 10^{22}$ &   2.95$\times 10^{22}$    &  1.735$\times 10^{21}$\\
$M_{\rm oc}$            &    6.728$\times 10^{22}$ &     2.084$\times 10^{23}$ &     2.351$\times 10^{23}$ &     2.428$\times 10^{23}$ &     2.446$\times 10^{23}$ &   2.213$\times 10^{23}$ &   2.93$\times 10^{22}$    &  2.423$\times 10^{23}$\\
$M_{\rm m}$             &    6.964$\times 10^{22}$ &     7.464$\times 10^{22}$ &     7.789$\times 10^{22}$ &     8.152$\times 10^{22}$ &     8.631$\times 10^{22}$ &   7.813$\times 10^{22}$ &   4.15$\times 10^{21}$    &  7.771$\times 10^{22}$\\
$M_{\rm cr}$            &    1.998$\times 10^{18}$ &     4.319$\times 10^{21}$ &     8.020$\times 10^{21}$ &     1.147$\times 10^{22}$ &     1.567$\times 10^{22}$ &   7.822$\times 10^{21}$ &   4.21$\times 10^{21}$    &  8.368$\times 10^{21}$\\                       
$M_{\icpoc}$            &    2.432$\times 10^{23}$ &     2.439$\times 10^{23}$ &     2.441$\times 10^{23}$ &     2.444$\times 10^{23}$ &     2.454$\times 10^{23}$ &   2.442$\times 10^{23}$ &   3.95$\times 10^{20}$    &  2.441$\times 10^{23}$\\
$M_{\mpcr}$             &    8.484$\times 10^{22}$ &     8.583$\times 10^{22}$ &     8.611$\times 10^{22}$ &     8.639$\times 10^{22}$ &     8.702$\times 10^{22}$ &   8.609$\times 10^{22}$ &   3.92$\times 10^{20}$    &  8.622$\times 10^{22}$\\
$M$                    &    3.301$\times 10^{23}$ &     3.301$\times 10^{23}$ &     3.301$\times 10^{23}$ &     3.301$\times 10^{23}$ &     3.302$\times 10^{23}$ &   3.301$\times 10^{23}$ &   1.93$\times 10^{19}$    &  3.301$\times 10^{23}$\\ 
\hline
  \end{tabular}
  \end{center}
  \tablecomments{Statistical properties of structural parameters of 1\,479 best-fit models (see text) extracted from about a million models of Mercury's interior generated with the method of \citet{hauc13}.  All of these models incorporate an Fe-Si core composition and no solid FeS layer.  Masses, radii, and densities are expressed in {\rm kg}, {\rm km}, and ${\rm kg\, m}^{-3}$, respectively.  Symbols are defined in Fig.~\ref{fig-layers}.  The last column describes a representative model, PRMM, with desirable structural properties.  Values for the inner core in PRMM are illustrative only.}
\end{table*}

\begin{figure}[htbp]
   \begin{center}
     \includegraphics[width=\columnwidth]{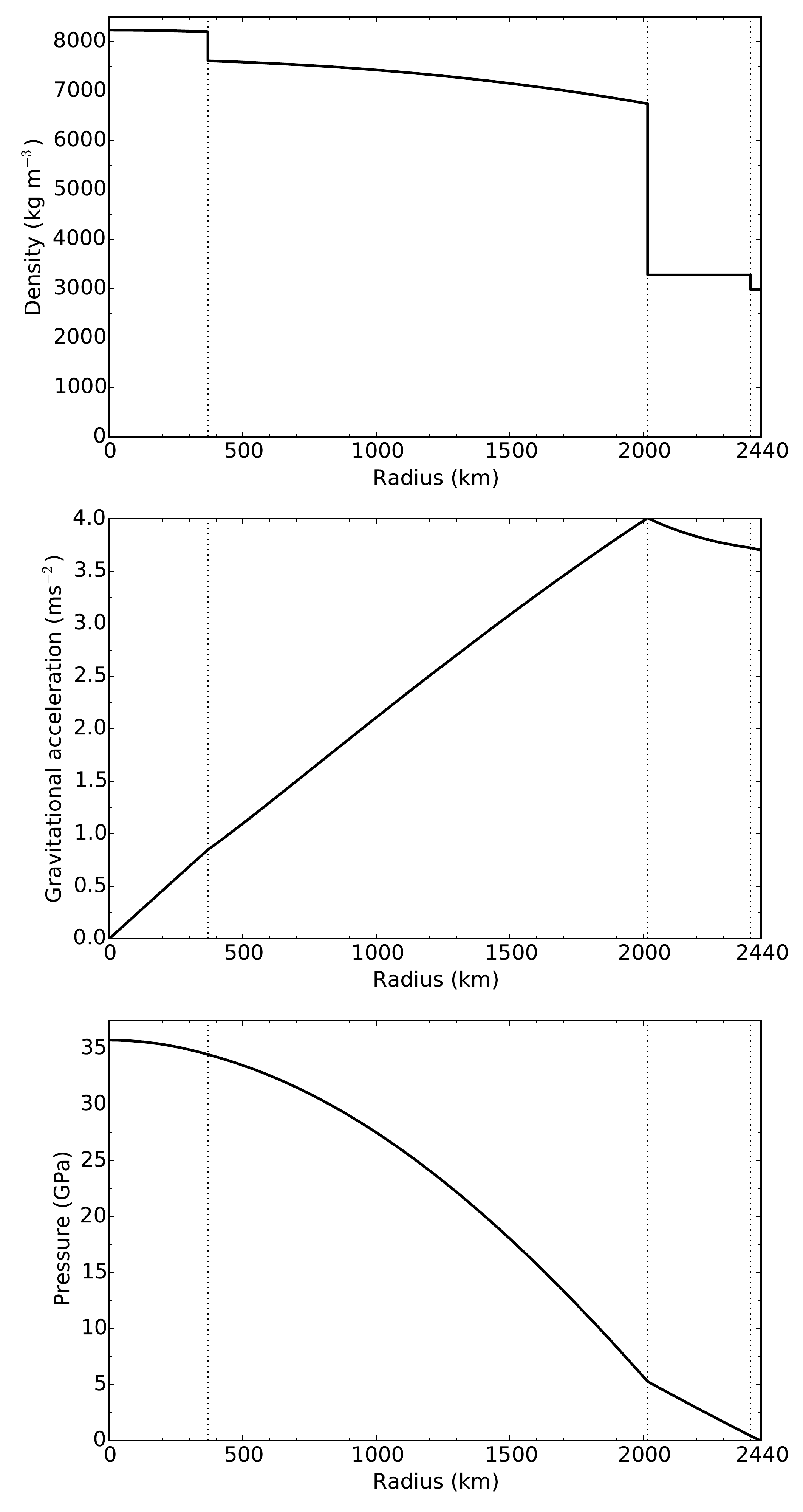}
     \caption{Illustration of the density, gravitational
             acceleration, and pressure corresponding to a model of
             Mercury's interior that closely matches the mass, radius,
             and moments of inertia of Mercury (PRMM).  This model
             also matches our preferred bounds on crustal thickness
             and density (Table~\ref{tab-summary}).  This model
             incorporates an Fe-Si core composition and no solid FeS layer.
             Inner core properties are merely illustrative and not
             representative.  Vertical dotted lines indicate
             transitions with increasing radius between inner and outer core, core and
             mantle, and mantle and crust.}
      \label{fig-profile}
   \end{center}
\end{figure}

We describe an example among the 1\,479 models in some detail
(Table~\ref{tab-stats} and Fig.~\ref{fig-profile}).  This model is
representative in the sense that its structural properties match
Mercury's mass, radius, and moments of inertia, as well as our preferred
bounds on crustal thickness and density.  However, we emphasize that
Mercury's inner core properties are unknown.  The inner core
properties of the chosen model are therefore illustrative and not
representative.  We also emphasize that our chosen model is no better
than any other model that fits the observational data.  The model does
have desirable structural properties, and, as such, it may be useful
for a variety of modeling tasks.  We refer to this model as the
Preliminary Reference Mercury Model (PRMM).

In PRMM, Mercury's mass is divided among inner core (0.5\%), outer
core (73.4\%), mantle (23.5\%), and crust (2.5\%).  The central
pressure is 35.77 GPa, and the pressure at the core-mantle boundary is
5.29 GPa.  Table~\ref{tab-prmm} lists the parameters that we used to
construct PRMM.

\begin{table*}
    \caption{Parameters used to construct PRMM.}
  \begin{center}
    \label{tab-prmm}
  \begin{tabular}{l l r l}
    \hline
      \hline
    Parameter & Symbol & Value & Units\\
    \hline
    Mass fraction of Si (below $R_{\rm oc}$)              & \mbox{$\chi$}$_{\rm Si}$  & 12.83              & wt \%\\    
Liquid Fe reference density                          & $\rho_{0,{\rm Fe,1}}$           & 6471.29            & ${\rm kg\, m}^{-3}$\\
Liquid Fe coefficient of thermal expansion           & $\alpha_{0,{\rm Fe,1}}$         & $9.2\times10^{-5}$  & ${\rm K}^{-1}$\\ 
Liquid Fe bulk modulus                               & $K_{0,{\rm Fe,1}}$              & 115.47              & GPa\\
Liquid Fe pressure derivative of bulk modulus        & $K'_{0,{\rm Fe,1}}$             & 4.93                & \\
Solid $\gamma$ Fe reference density                  & $\rho_{0,{\rm Fe,s}}$           & 7381.34            & ${\rm kg\, m}^{-3}$\\
Solid $\gamma$ Fe coefficient of thermal expansion   & $\alpha_{0,{\rm Fe,s}}$         & $6.4\times10^{-5}$  & ${\rm K}^{-1}$\\ 
Solid $\gamma$ Fe bulk modulus                       & $K_{0,{\rm Fe,s}}$              & 190.73              & GPa\\
Solid $\gamma$ Fe pressure derivative of bulk modulus& $K'_{0,{\rm Fe,s}}$             & 5.62                & \\
\hline
\end{tabular}
\end{center}
  \tablecomments{Temperature-dependent parameters are calculated for the value of the temperature at the core-mantle boundary $T_{\rm cmb} = 1945$ K.}
\end{table*}

PRMM was constructed with the benefit of Earth-based and MESSENGER
observations that were not available in earlier modeling efforts.
Salient differences between PRMM and pre-MESSENGER models include
narrower ranges of admissible structural parameter values compared
with the ranges considered by \citet{hard01}, \citet{vanh03}, and
\citet{rine08} and a core size that is substantially larger than the
core sizes assumed by \citet[][1\,660--1\,900 km]{sieg74},
\citet[][1\,840--1\,900 km]{stev83icar}, \citet[][$1\,860\pm80$ km]{spoh01},
and \citet[][1\,900 km]{breu07}.

\section{Implications}
\label{sec-implications}
\subsection{Thermal evolution}
\label{sec-thermal}

An accurate understanding of Mercury's thermal evolution requires
knowledge of the internal structure, because interior properties
dictate the processes and boundary conditions that have governed the
evolution.  The $\sim$400 km thickness of the silicate mantle and
crust has wide-ranging implications.  The thickness of this layer is a
fundamental control on both the vigor and ultimately the longevity of
mantle convection \citep[e.g.,][Chapter 19]{mich13,Tosi2013}.  The
vigor of the convection is described by the Rayleigh number, which is
the ratio of buoyancy forces to viscous forces in a fluid and scales
as the cube of the thickness the layer \citep[e.g.,][]{schu01}.
Pre-MESSENGER models typically invoked a mantle thickness of $\sim$600
km and therefore over-estimated the vigor of the convection by a
factor of a few.  The MESSENGER-derived value enables more accurate
calculations.  In particular, the thin mantle implies that convection
in Mercury's mantle has been less vigorous than previously thought and
may have completely ceased if the Rayleigh number fell below the
critical value for convection.  A detailed analysis of Mercury's
thermal evolution is given in Chapter 19.

\subsection{Surface geology}
Volcanism is intimately tied to mantle convection because
decompression melting is the primary source of magma in terrestrial
planets.  Mercury's crust, which is the product of perhaps the most
efficient crustal extraction among the inner planets
\citep{jame15,pado15}, was dominantly generated early in the planet's
history when radiogenic heat production was higher (Chapter 19).
Mercury's thin mantle limits the amount of heat transfer because of
the reduced vigor of convection and a possible transition to
conduction (Section \ref{sec-thermal}).  The reduced heat transfer
lowers the amount of volcanism, cooling, and ensuing global
contraction, all of which affect the geological evolution of the
surface.  In particular, the reduced heat transfer hypothesis is
consistent with observations of limited volcanism in the past
$\sim$3.5 billion years and an amount of radial contraction accommodated by thrust faulting
of no more
than 7 km \citep[][Chapters 10 and 19]{byrn14}.  Tectonic patterns
observed at the surface may be due to the interplay of tidal
despinning and global contraction (Chapter 10).  Surface composition
is also affected by mantle thickness, because the horizontal scale of
convection cells is similar to the thickness of the convecting layer.
Investigations of the source regions of surface volcanic material
indicate at least two separate sources \citep{char13}, consistent with
limited mixing of the mantle due to the small horizontal scales and
limited vigor of convection (Chapter 19).

\subsection{Capture in 3:2 resonance}
Mercury's distinctive 3:2 spin-orbit resonance was established at
least in part because of Mercury's internal structure.  The structure
of the interior and the nature of the interactions among inner core,
outer core, and mantle have a profound influence on the evolution of
the spin state and the response of the planet to external forces and
torques.  These processes dictate the overall tectonic and insolation
regimes that, in turn, have wide-ranging implications for a variety
of questions related to Mercury's shape, surface geology, thermal
regime, and even the presence of polar ice deposits.

The history of Mercury's spin-orbit configurations has been markedly
affected by the presence of a liquid core.  It has been suggested that
increased energy dissipation at a core-mantle interface would have led
to near-certain capture in specific spin-orbit
resonances~\citep{gold67,coun70,peal88}, although some models indicate
100\% capture probability in the 2:1 resonance~\citep{peal77}, which
would prevent evolution to the current configuration.  A solution to
this problem was found by \citet{corr04}, who showed that chaotic
variations in orbital eccentricity destabilize most spin-orbit
resonances and ultimately lead to a 55\% capture probability in the
3:2 resonance.  After spin state observations revealed Mercury's
liquid core, \citet{corr09} added core-mantle friction to their model.
They found capture probabilities of 32\% (2:1), 26\% (3:2), and 22\%
(5:2).  While capture in the 3:2 spin-orbit configuration is not the
most probable, the specific outcome depends on the particular
realization of orbital eccentricity evolution that Mercury
experienced.  The capture probability can be increased either if
Mercury's eccentricity reached very low values~\citep{corr09} or if
Mercury started in a retrograde spin configuration and became locked
in a synchronous state that was later destabilized by large
impacts~\citep{wiec12}.  Core-mantle friction also affects Mercury's
obliquity evolution, which itself can affect resonance capture
probabilities~\citep{corr10}.

The capture probability results depend on the choice of the tidal
torque formulation, which often relies on assumptions of constant time
lag or constant phase lag.  Models that incorporate a different
formulation based on a Darwin-Kaula expansion of the tidal torque
yield different capture probabilities~\citep{maka12,noye14} from
models that rely on a formulation with constant time lag or constant
phase lag.  The model of \citet{maka12} predicts 100\% capture
probability in the 3:2 resonance but does not include orbital
eccentricity variations.  The model of \citet{noye14} predicts capture
in a 2:1 or higher resonance unless Mercury was captured in the 3:2
resonance early in its evolution, i.e., before differentiation was
complete.
However, \citet{corr12} argued that large collisions destabilized all
spin-orbit resonances experienced early in Mercury's history and that
orbital eccentricity evolution dictated the final outcome.  According
to \citet{corr12}, the most probable outcome ($\sim$50\%) is capture
in the 3:2 resonance, regardless of the details of the tidal
formulation, core-mantle friction formulation, or collisional history.

Estimates for the timing of capture in the 3:2 resonance range from
very early~\citep[i.e., before differentiation was complete,][]{noye14}
to very late~\citep[i.e., $10^9$ y after formation,][]{tosi15}.

\subsection{Magnetic field generation}
Knowledge of Mercury's internal structure played a key role in solving
a long-standing puzzle related to the origin of the magnetic field.
The field that was detected by Mariner 10~\citep{ness74} appeared to
have an orientation similar to that of the spin axis.  For many years,
a dynamo mechanism involving motion in an electrically conducting
molten outer core was the preferred explanation for the origin of the
field~\citep{ness75,stev83rpp}, but alternative theories that do not
require a currently liquid core, such as remanent magnetism in the
crust, could not be ruled out~\citep{step76,ahar04}.  Because an
active dynamo was not the only possible mechanism for producing the
observed field, the detection of the magnetic field left the nature of
Mercury's core uncertain.  The unambiguous dynamical evidence provided
by libration measurements (Section~\ref{sec-librations}) indicated
that Mercury's outer librating shell is decoupled from the deep
interior and that Mercury's outer core must be molten.  Because a
liquid core is a necessary condition for dynamo action, the case for a
currently active dynamo was strengthened by the spin state
observations.  Magnetic field observations from MESSENGER's first two
flybys could not be unambiguously attributed to a dynamo
mechanism~\citep{ande08,ande10}.  After orbital insertion, however,
the case for a deep dynamo gradually became
incontrovertible~\citep[][Chapter 5]{ande12}.

\citet{stev83rpp,stev10} has shown that the existence of convection in
a partially molten core, rather than the vigor of that convection, is
the primary determinant of dynamo action.  He estimated that a fluid
layer thickness of order 100 km or more is required for sustaining
convection by compositional buoyancy in Mercury.  Given the $\sim$2000
km radius of the fluid outer core determined by Mercury's moments of
inertia, a convecting layer of sufficient depth can be easily
accommodated.  If it were not, the signature of an enormous inner core
would be detectable (Section~\ref{sec-inner}).  The lack of
information about the size of Mercury's inner core prevents a thorough
investigation of the working of the dynamo responsible for Mercury's
magnetic field.  Measurement of the inner core size is therefore an
important goal for future investigations.
Detailed discussions of Mercury's magnetic field and models for the
generation of that field over the history of the planet are given in
Chapters 5 and 19.

\section{Conclusions}

We have reviewed Mercury's rotational dynamics (Section~\ref{sec-rot}) and
showed how gravity (Section~\ref{sec-grav}) and spin
(Section~\ref{sec-spin}) observations can provide powerful bounds on
Mercury's internal structure.  We discussed the results of two- and
three-layer structural models (Sections~\ref{sec-2}), which provide a
good approximation to the results of more complex models.

Additional constraints derived from compositional studies
(Section~\ref{sec-comp}) enable the development of multi-layer models,
which admit a wide range of solutions (Section~\ref{sec-N}).  To
further constrain the range of possible models, we calculated the
tidal response of the planet and compared it with observations of the
$k_2$ Love number (Section~\ref{sec-tides}).  We have examined the
influence of an inner core on the spin state and the determination of
the moment of inertia (Section~\ref{sec-inner}), and we have presented
circumstantial evidence for a small
inner core.

We have described the statistical properties of 1\,479 interior models
that provide the best fit to the moment of inertia data.  We also
described a Preliminary Reference Mercury Model that incorporates all
existing constraints, including constraints on crustal density and
thickness (Section~\ref{sec-prmm}).  The description of radial
profiles of density, gravitational acceleration, and pressure will
prove useful for a variety of modeling tasks.

We have discussed the wide-ranging implications of Mercury's internal
structure on its thermal evolution, surface geology, capture in its
distinctive spin-orbit resonance, and magnetic field generation
(Section~\ref{sec-implications}).

\citet{peal76}'s ingenious procedure to determine the size and state
of Mercury's core permeates this work.  His insight allowed us to
quantify the properties of Mercury's core such that, at the time of
this writing, we know more about the core of Mercury than that of any
planet other than Earth.

Additional observations are necessary to place bounds on the size of
Mercury's inner core, either by improved measurements of longitudinal
librations or seismological observations.  The BepiColombo
mission~\citep[Chapter
  20]{nova02,balo02,jehn04,balo07,benk10,pfyf11,cica12} or a lander
mission~\citep{wu95} are expected to improve our knowledge of
Mercury's internal structure substantially.

\bibliography{int}\label{refs}

\begin{thebibliography}{165}
\expandafter\ifx\csname natexlab\endcsname\relax\def\natexlab#1{#1}\fi
\expandafter\ifx\csname selectlanguage\endcsname\relax
  \def\selectlanguage#1{\relax}\fi

\bibitem[\protect\citename{{Aharonson} {et~al.}, }2004]{ahar04}
{Aharonson}, O., {Zuber}, M.~T. and {Solomon}, S.~C. (2004).
\newblock Crustal remanence in an internally magnetized non-uniform shell: A
  possible source for Mercury's magnetic field?
\newblock {\em Earth Planet. Sci. Lett.}, {\bf 218}, 261--268.
\newblock doi: 10.1016/S0012-821X(03)00682-4.

\bibitem[\protect\citename{{Alterman} {et~al.}, }1959]{Alterman1959}
{Alterman}, Z., {Jarosch}, H. and {Pekeris}, C.~L. (1959).
\newblock {Oscillations of the Earth}.
\newblock {\em Proc. Roy. Soc. A}, {\bf 252}, 80--95.
\newblock doi: 10.1098/rspa.1959.0138.

\bibitem[\protect\citename{{Anderson} {et~al.}, }2008]{ande08}
{Anderson}, B.~J., {Acu{\~n}a}, M.~H., {Korth}, H., {Purucker}, M.~E.,
  {Johnson}, C.~L., {Slavin}, J.~A., {Solomon}, S.~C. and {McNutt}, R.~L., Jr.
  (2008).
\newblock The structure of Mercury's magnetic field from MESSENGER's first
  flyby.
\newblock {\em Science}, {\bf 321}, 82--85.
\newblock doi: 10.1126/science.1159081.

\bibitem[\protect\citename{{Anderson} {et~al.}, }2010]{ande10}
{Anderson}, B.~J., {Acu{\~n}a}, M.~H., {Korth}, H., {Slavin}, J.~A., {Uno}, H.,
  {Johnson}, C.~L., {Purucker}, M.~E., {Solomon}, S.~C., {Raines}, J.~M.,
  {Zurbuchen}, T.~H., {Gloeckler}, G. and {McNutt}, R.~L., Jr. (2010).
\newblock The magnetic field of Mercury.
\newblock {\em Space Sci. Rev.}, {\bf 152}, 307--339.
\newblock doi: 10.1007/978-1-4419-5901-0\_10.

\bibitem[\protect\citename{{Anderson} {et~al.}, }2012]{ande12}
{Anderson}, B.~J., {Johnson}, C.~L., {Korth}, H., {Winslow}, R.~M., {Borovsky},
  J.~E., {Purucker}, M.~E., {Slavin}, J.~A., {Solomon}, S.~C., {Zuber}, M.~T.
  and {McNutt}, R.~L., Jr. (2012).
\newblock {Low-degree structure in Mercury's planetary magnetic field}.
\newblock {\em J. Geophys. Res.}, {\bf 117}, E00L12.
\newblock doi: 10.1029/2012JE004159.

\bibitem[\protect\citename{{Anderson} {et~al.}, }1987]{ande87}
{Anderson}, J.~D., {Colombo}, G., {Esposito}, P.~B., {Lau}, E.~L. and {Trager},
  G.~B. (1987).
\newblock The mass, gravity field, and ephemeris of Mercury.
\newblock {\em Icarus}, {\bf 71}, 337--349.
\newblock doi: 10.1016/0019-1035(87)90033-9.

\bibitem[\protect\citename{{Archinal} {et~al.}, }2011]{arch11}
{Archinal}, B.~A., {A'Hearn}, M.~F., {Bowell}, E., {Conrad}, A., {Consolmagno},
  G.~J., {Courtin}, R., {Fukushima}, T., {Hestroffer}, D., {Hilton}, J.~L.,
  {Krasinsky}, G.~A., {Neumann}, G., {Oberst}, J., {Seidelmann}, P.~K.,
  {Stooke}, P., {Tholen}, D.~J., {Thomas}, P.~C. and {Williams}, I.~P. (2011).
\newblock {Report of the IAU Working Group on Cartographic Coordinates and
  Rotational Elements: 2009}.
\newblock {\em Celest. Mech. Dyn. Astron.}, {\bf 109}, 101--135.
\newblock doi: 10.1007/s10569-010-9320-4.

\bibitem[\protect\citename{{Ash} {et~al.}, }1971]{ash71}
{Ash}, M.~E., {Shapiro}, I.~I. and {Smith}, W.~B. (1971).
\newblock {The system of planetary masses}.
\newblock {\em Science}, {\bf 174}, 551--556.
\newblock doi: 10.1126/science.174.4009.551.

\bibitem[\protect\citename{{Baland} {et~al.}, }2017]{bala17}
{Baland}, R.-M., {Yseboodt}, M., {Rivoldini}, A. and {Van Hoolst}, T. (2017).
\newblock {Obliquity of Mercury: Influence of the precession of the pericenter
  and of tides}.
\newblock {\em Icarus}, {\bf 291}, 136--159.
\newblock doi: 10.1016/j.icarus.2017.03.020.

\bibitem[\protect\citename{{Balogh} and {Giampieri}, }2002]{balo02}
{Balogh}, A. and {Giampieri}, G. (2002).
\newblock Mercury: The planet and its orbit.
\newblock {\em Rep. Prog. Phys.}, {\bf 65}, 529--560.
\newblock doi: 10.1088/0034-4885/65/4/202.

\bibitem[\protect\citename{{Balogh} {et~al.}, }2007]{balo07}
{Balogh}, A., {Grard}, R., {Solomon}, S.~C., {Schulz}, R., {Langevin}, Y.,
  {Kasaba}, Y. and {Fujimoto}, M. (2007).
\newblock Missions to Mercury.
\newblock {\em Space Sci. Rev.}, {\bf 132}, 611--645.
\newblock doi: 10.1007/978-0-387-77539-5\_16.

\bibitem[\protect\citename{{Benkhoff} {et~al.}, }2010]{benk10}
{Benkhoff}, J., {van Casteren}, J., {Hayakawa}, H., {Fujimoto}, M., {Laakso},
  H., {Novara}, M., {Ferri}, P., {Middleton}, H.~R. and {Ziethe}, R. (2010).
\newblock BepiColombo -- Comprehensive exploration of Mercury: Mission overview
  and science goals.
\newblock {\em Planet. Space Sci.}, {\bf 58}, 2--20.
\newblock doi: 10.1016/j.pss.2009.09.020.

\bibitem[\protect\citename{{Bills} and {Comstock}, }2005]{bill05}
{Bills}, B.~G. and {Comstock}, R.~L. (2005).
\newblock {Forced obliquity variations of Mercury}.
\newblock {\em J. Geophys. Res.}, {\bf 110}, E04006.
\newblock doi: 10.1029/2003JE002116.

\bibitem[\protect\citename{{Bois}, }1995]{bois95}
{Bois}, E. (1995).
\newblock Proposed terminology for a general classification of rotational swing
  motions of the celestial solid bodies.
\newblock {\em Astron. Astrophys.}, {\bf 296}, 850--857.

\bibitem[\protect\citename{{Bois} and {Rambaux}, }2007]{bois07}
{Bois}, E. and {Rambaux}, N. (2007).
\newblock On the oscillations in Mercury's obliquity.
\newblock {\em Icarus}, {\bf 192}, 308--317.
\newblock doi: 10.1016/j.icarus.2007.07.015.

\bibitem[\protect\citename{{Borch} and {Green}, }1987]{Borch1987}
{Borch}, R.~S. and {Green}, H.~W., II. (1987).
\newblock {Dependence of creep in olivine on homologous temperature and its
  implications for flow in the mantle}.
\newblock {\em Nature}, {\bf 330}, 345--348.
\newblock doi: 10.1038/330345a0.

\bibitem[\protect\citename{{Breuer} {et~al.}, }2007]{breu07}
{Breuer}, D., {Hauck}, {S.~A., II}, {Buske}, M., {Pauer}, M. and {Spohn}, T.
  (2007).
\newblock Interior evolution of Mercury.
\newblock {\em Space Sci. Rev.}, {\bf 132}, 229--260.
\newblock doi: 10.1007/978-0-387-77539-5\_4.

\bibitem[\protect\citename{{Burbine} {et~al.}, }2002]{burb02}
{Burbine}, T.~H., {McCoy}, T.~J., {Nittler}, L.~R., {Benedix}, G.~K.,
  {Cloutis}, E.~A. and {Dickinson}, T.~L. (2002).
\newblock {Spectra of extremely reduced assemblages: Implications for Mercury}.
\newblock {\em Meteorit. Planet. Sci.}, {\bf 37}, 1233--1244.
\newblock doi: 10.1111/j.1945-5100.2002.tb00892.x.

\bibitem[\protect\citename{Byrne {et~al.}, }2014]{byrn14}
Byrne, P.~K., Klimczak, C., Celal~Sengor, A.~M., Solomon, S.~C., Watters, T.~R.
  and {Hauck}, {S.~A., II}. (2014).
\newblock Mercury's global contraction much greater than earlier estimates.
\newblock {\em Nature Geosci.}, {\bf 7}, 301--307.
\newblock doi: 10.1038/ngeo2097.

\bibitem[\protect\citename{{Cao} {et~al.}, }2014]{cao14}
{Cao}, H., {Aurnou}, J.~M., {Wicht}, J., {Dietrich}, W., {Soderlund}, K.~M. and
  {Russell}, C.~T. (2014).
\newblock A dynamo explanation for Mercury's anomalous magnetic field.
\newblock {\em Geophys. Res. Lett.}, {\bf 41}, 4127--4134.
\newblock doi: 10.1002/2014GL060196.

\bibitem[\protect\citename{{Cavanaugh} {et~al.}, }2007]{cava07}
{Cavanaugh}, J.~F., {Smith}, J.~C., {Sun}, X., {Bartels}, A.~E.,
  {Ramos-Izquierdo}, L., {Krebs}, D.~J., {McGarry}, J.~F., {Trunzo}, R.,
  {Novo-Gradac}, A.~M., {Britt}, J.~L., {Karsh}, J., {Katz}, R.~B., {Lukemire},
  A.~T., {Szymkiewicz}, R., {Berry}, D.~L., {Swinski}, J.~P., {Neumann}, G.~A.,
  {Zuber}, M.~T. and {Smith}, D.~E. (2007).
\newblock {The Mercury Laser Altimeter instrument for the MESSENGER mission}.
\newblock {\em Space Sci. Rev.}, {\bf 131}, 451--479.
\newblock doi: 10.1007/s11214-007-9273-4.

\bibitem[\protect\citename{{Chabot} {et~al.}, }2014]{chab14}
{Chabot}, N.~L., {Wollack}, E.~A., {Klima}, R.~L. and {Minitti}, M.~E. (2014).
\newblock {Experimental constraints on Mercury's core composition}.
\newblock {\em Earth Planet. Sci. Lett.}, {\bf 390}, 199--208.
\newblock doi: 10.1016/j.epsl.2014.01.004.

\bibitem[\protect\citename{{Charlier} {et~al.}, }2013]{char13}
{Charlier}, B., {Grove}, T.~L. and {Zuber}, M.~T. (2013).
\newblock Phase equilibria of ultramafic compositions on Mercury and the origin
  of the compositional dichotomy.
\newblock {\em Earth Planet. Sci. Lett.}, {\bf 363}, 50--60.
\newblock doi: 10.1016/j.epsl.2012.12.021.

\bibitem[\protect\citename{{Chen} {et~al.}, }2008]{chen08}
{Chen}, B., {Li}, J. and {Hauck}, {S.~A., II}. (2008).
\newblock {Non--ideal liquidus curve in the Fe-S system and Mercury's snowing
  core}.
\newblock {\em Geophys. Res. Lett.}, {\bf 35}, L07201.
\newblock doi: 10.1029/2008GL033311.

\bibitem[\protect\citename{{Cical{\`o}} and {Milani}, }2012]{cica12}
{Cical{\`o}}, S. and {Milani}, A. (2012).
\newblock Determination of the rotation of Mercury from satellite gravimetry.
\newblock {\em Mon. Not. Roy. Astron. Soc.}, {\bf 427}, 468--482.
\newblock doi: 10.1111/j.1365-2966.2012.21919.x.

\bibitem[\protect\citename{{Colombo}, }1965]{colo65}
{Colombo}, G. (1965).
\newblock Rotational period of the planet Mercury.
\newblock {\em Nature}, {\bf 208}, 575.
\newblock doi: 10.1038/208575a0.

\bibitem[\protect\citename{{Colombo}, }1966]{colo66}
{Colombo}, G. (1966).
\newblock Cassini's second and third laws.
\newblock {\em Astron. J.}, {\bf 71}, 891--896.
\newblock doi: 10.1007/978-94-010-3529-3\_2.

\bibitem[\protect\citename{{Correia} and {Laskar}, }2004]{corr04}
{Correia}, A.~C.~M. and {Laskar}, J. (2004).
\newblock Mercury's capture into the 3/2 spin-orbit resonance as a result of
  its chaotic dynamics.
\newblock {\em Nature}, {\bf 429}, 848--850.
\newblock doi: 10.1038/nature02609.

\bibitem[\protect\citename{{Correia} and {Laskar}, }2009]{corr09}
{Correia}, A.~C.~M. and {Laskar}, J. (2009).
\newblock Mercury's capture into the 3/2 spin-orbit resonance including the
  effect of core-mantle friction.
\newblock {\em Icarus}, {\bf 201}, 1--11.
\newblock doi: 10.1016/j.icarus.2008.12.034.

\bibitem[\protect\citename{{Correia} and {Laskar}, }2010]{corr10}
{Correia}, A.~C.~M. and {Laskar}, J. (2010).
\newblock Long-term evolution of the spin of Mercury. I. Effect of the
  obliquity and core-mantle friction.
\newblock {\em Icarus}, {\bf 205}, 338--355.
\newblock doi: 10.1016/j.icarus.2009.08.006.

\bibitem[\protect\citename{{Correia} and {Laskar}, }2012]{corr12}
{Correia}, A.~C.~M. and {Laskar}, J. (2012).
\newblock Impact cratering on Mercury: Consequences for the spin evolution.
\newblock {\em Astrophys. J.}, {\bf 751}, L43.
\newblock doi: 10.1088/2041-8205/751/2/L43.

\bibitem[\protect\citename{{Counselman} and {Shapiro}, }1970]{coun70}
{Counselman}, C.~C., III and {Shapiro}, I.~I. (1970).
\newblock Spin-orbit resonance of Mercury.
\newblock {\em Symp. Math.}, {\bf 3}, 121--169.

\bibitem[\protect\citename{{Davies} {et~al.}, }1980]{davi80}
{Davies}, M.~F., {Abalakin}, V.~K., {Duncombe}, R.~L., {Masursky}, H.,
  {Morando}, B., {Owen}, T.~C., {Seidelmann}, P.~K., {Sinclair}, A.~T.,
  {Wilkins}, G.~A. and {Cross}, C.~A. (1980).
\newblock Report of the IAU Working Group on Cartographic Coordinates and
  Rotational Elements of the Planets and Satellites.
\newblock {\em Celest. Mech.}, {\bf 22}, 205--230.
\newblock doi: 10.1007/BF01229508.

\bibitem[\protect\citename{{D'Hoedt} and {Lema{\^i}tre}, }2008]{dhoe08}
{D'Hoedt}, S. and {Lema{\^i}tre}, A. (2008).
\newblock Planetary long periodic terms in Mercury's rotation: A two
  dimensional adiabatic approach.
\newblock {\em Celest. Mech. Dyn. Astron.}, {\bf 101}, 127--139.
\newblock doi: 10.1007/s10569-007-9115-4.

\bibitem[\protect\citename{{D'Hoedt} {et~al.}, }2009]{dhoe09}
{D'Hoedt}, S., {Noyelles}, B., {Dufey}, J. and {Lemaitre}, A. (2009).
\newblock Determination of an instantaneous Laplace plane for Mercury's
  rotation.
\newblock {\em Adv. Space Res.}, {\bf 44}, 597--603.
\newblock doi: 10.1016/j.asr.2009.05.008.

\bibitem[\protect\citename{{Dufey} {et~al.}, }2008]{dufe08}
{Dufey}, J., {Lema{\^i}tre}, A. and {Rambaux}, N. (2008).
\newblock Planetary perturbations on Mercury's libration in longitude.
\newblock {\em Celest. Mech. Dyn. Astron.}, {\bf 101}, 141--157.
\newblock doi: 10.1007/s10569-008-9143-8.

\bibitem[\protect\citename{{Dumberry}, }2011]{dumb11}
{Dumberry}, M. (2011).
\newblock {The free librations of Mercury and the size of its inner core}.
\newblock {\em Geophys. Res. Lett.}, {\bf 38}, L16202.
\newblock doi: 10.1029/2011GL048277.

\bibitem[\protect\citename{{Dumberry} and {Rivoldini}, }2015]{dumb15}
{Dumberry}, M. and {Rivoldini}, A. (2015).
\newblock {Mercury's inner core size and core-crystallization regime}.
\newblock {\em Icarus}, {\bf 248}, 254--268.
\newblock doi: 10.1016/j.icarus.2014.10.038.

\bibitem[\protect\citename{{Dumberry} {et~al.}, }2013]{dumb13}
{Dumberry}, M., {Rivoldini}, A., {Van Hoolst}, T. and {Yseboodt}, M. (2013).
\newblock The role of Mercury's core density structure on its longitudinal
  librations.
\newblock {\em Icarus}, {\bf 225}, 62--74.
\newblock doi: 10.1016/j.icarus.2013.03.001.

\bibitem[\protect\citename{{Dziewonski} and {Anderson}, }1981]{dzie81}
{Dziewonski}, A.~M. and {Anderson}, D.~L. (1981).
\newblock Preliminary reference Earth model.
\newblock {\em Phys. Earth Planet. Inter.}, {\bf 25}, 297--356.
\newblock doi: 10.1016/0031-9201(81)90046-7.

\bibitem[\protect\citename{{Efroimsky}, }2012]{Efroimsky2012}
{Efroimsky}, M. (2012).
\newblock {Bodily tides near spin-orbit resonances}.
\newblock {\em Celest. Mech. Dyn. Astron.}, {\bf 112}, 283--330.
\newblock doi: 10.1007/s10569-011-9397-4.

\bibitem[\protect\citename{{Efroimsky} and {Lainey}, }2007]{Efroimsky2007}
{Efroimsky}, M. and {Lainey}, V. (2007).
\newblock {Physics of bodily tides in terrestrial planets and the appropriate
  scales of dynamical evolution}.
\newblock {\em J. Geophys. Res.}, {\bf 112}, E12003.
\newblock doi: 10.1029/2007JE002908.

\bibitem[\protect\citename{{Evans} {et~al.}, }2012]{evan12}
{Evans}, L.~G., {Peplowski}, P.~N., {Rhodes}, E.~A., {Lawrence}, D.~J.,
  {McCoy}, T.~J., {Nittler}, L.~R., {Solomon}, S.~C., {Sprague}, A.~L.,
  {Stockstill-Cahill}, K.~R., {Starr}, R.~D., {Weider}, S.~Z., {Boynton},
  W.~V., {Hamara}, D.~K. and {Goldsten}, J.~O. (2012).
\newblock {Major-element abundances on the surface of Mercury: Results from the
  MESSENGER Gamma-Ray Spectrometer}.
\newblock {\em J. Geophys. Res.}, {\bf 117}, E00L07.
\newblock doi: 10.1029/2012JE004178.

\bibitem[\protect\citename{{Fei} {et~al.}, }1995]{Fei1995}
{Fei}, Y., {Prewitt}, C.~T., {Mao}, H.-K. and {Bertka}, C.~M. (1995).
\newblock {Structure and density of FeS at high pressure and high temperature
  and the internal structure of Mars}.
\newblock {\em Science}, {\bf 268}, 1892--1894.
\newblock doi: 10.1126/science.268.5219.1892.

\bibitem[\protect\citename{{Genova} {et~al.}, }2013]{geno13}
{Genova}, A., {Iess}, L. and {Marabucci}, M. (2013).
\newblock {Mercury's gravity field from the first six months of MESSENGER
  data}.
\newblock {\em Planet. Space Sci.}, {\bf 81}, 55--64.
\newblock doi: 10.1016/j.pss.2013.02.006.

\bibitem[\protect\citename{{Gladman} {et~al.}, }1996]{glad96}
{Gladman}, B., {Dane Quinn}, D., {Nicholson}, P. and {Rand}, R. (1996).
\newblock Synchronous locking of tidally evolving satellites.
\newblock {\em Icarus}, {\bf 122}, 166--192.
\newblock doi: 10.1006/icar.1996.0117.

\bibitem[\protect\citename{{Goldreich} and {Peale}, }1966]{gold66aj}
{Goldreich}, P. and {Peale}, S. (1966).
\newblock Spin--orbit coupling in the solar system.
\newblock {\em Astron. J.}, {\bf 71}, 425--438.
\newblock doi: 10.1086/109947.

\bibitem[\protect\citename{{Goldreich} and {Peale}, }1967]{gold67}
{Goldreich}, P. and {Peale}, S. (1967).
\newblock Spin-orbit coupling in the solar system. II. The resonant rotation of
  Venus.
\newblock {\em Astron. J.}, {\bf 72}, 662--668.
\newblock doi: 10.1086/110289.

\bibitem[\protect\citename{{Grott} {et~al.}, }2011]{grot11}
{Grott}, M., {Breuer}, D. and {Laneuville}, M. (2011).
\newblock {Thermo-chemical evolution and global contraction of Mercury}.
\newblock {\em Earth Planet. Sci. Lett.}, {\bf 307}, 135--146.
\newblock doi: 10.1016/j.epsl.2011.04.040.

\bibitem[\protect\citename{{Harder} and {Schubert}, }2001]{hard01}
{Harder}, H. and {Schubert}, G. (2001).
\newblock Sulfur in Mercury's core?
\newblock {\em Icarus}, {\bf 151}, 118--122.
\newblock doi: 10.1006/icar.2001.6586.

\bibitem[\protect\citename{{Hauck} {et~al.}, }2007]{hauc07}
{Hauck}, {S.~A., II}, {Solomon}, S.~C. and {Smith}, D.~A. (2007).
\newblock Predicted recovery of Mercury's internal structure by MESSENGER.
\newblock {\em Geophys. Res. Lett.}, {\bf 34}, L18201.
\newblock doi: 10.1029/2007GL030793.

\bibitem[\protect\citename{{Hauck} {et~al.}, }2013]{hauc13}
{Hauck}, {S.~A., II}, {Margot}, J.~L., {Solomon}, S.~C., {Phillips}, R.~J.,
  {Johnson}, C.~L., {Lemoine}, F.~G., {Mazarico}, E., {McCoy}, T.~J.,
  {Padovan}, S., {Peale}, S.~J., {Perry}, M.~E., {Smith}, D.~E. and {Zuber},
  M.~T. (2013).
\newblock The curious case of Mercury's internal structure.
\newblock {\em J. Geophys. Res. Planets}, {\bf 118}, 1204--1220.
\newblock doi: 10.1002/jgre.20091.

\bibitem[\protect\citename{{Hawkins} {et~al.}, }2007]{hawk07}
{Hawkins}, S.~E., III, {Boldt}, J.~D., {Darlington}, E.~H., {Espiritu}, R.,
  {Gold}, R.~E., {Gotwols}, B., {Grey}, M.~P., {Hash}, C.~D., {Hayes}, J.~R.,
  {Jaskulek}, S.~E., {Kardian}, C.~J., {Keller}, M.~R., {Malaret}, E.~R.,
  {Murchie}, S.~L., {Murphy}, P.~K., {Peacock}, K., {Prockter}, L.~M.,
  {Reiter}, R.~A., {Robinson}, M.~S., {Schaefer}, E.~D., {Shelton}, R.~G.,
  {Sterner}, R.~E., II, {Taylor}, H.~W., {Watters}, T.~R. and {Williams}, B.~D.
  (2007).
\newblock {The Mercury Dual Imaging System on the MESSENGER spacecraft}.
\newblock {\em Space Sci. Rev.}, {\bf 131}, 247--338.
\newblock doi: 10.1007/s11214-007-9266-3.

\bibitem[\protect\citename{{Hirschmann}, }2000]{Hirschmann2000}
{Hirschmann}, M.~M. (2000).
\newblock {Mantle solidus: Experimental constraints and the effects of
  peridotite composition}.
\newblock {\em Geochem. Geophys. Geosyst.}, {\bf 1}, 1042--1068.
\newblock doi: 10.1029/2000GC000070.

\bibitem[\protect\citename{Hofmeister and Mao, }{2003}]{Hofmeister2003}
Hofmeister, A.~M. and Mao, H.~K. ({2003}).
\newblock {Pressure derivatives of shear and bulk moduli from the thermal
  Gruneisen parameter and volume-pressure data}.
\newblock {\em {Geochim. Cosmochim. Acta}}, {\bf {67}}, 1207--1227.
\newblock doi: 10.1016/S0016-7037(02)01289-9.

\bibitem[\protect\citename{Holin, }1988]{holi88}
Holin, I.~V. (1988).
\newblock Space-time coherence of a signal diffusely scattered by an
  arbitrarily moving surface in the case of monochromatic sounding.
\newblock {\em Izvestiya Vysshikh Uchebnykh Zavedenii, Radiofizika}, {\bf
  31}(5), 515--518.

\bibitem[\protect\citename{Holin, }1992]{holi92}
Holin, I.~V. (1992).
\newblock Accuracy of body-rotation-parameter measurement with monochromatic
  illumination and two-element reception.
\newblock {\em Izvestiya Vysshikh Uchebnykh Zavedenii, Radiofizika}, {\bf
  35}(5), 433--439.
\newblock doi: 10.1007/BF01038312.

\bibitem[\protect\citename{{Howard} {et~al.}, }1974]{howa74}
{Howard}, H.~T., {Tyler}, G.~L., {Esposito}, P.~B., {Anderson}, J.~D.,
  {Reasenberg}, R.~D., {Shapiro}, I.~I., {Fjeldbo}, G., {Kliore}, A.~J.,
  {Levy}, G.~S., {Brunn}, D.~L., {Dickinson}, R., {Edelson}, R.~E., {Martin},
  W.~L., {Postal}, R.~B., {Seidel}, B., {Sesplaukis}, T.~T., {Shirley}, D.~L.,
  {Stelzried}, C.~T., {Sweetnam}, D.~N., {Wood}, G.~E. and {Zygielbaum}, A.~I.
  (1974).
\newblock Mercury: Results on mass, radius, ionosphere, and atmosphere from
  Mariner 10 dual--frequency radio signals.
\newblock {\em Science}, {\bf 185}, 179--180.
\newblock doi: 10.1126/science.185.4146.179.

\bibitem[\protect\citename{{Iess} {et~al.}, }2012]{Iess2012}
{Iess}, L., {Jacobson}, R.~A., {Ducci}, M., {Stevenson}, D.~J., {Lunine},
  J.~I., {Armstrong}, J.~W., {Asmar}, S.~W., {Racioppa}, P., {Rappaport}, N.~J.
  and {Tortora}, P. (2012).
\newblock {The tides of Titan}.
\newblock {\em Science}, {\bf 337}, 457--459.
\newblock doi: 10.1126/science.1219631.

\bibitem[\protect\citename{{Jackson} {et~al.}, }2010]{Jackson2010}
{Jackson}, I., {Faul}, U.~H., {Suetsugu}, D., {Bina}, C., {Inoue}, T. and
  {Jellinek}, M. (2010).
\newblock {Grainsize-sensitive viscoelastic relaxation in olivine: Towards a
  robust laboratory-based model for seismological application}.
\newblock {\em Phys. Earth Planet. Inter.}, {\bf 183}, 151--163.
\newblock doi: 10.1016/j.pepi.2010.09.005.

\bibitem[\protect\citename{{James} {et~al.}, }2015]{jame15}
{James}, P.~B., {Zuber}, M.~T., {Phillips}, R.~J. and {Solomon}, S.~C. (2015).
\newblock Support of long-wavelength topography on Mercury inferred from
  MESSENGER measurements of gravity and topography.
\newblock {\em J. Geophys. Res. Planets}, {\bf 120}, 287--310.
\newblock doi: 10.1002/2014JE004713.

\bibitem[\protect\citename{{Jehn} {et~al.}, }2004]{jehn04}
{Jehn}, R., {Corral}, C. and {Giampieri}, G. (2004).
\newblock Estimating Mercury's 88-day libration amplitude from orbit.
\newblock {\em Planet. Space Sci.}, {\bf 52}, 727--732.
\newblock doi: 10.1016/j.pss.2003.12.012.

\bibitem[\protect\citename{{Kaufmann} and {Lambeck}, }2000]{Kaufmann2000}
{Kaufmann}, G. and {Lambeck}, K. (2000).
\newblock {Mantle dynamics, postglacial rebound and the radial viscosity
  profile}.
\newblock {\em Phys. Earth Planet. Inter.}, {\bf 121}, 301--324.
\newblock doi: 10.1016/S0031-9201(00)00174-6.

\bibitem[\protect\citename{{Kaula}, }2000]{kaul00}
{Kaula}, W.~M. (2000).
\newblock {\em Theory of Satellite Geodesy: Applications of Satellites to
  Geodesy}.
\newblock Dover Publications, Mineola, NY.

\bibitem[\protect\citename{{Klaasen}, }1976]{klaa76}
{Klaasen}, K.~P. (1976).
\newblock Mercury's rotation axis and period.
\newblock {\em Icarus}, {\bf 28}, 469--478.
\newblock doi: 10.1016/0019-1035(76)90120-2.

\bibitem[\protect\citename{{Knibbe} and {van Westrenen}, }2015]{knib15}
{Knibbe}, J.~S. and {van Westrenen}, W. (2015).
\newblock {The interior configuration of planet Mercury constrained by moment
  of inertia and planetary contraction}.
\newblock {\em J. Geophys. Res. Planets}, {\bf 120}, 1904--1923.
\newblock doi: 10.1002/2015JE004908.

\bibitem[\protect\citename{{Koning} and {Dumberry}, }2013]{koni13}
{Koning}, A. and {Dumberry}, M. (2013).
\newblock {Internal forcing of Mercury's long period free librations}.
\newblock {\em Icarus}, {\bf 223}, 40--47.
\newblock doi: 10.1016/j.icarus.2012.11.022.

\bibitem[\protect\citename{{Konopliv} and {Yoder}, }1996]{Konopliv1996}
{Konopliv}, A.~S. and {Yoder}, C.~F. (1996).
\newblock Venusian $k_{2}$ tidal Love number from Magellan and PVO tracking
  data.
\newblock {\em Geophys. Res. Lett.}, {\bf 23}, 1857--1860.
\newblock doi: 10.1029/96GL01589.

\bibitem[\protect\citename{Kuwayama and Hirose, }2004]{kuwa04}
Kuwayama, Y. and Hirose, K. (2004).
\newblock {Phase relations in the system Fe-FeSi at 21 GPa}.
\newblock {\em Am. Mineral.}, {\bf 89}, 273–--276.
\newblock doi: 10.2138/am-2004-2-303.

\bibitem[\protect\citename{{Makarov}, }2012]{maka12}
{Makarov}, V.~V. (2012).
\newblock Conditions of passage and entrapment of terrestrial planets in
  spin-orbit resonances.
\newblock {\em Astrophys. J.}, {\bf 752}, 73--80.
\newblock doi: 10.1088/0004-637X/752/1/73.

\bibitem[\protect\citename{{Malavergne} {et~al.}, }2010]{mala10}
{Malavergne}, V., {Toplis}, M.~J., {Berthet}, S. and {Jones}, J. (2010).
\newblock {Highly reducing conditions during core formation on Mercury:
  Implications for internal structure and the origin of a magnetic field}.
\newblock {\em Icarus}, {\bf 206}, 199--209.
\newblock doi: 10.1016/j.icarus.2009.09.001.

\bibitem[\protect\citename{{Margot}, }2009]{marg09cmda}
{Margot}, J.~L. (2009).
\newblock A Mercury orientation model including non-zero obliquity and
  librations.
\newblock {\em Celest. Mech. Dyn. Astron.}, {\bf 105}, 329--336.
\newblock doi: 10.1007/s10569-009-9234-1.

\bibitem[\protect\citename{{Margot} {et~al.}, }2007]{marg07}
{Margot}, J.~L., {Peale}, S.~J., {Jurgens}, R.~F., {Slade}, M.~A. and {Holin},
  I.~V. (2007).
\newblock {Large longitude libration of Mercury reveals a molten core}.
\newblock {\em Science}, {\bf 316}, 710--714.
\newblock doi: 10.1126/science.1140514.

\bibitem[\protect\citename{{Margot} {et~al.}, }2012]{marg12jgr}
{Margot}, J.~L., {Peale}, S.~J., {Solomon}, S.~C., {Hauck}, {S.~A., II},
  {Ghigo}, F.~D., {Jurgens}, R.~F., {Yseboodt}, M., {Giorgini}, J.~D.,
  {Padovan}, S. and {Campbell}, D.~B. (2012).
\newblock {Mercury's moment of inertia from spin and gravity data}.
\newblock {\em J. Geophys. Res.}, {\bf 117}, E00L09.
\newblock doi: 10.1029/2012JE004161.

\bibitem[\protect\citename{{Matsuyama} and {Nimmo}, }2009]{mats09}
{Matsuyama}, I. and {Nimmo}, F. (2009).
\newblock {Gravity and tectonic patterns of Mercury: Effect of tidal
  deformation, spin-orbit resonance, nonzero eccentricity, despinning, and
  reorientation}.
\newblock {\em J. Geophys. Res.}, {\bf 114}, E01010.
\newblock doi: 10.1029/2008JE003252.

\bibitem[\protect\citename{{Mazarico} {et~al.}, }2014]{maza14}
{Mazarico}, E., {Genova}, A., {Goossens}, S., {Lemoine}, F.~G., {Neumann},
  G.~A., {Zuber}, M.~T., {Smith}, D.~E. and {Solomon}, S.~C. (2014).
\newblock The gravity field, orientation, and ephemeris of Mercury from
  MESSENGER observations after three years in orbit.
\newblock {\em J. Geophys. Res. Planets}, {\bf 119}, 2417--2436.
\newblock doi: 10.1002/2014JE004675.

\bibitem[\protect\citename{{McCubbin} {et~al.}, }2012]{mccu12}
{McCubbin}, F.~M., {Riner}, M.~A., {Vander Kaaden}, K.~E. and {Burkemper},
  L.~K. (2012).
\newblock {Is Mercury a volatile--rich planet?}
\newblock {\em Geophys. Res. Lett.}, {\bf 39}, L09202.
\newblock doi: 10.1029/2012GL051711.

\bibitem[\protect\citename{{Michel} {et~al.}, }2013]{mich13}
{Michel}, N.~C., {Hauck}, {S.~A., II}, {Solomon}, S.~C., {Phillips}, R.~J.,
  {Roberts}, J.~H. and {Zuber}, M.~T. (2013).
\newblock {Thermal evolution of Mercury as constrained by MESSENGER
  observations}.
\newblock {\em J. Geophys. Res. Planets}, {\bf 118}, 1033--1044.
\newblock doi: 10.1002/jgre.20049.

\bibitem[\protect\citename{{Mohr} {et~al.}, }2016]{mohr16}
{Mohr}, P.~J., {Newell}, D.~B. and {Taylor}, B.~N. (2016).
\newblock {CODATA recommended values of the fundamental physical constants:
  2014}.
\newblock {\em Rev. Mod. Phys.}, {\bf 88}(3), 035009.
\newblock doi: 10.1103/RevModPhys.88.035009.

\bibitem[\protect\citename{{Moore} and {Schubert}, }2000]{Moore2000}
{Moore}, W.~B. and {Schubert}, G. (2000).
\newblock {Note: The tidal response of Europa}.
\newblock {\em Icarus}, {\bf 147}, 317--319.
\newblock doi: 10.1006/icar.2000.6460.

\bibitem[\protect\citename{{Moore} and {Schubert}, }2003]{Moore2003}
{Moore}, W.~B. and {Schubert}, G. (2003).
\newblock {The tidal response of Ganymede and Callisto with and without liquid
  water oceans}.
\newblock {\em Icarus}, {\bf 166}, 223--226.
\newblock doi: 10.1016/j.icarus.2003.07.001.

\bibitem[\protect\citename{{Morard} and {Katsura}, }2010]{mora10}
{Morard}, G. and {Katsura}, T. (2010).
\newblock {Pressure--temperature cartography of Fe-S-Si immiscible system}.
\newblock {\em Geochim. Cosmochim. Acta}, {\bf 74}, 3659--3667.
\newblock doi: 10.1016/j.gca.2010.03.025.

\bibitem[\protect\citename{{Morgan} and {Anders}, }1980]{Morgan1980}
{Morgan}, J.~W. and {Anders}, E. (1980).
\newblock {Chemical composition of Earth, Venus, and Mercury}.
\newblock {\em Proc. Nat. Acad. Sci.}, {\bf 77}, 6973--6977.
\newblock doi: 10.1073/pnas.77.12.6973.

\bibitem[\protect\citename{{Naidu} and {Margot}, }2015]{naid15}
{Naidu}, S.~P. and {Margot}, J.~L. (2015).
\newblock Near-Earth asteroid satellite spins under spin-orbit coupling.
\newblock {\em Astron. J.}, {\bf 149}, 80--90.
\newblock doi: 10.1088/0004-6256/149/2/80.

\bibitem[\protect\citename{{Namur} {et~al.}, }2016a]{namu16}
{Namur}, O., {Collinet}, M., {Charlier}, B., {Grove}, T.~L., {Holtz}, F. and
  {McCammon}, C. (2016a).
\newblock {Melting processes and mantle sources of lavas on Mercury}.
\newblock {\em Earth Planet. Sc. Lett.}, {\bf 439}, 117--128.
\newblock doi: 10.1016/j.epsl.2016.01.030.

\bibitem[\protect\citename{{Namur} {et~al.}, }2016b]{namu16core}
{Namur}, O., {Charlier}, B., {Holtz}, F., {Cartier}, C. and {McCammon}, C.
  (2016b).
\newblock {Sulfur solubility in reduced mafic silicate melts: Implications for
  the speciation and distribution of sulfur on Mercury}.
\newblock {\em Earth Planet. Sc. Lett.}, {\bf 448}, 102--114.
\newblock doi: 10.1016/j.epsl.2016.05.024.

\bibitem[\protect\citename{{Ness} {et~al.}, }1974]{ness74}
{Ness}, N.~F., {Behannon}, K.~W., {Lepping}, R.~P., {Whang}, Y.~C. and
  {Schatten}, K.~H. (1974).
\newblock Magnetic field observations near Mercury: Preliminary results from
  Mariner 10.
\newblock {\em Science}, {\bf 185}, 151--160.
\newblock doi: 10.1126/science.185.4146.151.

\bibitem[\protect\citename{{Ness} {et~al.}, }1975]{ness75}
{Ness}, N.~F., {Behannon}, K.~W., {Lepping}, R.~P. and {Whang}, Y.~C. (1975).
\newblock The magnetic field of Mercury. I.
\newblock {\em J. Geophys. Res.}, {\bf 80}, 2708--2716.
\newblock doi: 10.1017/S1539299600002562.

\bibitem[\protect\citename{{Nimmo} and {Faul}, }2013]{Nimmo2013}
{Nimmo}, F. and {Faul}, U.~H. (2013).
\newblock {Dissipation at tidal and seismic frequencies in a melt-free,
  anhydrous Mars}.
\newblock {\em J. Geophys. Res. Planets}, {\bf 118}, 2558--2569.
\newblock doi: 10.1002/2013JE004499.

\bibitem[\protect\citename{{Nimmo} {et~al.}, }2012]{Nimmo2012}
{Nimmo}, F., {Faul}, U.~H. and {Garnero}, E.~J. (2012).
\newblock {Dissipation at tidal and seismic frequencies in a melt-free Moon}.
\newblock {\em J. Geophys. Res.}, {\bf 117}, E09005.
\newblock doi: 10.1029/2012JE004160.

\bibitem[\protect\citename{{Nittler} {et~al.}, }2011]{nitt11}
{Nittler}, L.~R., {Starr}, R.~D., {Weider}, S.~Z., {McCoy}, T.~J., {Boynton},
  W.~V., {Ebel}, D.~S., {Ernst}, C.~M., {Evans}, L.~G., {Goldsten}, J.~O.,
  {Hamara}, D.~K., {Lawrence}, D.~J., {McNutt}, R.~L., Jr., {Schlemm}, C.~E.,
  {Solomon}, S.~C. and {Sprague}, A.~L. (2011).
\newblock {The major--element composition of Mercury's surface from MESSENGER
  X-ray spectrometry}.
\newblock {\em Science}, {\bf 333}, 1847--1850.
\newblock doi: 10.1126/science.1211567.

\bibitem[\protect\citename{{Novara}, }2002]{nova02}
{Novara}, M. (2002).
\newblock The BepiColombo ESA cornerstone mission to Mercury.
\newblock {\em Acta Astronaut.}, {\bf 51}, 387--395.
\newblock doi: 10.1016/S0094-5765(02)00065-6.

\bibitem[\protect\citename{{Noyelles} and {Lhotka}, }2013]{noye13}
{Noyelles}, B. and {Lhotka}, C. (2013).
\newblock The influence of orbital dynamics, shape and tides on the obliquity
  of Mercury.
\newblock {\em Adv. Space Res.}, {\bf 52}, 2085--2101.
\newblock doi: 10.1016/j.asr.2013.09.024.

\bibitem[\protect\citename{{Noyelles} {et~al.}, }2014]{noye14}
{Noyelles}, B., {Frouard}, J., {Makarov}, V.~V. and {Efroimsky}, M. (2014).
\newblock Spin-orbit evolution of Mercury revisited.
\newblock {\em Icarus}, {\bf 241}, 26--44.
\newblock doi: 10.1016/j.icarus.2014.05.045.

\bibitem[\protect\citename{Padovan {et~al.}, }2014]{pado14}
Padovan, S., {Margot}, J.~L., {Hauck}, {S.~A., II}, Moore, B. and Solomon,
  S.~C. (2014).
\newblock The tides of Mercury and possible implications for its interior
  structure.
\newblock {\em J. Geophys. Res. Planets}, {\bf 119}, 850--866.
\newblock doi: 10.1002/2013JE004459.

\bibitem[\protect\citename{{Padovan} {et~al.}, }2015]{pado15}
{Padovan}, S., {Wieczorek}, M.~A., {Margot}, J.~L., {Tosi}, N. and {Solomon},
  S.~C. (2015).
\newblock Thickness of the crust of Mercury from geoid-to-topography ratios.
\newblock {\em Geophys. Res. Lett.}, {\bf 42}, 1029--1038.
\newblock doi: 10.1002/2014GL062487.

\bibitem[\protect\citename{{Peale}, }1969]{peal69}
{Peale}, S.~J. (1969).
\newblock Generalized Cassini's laws.
\newblock {\em Astron. J.}, {\bf 74}, 483--489.
\newblock doi: 10.1086/110825.

\bibitem[\protect\citename{{Peale}, }1972]{peal72}
{Peale}, S.~J. (1972).
\newblock Determination of parameters related to the interior of Mercury.
\newblock {\em Icarus}, {\bf 17}, 168--173.
\newblock doi: 10.1016/0019-1035(72)90052-8.

\bibitem[\protect\citename{{Peale}, }1976]{peal76}
{Peale}, S.~J. (1976).
\newblock Does Mercury have a molten core?
\newblock {\em Nature}, {\bf 262}, 765--766.
\newblock doi: 10.1038/262765a0.

\bibitem[\protect\citename{{Peale}, }1981]{peal81}
{Peale}, S.~J. (1981).
\newblock Measurement accuracies required for the determination of a Mercurian
  liquid core.
\newblock {\em Icarus}, {\bf 48}, 143--145.
\newblock doi: 10.1016/0019-1035(81)90160-3.

\bibitem[\protect\citename{{Peale}, }1988]{peal88}
{Peale}, S.~J. (1988).
\newblock The rotational dynamics of Mercury and the state of its core. In {\em
  Mercury}, ed. {Vilas}, F., {Chapman}, C.~R. and {Matthews}, M.~S.
\newblock {pp.  461--493}.
\newblock Tucson, AZ: University of Arizona Press.

\bibitem[\protect\citename{{Peale}, }2005]{peal05}
{Peale}, S.~J. (2005).
\newblock The free precession and libration of Mercury.
\newblock {\em Icarus}, {\bf 178}, 4--18.
\newblock doi: 10.1016/j.icarus.2005.03.017.

\bibitem[\protect\citename{{Peale}, }2006]{peal06}
{Peale}, S.~J. (2006).
\newblock The proximity of Mercury's spin to Cassini state 1 from adiabatic
  invariance.
\newblock {\em Icarus}, {\bf 181}, 338--347.
\newblock doi: 10.1016/j.icarus.2005.10.006.

\bibitem[\protect\citename{{Peale} and {Boss}, }1977]{peal77}
{Peale}, S.~J. and {Boss}, A.~P. (1977).
\newblock A spin--orbit constraint on the viscosity of a Mercurian liquid core.
\newblock {\em J. Geophys. Res.}, {\bf 82}, 743--749.
\newblock doi: 10.1029/JB082i005p00743.

\bibitem[\protect\citename{{Peale} {et~al.}, }2002]{peal02}
{Peale}, S.~J., {Phillips}, R.~J., {Solomon}, S.~C., {Smith}, D.~E. and
  {Zuber}, M.~T. (2002).
\newblock A procedure for determining the nature of Mercury's core.
\newblock {\em Meteorit. Planet. Sci.}, {\bf 37}, 1269--1283.
\newblock doi: 10.1111/j.1945-5100.2002.tb00895.x.

\bibitem[\protect\citename{{Peale} {et~al.}, }2007]{peal07}
{Peale}, S.~J., {Yseboodt}, M. and {Margot}, J.~L. (2007).
\newblock Long-period forcing of Mercury's libration in longitude.
\newblock {\em Icarus}, {\bf 187}, 365--373.
\newblock doi: 10.1016/j.icarus.2006.10.028.

\bibitem[\protect\citename{{Peale} {et~al.}, }2009]{peal09}
{Peale}, S.~J., {Margot}, J.~L. and {Yseboodt}, M. (2009).
\newblock Resonant forcing of Mercury's libration in longitude.
\newblock {\em Icarus}, {\bf 199}, 1--8.
\newblock doi: 10.1016/j.icarus.2008.09.002.

\bibitem[\protect\citename{{Peale} {et~al.}, }2014]{peal14}
{Peale}, S.~J., {Margot}, J.~L., {Hauck}, {S.~A., II} and {Solomon}, S.~C.
  (2014).
\newblock Effect of core-mantle and tidal torques on Mercury's spin axis
  orientation.
\newblock {\em Icarus}, {\bf 231}, 206--220.
\newblock doi: 10.1016/j.icarus.2013.12.007.

\bibitem[\protect\citename{{Peale} {et~al.}, }2016]{peal16}
{Peale}, S.~J., {Margot}, J.~L., {Hauck}, {S.~A., II} and {Solomon}, S.~C.
  (2016).
\newblock {Consequences of a solid inner core on Mercury's spin configuration}.
\newblock {\em Icarus}, {\bf 264}, 443--455.
\newblock doi: 10.1016/j.icarus.2015.09.024.

\bibitem[\protect\citename{{Peplowski} {et~al.}, }2012]{Peplowski2012}
{Peplowski}, P.~N., {Lawrence}, D.~J., {Rhodes}, E.~A., {Sprague}, A.~L.,
  {McCoy}, T.~J., {Denevi}, B.~W., {Evans}, L.~G., {Head}, J.~W., {Nittler},
  L.~R., {Solomon}, S.~C., {Stockstill-Cahill}, K.~R. and {Weider}, S.~Z.
  (2012).
\newblock {Variations in the abundances of potassium and thorium on the surface
  of Mercury: Results from the MESSENGER Gamma-Ray Spectrometer}.
\newblock {\em J. Geophys. Res.}, {\bf 117}, E00L04.
\newblock doi: 10.1029/2012JE004141.

\bibitem[\protect\citename{{Perry} {et~al.}, }2015]{perr15}
{Perry}, M.~E., {Neumann}, G.~A., {Phillips}, R.~J., {Barnouin}, O.~S.,
  {Ernst}, C.~M., {Kahan}, D.~S., {Solomon}, S.~C., {Zuber}, M.~T., {Smith},
  D.~E., {Hauck}, {S.~A., II}, {Peale}, S.~J., {Margot}, J.~L., {Mazarico}, E.,
  {Johnson}, C.~L., {Gaskell}, R.~W., {Roberts}, J.~H., {McNutt}, R.~L., Jr.
  and {Oberst}, J. (2015).
\newblock {The low-degree shape of Mercury}.
\newblock {\em Geophys. Res. Lett.}, {\bf 42}, 6951--6958.
\newblock doi: 10.1002/2015GL065101.

\bibitem[\protect\citename{Pettengill and Dyce, }1965]{pett65}
Pettengill, G.~H. and Dyce, R.~B. (1965).
\newblock A radar determination of the rotation of the planet Mercury.
\newblock {\em Nature}, {\bf 206}, 1240.
\newblock doi: 10.1038/2061240a0.

\bibitem[\protect\citename{{Pfyffer} {et~al.}, }2011]{pfyf11}
{Pfyffer}, G., {Van Hoolst}, T. and {Dehant}, V. (2011).
\newblock Librations and obliquity of Mercury from the BepiColombo
  radio-science and camera experiments.
\newblock {\em Planet. Space Sci.}, {\bf 59}, 848--861.
\newblock doi: 10.1016/j.pss.2011.03.017.

\bibitem[\protect\citename{{Poirier}, }2000]{poir00}
{Poirier}, J.-P. (2000).
\newblock {\em {Introduction to the physics of the Earth}}. 2 edn.
\newblock {Cambridge University Press}.

\bibitem[\protect\citename{{Rambaux} {et~al.}, }2007]{ramb07}
{Rambaux}, N., {Van Hoolst}, T., {Dehant}, V. and {Bois}, E. (2007).
\newblock {Inertial core--mantle coupling and libration of Mercury}.
\newblock {\em Astron. Astrophys.}, {\bf 468}, 711--719.
\newblock doi: 10.1051/0004-6361:20053974.

\bibitem[\protect\citename{{Riner} {et~al.}, }2008]{rine08}
{Riner}, M.~A., {Bina}, C.~R., {Robinson}, M.~S. and {Desch}, S.~J. (2008).
\newblock Internal structure of Mercury: Implications of a molten core.
\newblock {\em J. Geophys. Res.}, {\bf 113}, E08013.
\newblock doi: 10.1029/2007JE002993.

\bibitem[\protect\citename{{Rivoldini} and {Van Hoolst}, }2013]{rivo13}
{Rivoldini}, A. and {Van Hoolst}, T. (2013).
\newblock The interior structure of Mercury constrained by the low-degree
  gravity field and the rotation of Mercury.
\newblock {\em Earth Planet. Sci. Lett.}, {\bf 377}, 62--72.
\newblock doi: 10.1016/j.epsl.2013.07.021.

\bibitem[\protect\citename{{Rivoldini} {et~al.}, }2009]{rivo09}
{Rivoldini}, A., {Van Hoolst}, T. and {Verhoeven}, O. (2009).
\newblock The interior structure of Mercury and its core sulfur content.
\newblock {\em Icarus}, {\bf 201}, 12--30.
\newblock doi: 10.1016/j.icarus.2008.12.020.

\bibitem[\protect\citename{{Robinson} and {Taylor}, }2001]{robi01}
{Robinson}, M.~S. and {Taylor}, G.~J. (2001).
\newblock {Ferrous oxide in Mercury's crust and mantle}.
\newblock {\em Meteorit. Planet. Sci.}, {\bf 36}, 841--847.
\newblock doi: 10.1111/j.1945-5100.2001.tb01921.x.

\bibitem[\protect\citename{{Sabadini} and {Vermeersen}, }2004]{Sabadini2004}
{Sabadini}, R. and {Vermeersen}, B. (2004).
\newblock {\em Global Dynamics of the Earth: Applications of Normal Mode
  Relaxation Theory to Solid-Earth Geophysics}.
\newblock Dordrecht, The Netherlands: Kluwer Academic Publishers.

\bibitem[\protect\citename{{Sanloup} and {Fei}, }2004]{sanl04}
{Sanloup}, C. and {Fei}, Y. (2004).
\newblock {Closure of the Fe-S-Si liquid miscibility gap at high pressure}.
\newblock {\em Phys. Earth Planet. Inter.}, {\bf 147}, 57--65.
\newblock doi: 10.1016/j.pepi.2004.06.008.

\bibitem[\protect\citename{{Schubert} {et~al.}, }1988]{schu88}
{Schubert}, G., {Ross}, M.~N., {Stevenson}, D.~J. and {Spohn}, T. (1988).
\newblock Mercury's thermal history and the generation of its magnetic field.
  In {\em Mercury}, ed. {Vilas}, F., {Chapman}, C.~R. and {Matthews}, M.~S.
\newblock {pp.  429--460}.
\newblock Tucson, AZ: University of Arizona Press.

\bibitem[\protect\citename{Schubert {et~al.}, }2001]{schu01}
Schubert, G., Turcotte, D.L. and Olson, P. (2001).
\newblock {\em Mantle Convection in the Earth and Planets}.
\newblock Cambridge University Press.

\bibitem[\protect\citename{{Siegfried} and {Solomon}, }1974]{sieg74}
{Siegfried}, R.~W., II and {Solomon}, S.~C. (1974).
\newblock Mercury: Internal structure and thermal evolution.
\newblock {\em Icarus}, {\bf 23}, 192--205.
\newblock doi: 10.1016/0019-1035(74)90005-0.

\bibitem[\protect\citename{Siivola and Schmid, }2007]{Siivola2007}
Siivola, J. and Schmid, R. (2007).
\newblock List of mineral abbreviations, {R}ecommendations by the {I}{U}{G}{S}
  {S}ubcommission on the {S}ystematics of {M}etamorphic {R}ocks.
\newblock {\em Electronic Source:
  http://www.bgs.ac.uk/scmr/docs/papers/paper\_12.pdf}.

\bibitem[\protect\citename{{Smith} {et~al.}, }2010]{smit10}
{Smith}, D.~E., {Zuber}, M.~T., {Phillips}, R.~J., {Solomon}, S.~C., {Neumann},
  G.~A., {Lemoine}, F.~G., {Peale}, S.~J., {Margot}, {J.~L.}, {Torrence},
  M.~H., {Talpe}, M.~J., {Head}, J.~W., {Hauck}, {S.~A., II}, {Johnson}, C.~L.,
  {Perry}, M.~E., {Barnouin}, O.~S., {McNutt}, R.~L., Jr. and {Oberst}, J.
  (2010).
\newblock The equatorial shape and gravity field of Mercury from MESSENGER
  flybys 1 and 2.
\newblock {\em Icarus}, {\bf 209}, 88--100.
\newblock doi: 10.1016/j.icarus.2010.04.007.

\bibitem[\protect\citename{{Smith} {et~al.}, }2012]{smit12}
{Smith}, D.~E., {Zuber}, M.~T., {Phillips}, R.~J., {Solomon}, S.~C., {Hauck},
  {S.~A., II}, {Lemoine}, F.~G., {Mazarico}, E., {Neumann}, G.~A., {Peale},
  S.~J., {Margot}, J.~L., {Johnson}, C.~L., {Torrence}, M.~H., {Perry}, M.~E.,
  {Rowlands}, D.~D., {Goossens}, S., {Head}, J.~W. and {Taylor}, A.~H. (2012).
\newblock Gravity field and internal structure of Mercury from MESSENGER.
\newblock {\em Science}, {\bf 336}, 214--217.
\newblock doi: 10.1126/science.1218809.

\bibitem[\protect\citename{Smyth and McCormick, }1995]{smyt95}
Smyth, J.~R. and McCormick, T.~C. (1995).
\newblock Crystallographic data for minerals. In {\em Mineral Physics and
  Crystallography: A Handbook of Physical Constants}, ed. Ahrens, T.~J.
\newblock {pp.  1--17}.
\newblock Washington, D.C.: American Geophysical Union.

\bibitem[\protect\citename{{Solomon} {et~al.}, }2001]{solo01}
{Solomon}, S.~C., {McNutt}, R.~L., Jr., {Gold}, R.~E., {Acu{\~n}a}, M.~H.,
  {Baker}, D.~N., {Boynton}, W.~V., {Chapman}, C.~R., {Cheng}, A.~F.,
  {Gloeckler}, G., {Head}, III, J.~W., {Krimigis}, S.~M., {McClintock}, W.~E.,
  {Murchie}, S.~L., {Peale}, S.~J., {Phillips}, R.~J., {Robinson}, M.~S.,
  {Slavin}, J.~A., {Smith}, D.~E., {Strom}, R.~G., {Trombka}, J.~I. and
  {Zuber}, M.~T. (2001).
\newblock {The MESSENGER mission to Mercury: Scientific objectives and
  implementation}.
\newblock {\em Planet. Space Sci.}, {\bf 49}, 1445--1465.
\newblock doi: 10.1016/S0032-0633(01)00085-X.

\bibitem[\protect\citename{{Spohn} {et~al.}, }2001]{spoh01}
{Spohn}, T., {Sohl}, F., {Wieczerkowski}, K. and {Conzelmann}, V. (2001).
\newblock The interior structure of Mercury: What we know, what we expect from
  BepiColombo.
\newblock {\em Planet. Space Sci.}, {\bf 49}, 1561--1570.
\newblock doi: 10.1016/S0032-0633(01)00093-9.

\bibitem[\protect\citename{{Stark} {et~al.}, }2015a]{star15grl}
{Stark}, A., Oberst, J., Preusker, F., Peale, S.~J., Margot, J.~L., Phillips,
  R.~J., Neumann, G.~A., Smith, D.~E., Zuber, M.~T. and Solomon, S.~C. (2015a).
\newblock {First MESSENGER orbital observations of Mercury's librations}.
\newblock {\em Geophys. Res. Lett.}, {\bf 42}, 7881--7889.
\newblock doi: 10.1002/2015GL065152.

\bibitem[\protect\citename{{Stark} {et~al.}, }2015b]{star15cmda}
{Stark}, A., {Oberst}, J. and {Hussmann}, H. (2015b).
\newblock {Mercury's resonant rotation from secular orbital elements}.
\newblock {\em Celest. Mech. Dyn. Astron.}, {\bf 123}, 263--277.
\newblock doi: 10.1007/s10569-015-9633-4.

\bibitem[\protect\citename{{Stark} {et~al.}, }2015c]{star15pss}
{Stark}, A., {Oberst}, J., {Preusker}, F., {Gwinner}, K., {Peale}, S.~J.,
  {Margot}, J.~L., {Phillips}, R.~J., {Zuber}, M.~T. and {Solomon}, S.~C.
  (2015c).
\newblock {Mercury's rotational parameters from MESSENGER image and laser
  altimeter data: A feasibility study}.
\newblock {\em Planet. Space Sci.}, {\bf 117}, 64--72.
\newblock doi: 10.1016/j.pss.2015.05.006.

\bibitem[\protect\citename{{Stephenson}, }1976]{step76}
{Stephenson}, A. (1976).
\newblock Crustal remanence and the magnetic moment of Mercury.
\newblock {\em Earth Planet. Sci. Lett.}, {\bf 28}, 454--458.
\newblock doi: 10.1016/0012-821X(76)90206-5.

\bibitem[\protect\citename{{Stevenson}, }1983]{stev83rpp}
{Stevenson}, D.~J. (1983).
\newblock Planetary magnetic fields.
\newblock {\em Rep. Prog. Phys.}, {\bf 46}, 555--620.
\newblock doi: 10.1016/S0012-821X(02)01126-3.

\bibitem[\protect\citename{{Stevenson}, }2010]{stev10}
{Stevenson}, D.~J. (2010).
\newblock Planetary magnetic fields: Achievements and prospects.
\newblock {\em Space Sci. Rev.}, {\bf 152}, 651--664.
\newblock doi: 10.1007/978-1-4419-5901-0\_20.

\bibitem[\protect\citename{{Stevenson} {et~al.}, }1983]{stev83icar}
{Stevenson}, D.~J., {Spohn}, T. and {Schubert}, G. (1983).
\newblock Magnetism and thermal evolution of the terrestrial planets.
\newblock {\em Icarus}, {\bf 54}, 466--489.
\newblock doi: 10.1016/0019-1035(83)90241-5.

\bibitem[\protect\citename{Taylor and Scott, }2003]{Taylor2003}
Taylor, G.J. and Scott, E.R.D. (2003).
\newblock Mercury. In {\em Treatise on Geochemistry}, ed. Holland, H.~D. and
  Turekian, K.~K.
\newblock {pp.  477--485}.
\newblock Oxford: Pergamon.

\bibitem[\protect\citename{{Tosi} {et~al.}, }2013]{Tosi2013}
{Tosi}, N., {Grott}, M., {Plesa}, A.-C. and {Breuer}, D. (2013).
\newblock {Thermochemical evolution of Mercury's interior}.
\newblock {\em J. Geophys. Res. Planets}, {\bf 118}, 2474--2487.
\newblock doi: 10.1002/jgre.20168.

\bibitem[\protect\citename{{Tosi} {et~al.}, }2015]{tosi15}
{Tosi}, N., {{\v C}adek}, O., {B{\v e}hounkov{\'a}}, M., {K{\'a}{\AA}ov{\'a}},
  M., {Plesa}, A.-C., {Grott}, M., {Breuer}, D., {Padovan}, S. and {Wieczorek},
  M.~A. (2015).
\newblock Mercury's low-degree geoid and topography controlled by
  insolation-driven elastic deformation.
\newblock {\em Geophys. Res. Lett.}, {\bf 42}, 7327--7335.
\newblock doi: 10.1002/2015GL065314.

\bibitem[\protect\citename{{Turcotte} and {Schubert}, }2002]{Turcotte2002}
{Turcotte}, D.~L. and {Schubert}, G. (2002).
\newblock {\em Geodynamics}. $2^{\textrm{nd}}$ edn.
\newblock Cambridge, UK: Cambridge University Press.

\bibitem[\protect\citename{{Van Hoolst} and {Jacobs}, }2003]{vanh03}
{Van Hoolst}, T. and {Jacobs}, C. (2003).
\newblock {Mercury's tides and interior structure}.
\newblock {\em J. Geophys. Res.}, {\bf 108}, 5121--5136.
\newblock doi: 10.1029/2003JE002126.

\bibitem[\protect\citename{{Van Hoolst} {et~al.}, }2007]{vanh07}
{Van Hoolst}, T., {Sohl}, F., {Holin}, I., {Verhoeven}, O., {Dehant}, V. and
  {Spohn}, T. (2007).
\newblock {Mercury's interior structure, rotation, and tides}.
\newblock {\em Space Sci. Rev.}, {\bf 132}, 203--227.
\newblock doi: 10.1007/s11214-007-9202-6.

\bibitem[\protect\citename{{Van Hoolst} {et~al.}, }2012]{vanh12}
{Van Hoolst}, T., {Rivoldini}, A., {Baland}, R.-M. and {Yseboodt}, M. (2012).
\newblock The effect of tides and an inner core on the forced longitudinal
  libration of Mercury.
\newblock {\em Earth Planet. Sci. Lett.}, {\bf 333}, 83--90.
\newblock doi: 10.1016/j.epsl.2012.04.014.

\bibitem[\protect\citename{{Vander Kaaden} and {McCubbin},
  }2016]{VanderKaaden2016}
{Vander Kaaden}, K.~E. and {McCubbin}, F.~M. (2016).
\newblock {The origin of boninites on Mercury: An experimental study of the
  northern volcanic plains lavas}.
\newblock {\em {Geochim. Cosmochim. Acta}}, {\bf 173}, 246--263.
\newblock doi: 10.1016/j.gca.2015.10.016.

\bibitem[\protect\citename{{Veasey} and {Dumberry}, }2011]{veas11}
{Veasey}, M. and {Dumberry}, M. (2011).
\newblock {The influence of Mercury's inner core on its physical libration}.
\newblock {\em Icarus}, {\bf 214}, 265--274.
\newblock doi: 10.1016/j.icarus.2011.04.025.

\bibitem[\protect\citename{{Verma} and {Margot}, }2016]{verm16}
{Verma}, A.~K. and {Margot}, J.~L. (2016).
\newblock {Mercury's gravity, tides, and spin from MESSENGER radio science
  data}.
\newblock {\em J. Geophys. Res. Planets}, {\bf 121}, 1627--1640.
\newblock doi: 10.1002/2016JE005037.

\bibitem[\protect\citename{{Wasson}, }1988]{wass88}
{Wasson}, J.~T. (1988).
\newblock The building stones of the planets. In {\em Mercury}, ed. {Vilas},
  F., {Chapman}, C.~R. and {Matthews}, M.~S.
\newblock {pp.  622--650}.
\newblock Tucson, AZ: University of Arizona Press.

\bibitem[\protect\citename{{Watt} {et~al.}, }1976]{watt76}
{Watt}, J.~P., {Davies}, G.~F. and {O'Connell}, R.~J. (1976).
\newblock The elastic properties of composite materials.
\newblock {\em Rev. Geophys. Space Phys.}, {\bf 14}, 541--563.
\newblock doi: 10.1029/RG014i004p00541.

\bibitem[\protect\citename{{Weider} {et~al.}, }2014]{weid14}
{Weider}, S.~Z., {Nittler}, L.~R., {Starr}, R.~D., {McCoy}, T.~J. and
  {Solomon}, S.~C. (2014).
\newblock {Variations in the abundance of iron on Mercury's surface from
  MESSENGER X-Ray Spectrometer observations}.
\newblock {\em Icarus}, {\bf 235}, 170--186.
\newblock doi: 10.1016/j.icarus.2014.03.002.

\bibitem[\protect\citename{{Weider} {et~al.}, }2015]{weid15}
{Weider}, S.~Z., {Nittler}, L.~R., {Starr}, R.~D., {Crapster-Pregont}, E.~J.,
  {Peplowski}, P.~N., {Denevi}, B.~W., {Head}, J.~W., {Byrne}, P.~K., {Hauck},
  {S.~A., II}, {Ebel}, D.~S. and {Solomon}, S.~C. (2015).
\newblock {Evidence for geochemical terranes on Mercury: Global mapping of
  major elements with MESSENGER's X-Ray Spectrometer}.
\newblock {\em Earth Planet. Sci. Lett.}, {\bf 416}, 109--120.
\newblock doi: 10.1016/j.epsl.2015.01.023.

\bibitem[\protect\citename{{Wieczorek} {et~al.}, }2012]{wiec12}
{Wieczorek}, M.~A., {Correia}, A.~C.~M., {Le Feuvre}, M., {Laskar}, J. and
  {Rambaux}, N. (2012).
\newblock Mercury's spin--orbit resonance explained by initial retrograde and
  subsequent synchronous rotation.
\newblock {\em Nature Geosci.}, {\bf 5}, 18--21.
\newblock doi: 10.1038/ngeo1350.

\bibitem[\protect\citename{{Wieczorek} {et~al.}, }2013]{wiec13}
{Wieczorek}, M.~A., {Neumann}, G.~A., {Nimmo}, F., {Kiefer}, W.~S., {Taylor},
  G.~J., {Melosh}, H.~J., {Phillips}, R.~J., {Solomon}, S.~C., {Andrews-Hanna},
  J.~C., {Asmar}, S.~W., {Konopliv}, A.~S., {Lemoine}, F.~G., {Smith}, D.~E.,
  {Watkins}, M.~M., {Williams}, J.~G. and {Zuber}, M.~T. (2013).
\newblock {The crust of the Moon as seen by GRAIL}.
\newblock {\em Science}, {\bf 339}, 671--675.
\newblock doi: 10.1126/science.1231530.

\bibitem[\protect\citename{{Williams}, }1994]{will94}
{Williams}, J.~G. (1994).
\newblock {Contributions to the Earth's obliquity rate, precession, and
  nutation}.
\newblock {\em Astron. J.}, {\bf 108}, 711--724.
\newblock doi: 10.1086/117108.

\bibitem[\protect\citename{{Williams} {et~al.}, }1996]{will96}
{Williams}, J.~G., {Newhall}, X.~X. and {Dickey}, J.~O. (1996).
\newblock {Lunar moments, tides, orientation, and coordinate frames}.
\newblock {\em Planet. Space Sci.}, {\bf 44}, 1077--1080.
\newblock doi: 10.1016/0032-0633(95)00154-9.

\bibitem[\protect\citename{{Williams} {et~al.}, }2001]{will01moon}
{Williams}, J.~G., {Boggs}, D.~H., {Yoder}, C.~F., {Ratcliff}, J.~T. and
  {Dickey}, J.~O. (2001).
\newblock Lunar rotational dissipation in solid body and molten core.
\newblock {\em J. Geophys. Res.}, {\bf 106}, 27933--27968.
\newblock doi: 10.1029/2000JE001396.

\bibitem[\protect\citename{{Wolf}, }1994]{Wolf1994}
{Wolf}, D. (1994).
\newblock {Lam\'e's problem of gravitational viscoelasticity: The isochemical,
  incompressible planet}.
\newblock {\em Geophys. J. Int.}, {\bf 116}, 321--348.
\newblock doi: 10.1111/j.1365--246X.1994.tb01801.x.

\bibitem[\protect\citename{{Wu} {et~al.}, }1995]{wu95}
{Wu}, X., {Bender}, P.~L. and {Rosborough}, G.~W. (1995).
\newblock Probing the interior structure of Mercury from an orbiter plus single
  lander.
\newblock {\em J. Geophys. Res.}, {\bf 100}, 1515--1525.
\newblock doi: 10.1029/94JE02833.

\bibitem[\protect\citename{{Yoder}, }1981]{yode81}
{Yoder}, C.~F. (1981).
\newblock The free librations of a dissipative moon.
\newblock {\em Phil. Trans. Roy. Soc. A}, {\bf 303}, 327--338.
\newblock doi: 10.1098/rsta.1981.0206.

\bibitem[\protect\citename{{Yoder} {et~al.}, }2003]{Yoder2003}
{Yoder}, C.~F., {Konopliv}, A.~S., {Yuan}, D.~N., {Standish}, E.~M. and
  {Folkner}, W.~M. (2003).
\newblock {Fluid core size of Mars from detection of the solar tide}.
\newblock {\em Science}, {\bf 300}, 299--303.
\newblock doi: 10.1126/science.1079645.

\bibitem[\protect\citename{{Yseboodt} and {Margot}, }2006]{yseb06}
{Yseboodt}, M. and {Margot}, J.~L. (2006).
\newblock Evolution of Mercury's obliquity.
\newblock {\em Icarus}, {\bf 181}, 327--337.
\newblock doi: 10.1016/j.icarus.2005.11.024.

\bibitem[\protect\citename{{Yseboodt} {et~al.}, }2010]{yseb10}
{Yseboodt}, M., {Margot}, {J.~L.} and {Peale}, S.~J. (2010).
\newblock Analytical model of the long-period forced longitude librations of
  Mercury.
\newblock {\em Icarus}, {\bf 207}, 536--544.
\newblock doi: 10.1016/j.icarus.2009.12.020.

\bibitem[\protect\citename{{Yseboodt} {et~al.}, }2013]{yseb13}
{Yseboodt}, M., {Rivoldini}, A., {Van Hoolst}, T. and {Dumberry}, M. (2013).
\newblock Influence of an inner core on the long-period forced librations of
  Mercury.
\newblock {\em Icarus}, {\bf 226}, 41--51.
\newblock doi: 10.1016/j.icarus.2013.05.011.

\bibitem[\protect\citename{{Zolotov} {et~al.}, }2013]{zolo13}
{Zolotov}, M.~Yu., {Sprague}, A.~L., {Hauck}, {S.~A., II}, {Nittler}, L.~R.,
  {Solomon}, S.~C. and {Weider}, S.~Z. (2013).
\newblock {The redox state, FeO content, and origin of sulfur-rich magmas on
  Mercury}.
\newblock {\em J. Geophys. Res. Planets}, {\bf 118}, 138--146.
\newblock doi: 10.1029/2012JE004274.

\bibitem[\protect\citename{{Zuber} {et~al.}, }2012]{zube12}
{Zuber}, M.~T., {Smith}, D.~E., {Phillips}, R.~J., {Solomon}, S.~C., {Neumann},
  G.~A., {Hauck}, {S.~A., II}, {Peale}, S.~J., {Barnouin}, O.~S., {Head},
  J.~W., {Johnson}, C.~L., {Lemoine}, F.~G., {Mazarico}, E., {Sun}, X.,
  {Torrence}, M.~H., {Freed}, A.~M., {Klimczak}, C., {Margot}, J.~L., {Oberst},
  J., {Perry}, M.~E., {McNutt}, R.~L., Jr., {Balcerski}, J.~A., {Michel}, N.,
  {Talpe}, M.~J. and {Yang}, D. (2012).
\newblock {Topography of the northern hemisphere of Mercury from MESSENGER
  laser altimetry}.
\newblock {\em Science}, {\bf 336}, 217--220.
\newblock doi: 10.1126/science.1218805.

\end{thebibliography}
\bibliographystyle{cup}

\end{document}